\documentclass[12pt]{article}
\usepackage{amssymb,amsmath}
\usepackage{graphicx}
\usepackage{ytableau}
\usepackage{comment}
\usepackage[vcentermath]{youngtab}
\usepackage{hyperref}
\usepackage{tikz}

\DeclareMathOperator*{\Tr}{{\rm Tr}}

\numberwithin{equation}{section}

\begin{document}

\thispagestyle{empty}
\begin{flushright}
RUP-25-4
\end{flushright}
\vskip1cm
\begin{center}
{\bf {\LARGE $\mathcal{N}=4$ line defect correlators of type BCD}}

\vskip1.5cm

Yasuyuki Hatsuda$^a$\footnote{yhatsuda@rikkyo.ac.jp}, 

\bigskip
$^a$
{\it Department of Physics, Rikkyo University, \\
Toshima, Tokyo 171-8501, Japan
}

\bigskip
\bigskip

Hai Lin$^b$\footnote{hailinhl@seu.edu.cn}

\bigskip
\bigskip

and 

\bigskip
\bigskip

Tadashi Okazaki$^b$\footnote{tokazaki@seu.edu.cn}

\bigskip
$^b$
{\it Shing-Tung Yau Center of Southeast University,\\
Yifu Architecture Building, No.2 Sipailou, Xuanwu district, \\
Nanjing, Jiangsu, 210096, China
}

\end{center}

\vskip1.25cm
\begin{abstract}
We analyze the line defect correlation functions 
decorating the supersymmetric indices of $\mathcal{N}=4$ super Yang-Mills theories with orthogonal and symplectic gauge groups. 
We obtain exact closed-form expressions for the correlators by means of combinatorial methods associated with the group characters, 
including the generalized Murnaghan-Nakayama rules. 
The exact results for the large $N$ correlators lead to 
the gravity indices for the gravity dual spectrum of the fluctuations of the fundamental string wrapping $AdS_2$ 
and those of the fat string, the D5-brane wrapping $AdS_2$ $\times$ $\mathbb{RP}^4$. 
\end{abstract}

\newpage
\setcounter{tocdepth}{3}
\tableofcontents

\section{Introduction and conclusion}
Wilson line operators and 't Hooft line operators describe 
probe point-like electric and magnetic charged particles and insert electrically and magnetically charged sources respectively. 
In $\mathcal{N}=4$ super Yang-Mills (SYM) theory, 
the Wilson line is labeled by a representation $\mathcal{R}$ of gauge group $G$ 
or equivalently a weight of the associated Lie algebra $\mathfrak{g}$ of $G$, 
while the 't Hooft line is characterized by an element of the cocharacter lattice of $G$ 
or equivalently a weight of $\mathfrak{g}^{\vee}$, the Langlands dual (GNO dual \cite{Goddard:1976qe}) of $\mathfrak{g}$. 
More generally, there exist dyonic line operators that carry both electric and magnetic charges 
which are labeled by a pair of weights in $\mathfrak{g}$ and $\mathfrak{g}^{\vee}$. 

It is conjectured that these line operators map to each other under the action of S-duality \cite{Kapustin:2005py}. 
Since the choice of the line operators can distinguish the gauge theories based on the same Lie algebra $\mathfrak{g}$, 
there exist non-trivial S-duality orbits of $\mathcal{N}=4$ SYM theories \cite{Aharony:2013hda}. 

In addition, as $\mathcal{N}=4$ SYM theories can be realized in Type IIB superstring theory, 
these line operators are conjectured to be holographically equivalent to certain brane configurations in Type IIB string theory. 
For example, the Wilson line operators transforming in the fundamental, symmetric and antisymmetric representations are dual to a fundamental string \cite{Maldacena:1998im,Rey:1998ik}, 
a D3-brane \cite{Drukker:2005kx,Gomis:2006sb,Gomis:2006im,Rodriguez-Gomez:2006fmx,Yamaguchi:2007ps} 
and a D5-brane \cite{Yamaguchi:2006tq,Gomis:2006sb,Rodriguez-Gomez:2006fmx,Hartnoll:2006hr} respectively. 
In the case with orthogonal gauge groups, there exist the Wilson line operators in the spinor representation. 
It is conjectured to be dual to a \textit{fat string} as a D5-brane wrapping $AdS_2$ $\times$ $\mathbb{RP}^4$ in Type IIB string theory on $AdS_5\times \mathbb{RP}^5$ \cite{Witten:1998xy} 
(see \cite{Fiol:2014fla,Giombi:2020kvo,Okuyama:2022lbn,Zhang:2023yus} for more recent analysis). 
The structure of the S-duality orbits in the presence of line operators is also examined in Type IIB string theory \cite{Bergman:2022otk}. 

In order to test the S-duality relation and the gravity duals associated with the line operators in $\mathcal{N}=4$ SYM theories, 
there is a useful observable, which can be obtained by decorating the supersymmetric indices with the line defect operators. 
While the ordinary supersymmetric indices can be defined as certain supersymmetric partition functions on $S^1$ $\times$ $S^3$, 
such correlators can be obtained by introducing the line operators that wrap the $S^1$ and localize at points along a great circle on the $S^3$ 
(see e.g. \cite{Dimofte:2011py,Gang:2012yr,Drukker:2015spa,Cordova:2016uwk,Hatsuda:2023iwi}). 
The line defect correlators can be also defined as a trace over the modified Hilbert space on the $S^3$ due to the inserted line defects 
so that they count the BPS local operators living at the junctions of the line defects. 
The line defect correlators are independent of the continuous coupling constant and the distances between the line defects. 
They can be used to test S-duality of the line defects as they should be equal for pairs of the S-dual line defects. 
In fact, S-duality between the Wilson line and the 't Hooft line for the minuscule representation is numerically tested in \cite{Gang:2012yr}. 
Furthermore, according to the holography, they can encode the gravity spectra of quantum fluctuations around the dual supergravity backgrounds. 
For $G=U(N)$, the large $N$ line defect indices for the Wilson line operators in various representations are obtained from the gauge theory 
\cite{Gang:2012yr,Drukker:2015spa,Hatsuda:2023iwi,Hatsuda:2023imp,Hatsuda:2023iof}. 

In this paper we investigate exact closed-form expressions for the line defect correlation functions for 4d $\mathcal{N}=4$ SYM theories with orthogonal and symplectic gauge groups that decorate the supersymmetric indices. 
The theories can be realized in Type IIB string theory as world-volume theories for a stack of $N$ D3-branes in the background of an O3-plane. 
We obtain several exact formulae for the line defect correlators by means of combinatorial method associated with the group characters, 
including the generalized Murnaghan-Nakayama rules which expand the group characters in terms of the power sum symmetric functions. 
The correlators can enumerate the BPS local operators as their expansion coefficients encode the number of BPS local operators living at the junction of line operators. 
For finite rank gauge theories, the half-BPS limits of the normalized multi-point functions have interesting combinatorial interpretation, 
including the multiplicities of certain representations in the tensor powers of the representations in which the inserted Wilson lines transform. 
In the presence of the O3$^-$-plane, the theory has the disconnected gauge group $O(2N)$ \cite{Garcia-Etxebarria:2015wns,Aharony:2016kai}, 
for which very little attention has been paid to the line operators. 
We compute the line defect correlators to obtain exact results of the half-BPS limits. 
We also demonstrate that S-duality of Wilson line in the fundamental representation 
and the 't Hooft line of the magnetic charge $B=(1,0^{N-1})$ can be generalized to the disconnected $O(2N)$ gauge theories 
by checking the precise matching of their correlators. 

On the other hand, in the large $N$ limits, the line defect correlators can capture the spectra of the excitations on the holographically dual $AdS_2$ geometries 
embedded in $AdS_5\times \mathbb{RP}^5$ \cite{Witten:1998xy}. 
We find that the large $N$ normalized two-point function of the fundamental Wilson lines is equivalent to that for unitary gauge theory as computed in \cite{Gang:2012yr,Hatsuda:2023iwi}. 
This indicates that the BPS spectrum of the fluctuation modes of the fundamental string wrapping $AdS_2$ in $AdS_5$ $\times$ $\mathbb{RP}^5$ 
is comparable with that of the fundamental string wrapped on $AdS_2$ in $AdS_5$ $\times$ $S^5$ which is analyzed in the Green-Schwarz formalism \cite{Drukker:2000ep}. 
Also, we find the exact forms of the large $N$ normalized two-point function of the spinor Wilson lines for orthogonal gauge theories 
that is expected to encode the fluctuation modes of the D5-brane wrapping $AdS_2\times \mathbb{RP}^4$. 
In the half-BPS limit, it precisely coincides with the generating function for the partitions into distinct parts, 
while the large $N$ normalized two-point function of the antisymmetric Wilson lines for $U(N)$ gauge theory 
associated with the D5-brane wrapping $AdS_2\times S^4$ agrees with the generating function for the partitions \cite{Hatsuda:2023iwi}. 
We also discuss another approach to the half-BPS line defect index for the spinor Wilson lines 
by analyzing quadratic fluctuations of the D5-brane action around the classical solution in a similar manner as \cite{Faraggi:2011bb,Faraggi:2011ge}. 

\subsection{Structure}
The organization of the paper is as follows. 
In section \ref{sec_branesetup} we review the brane setup of line operators in $\mathcal{N}=4$ SYM theories 
of orthogonal or symplectic gauge groups in Type IIB string theory and the associated gravity dual geometries. 
In section \ref{sec_lineindex} we review the line defect correlation functions that decorate supersymmetric indices for $\mathcal{N}=4$ SYM theories. 
In section \ref{sec_Btype}, section \ref{sec_Ctype} and section \ref{sec_Dtype} 
we examine the line defect correlation functions for $\mathcal{N}=4$ SYM theories of gauge groups based on 
the Lie algebras $\mathfrak{so}(2N+1)$, $\mathfrak{usp}(2N)$ and $\mathfrak{so}(2N)$ respectively.  
In particular, we find exact results for the two-point functions of the fundamental Wilson lines and those of the spinor Wilson lines for each of gauge groups in the half-BPS limit. 
In section \ref{sec_largeN} the exact closed-form of the large $N$ limits of the line defect correlators are derived by means of the combinatorial method. 
Consequently, we obtain the single particle gravity indices for the holographically dual geometries 
including the fundamental string and the fat string propagating in $AdS_5\times \mathbb{RP}^5$ of Type IIB string theory. 
In Appendix \ref{app:character} several character formulae which are useful to compute the line defect correlators are summarized. 

\subsection{Future works}

\begin{itemize}

\item 
It would be interesting to explore the exact closed-form expressions 
for the line defect correlators of more general line operators, including the Wilson lines in higher rank representations and the 't Hooft lines associated with the non-minuscule representations for orthogonal and symplectic gauge groups. In the case with unitary gauge theories, the exact formulae can be derived by means of the Fermi-gas method \cite{Hatsuda:2023iwi} based on the Frobenius determinant formula \cite{Frobenius:1882uber} or Fay's trisecant identity \cite{MR0335789}. While the Fermi-gas approach for the unflavored Schur indices of the orthogonal and symplectic gauge theories were examined in \cite{Du:2023kfu}, it would be nice to find the determinant formula of BCD-type that can be applicable to the flavored indices of orthogonal or symplectic gauge theories.\footnote{We appreciate Masatoshi Noumi for useful discussion on the related topics. }
For the non-minuscule representations the correlators of the 't Hooft lines will be generalized by taking account into the monopole bubbling index 
(see \cite{Hayashi:2020ofu} for the monopole index for orthogonal and symplectic gauge theories). 

\item 
The large $N$ limits of the line defect correlators of the charged Wilson lines in $\mathcal{N}=4$ $U(N)$ gauge theory follow the factorization property \cite{Hatsuda:2023iwi,Hatsuda:2023imp}. 
It enables us to compute various large $N$ line defect correlators of the Wilson lines. 
In particular, in the case with a quadratic area growth of boxes of the associated Young diagram, 
the gravity dual is the fully backreacted bubbling geometry. 
The degeneracy of the BPS states has a combinatorial interpretation in terms of the general necklace polynomial \cite{Hatsuda:2023iof}. 
It is tempting to generalize the results to the orientifold bubbling geometries dual to the orthogonal and symplectic gauge theories. 

\item 
The S-duality relation of the half-BPS boundary conditions and interfaces 
in $\mathcal{N}=4$ SYM theories of orthogonal and symplectic gauge groups are proposed in \cite{Gaiotto:2008ak} 
and they are confirmed as precise matching of the half-indices \cite{Hatsuda:2024lcc}. 
They can be further decorated by the line defect operators. 
We hope to report the detailed analysis of such line defect half-indices in the upcoming work. 

\item It is pointed out that the line defect correlators admit intriguing giant graviton expansions. 
See e.g. \cite{Imamura:2024lkw,Imamura:2024pgp,Beccaria:2024oif,Hatsuda:2024uwt,Imamura:2024zvw} and related references, 
for the cases of antisymmetric Wilson lines and symmetric Wilson lines in $\mathcal{N}=4$ $U(N)$ gauge theory. 
It would be good to investigate the giant graviton expansions for the line defect correlators in orthogonal and symplectic gauge theories in detail. 

\item In this paper, we have evaluated line defect correlators with various representations, which count the BPS local operators in the presence of line defects. 
On the other hand, there are direct methods which can label and construct the local operators using the combinatorial methods e.g. in 
\cite{Lewis-Brown:2018dje,deMelloKoch:2024sdf,Berenstein:2022srd}. 
It would be good to generalize these methods to incorporate the labeling of the local operators taking into account the gauge invariants in the presence of the line defects. 
We hope that we can simplify the enumeration of gauge-invariant operators in AdS/CFT and also bridge representation theory and algebraic combinatorics with various non-perturbative string theory dynamics. 

\end{itemize}

\section{Brane configuration}
\label{sec_branesetup}

\subsection{Brane construction}
Let us consider a stack of D3-branes together with one type of O3-planes which extend along the $0126$ directions in Type IIB string theory. 
The low-energy effective world-volume theories on D3-branes are identified with 4d $\mathcal{N}=4$ SYM theories of orthogonal or symplectic gauge group. 

There exist four types of O3-planes which are distinguished by two $\mathbb{Z}_2$-valued discrete fluxes, 
or discrete torsions $\theta_{RR}$ and $\theta_{NS}$ for Ramond-Ramond (RR) and Neveu-Schwarz (NS) $2$-forms. 
The $\widetilde{\textrm{O3}}^{-}$-plane has the non-trivial RR flux and $1/4$ unit of D3-brane charge. 
In this case one finds gauge theory based on the gauge algebra $\mathfrak{so}(2N+1)$. 
It is invariant under the $T$ transformation of $SL(2,\mathbb{Z})$ S-duality in Type IIB string theory 
but transforms into the O3$^+$-plane with non-trivial NS flux under the $S$ transformation. 
The theory of $N$ D3-branes in the presence of the O3$^+$-plane has gauge group associated with the Lie algebra $\mathfrak{usp}(2N)$. 
Under the $T$ transformation the O3$^+$-plane maps to the $\widetilde{\textrm{O3}}^+$-plane with non-trivial RR and NS fluxes. 
The theory of $N$ D3-branes with the $\widetilde{\textrm{O3}}^+$-plane is gauge theory with symplectic gauge group, however, it has a unit of theta-angle. 
The $\widetilde{\textrm{O3}}^+$-plane is invariant under the $S$ transformation. 
The O3$^-$-plane has no non-trivial discrete fluxes and it carries $-1/4$ unit of D3-brane charge. 
The theory of $N$ D3-branes in the background of the O3$^-$-plane has gauge group associated with the Lie algebra $\mathfrak{so}(2N)$. 
It is invariant under the $S$ transformation. 
To summarize, four distinct O3-planes and SYM theories with orthogonal or symplectic gauge group are given by
\begin{align}
\label{O3_SYM}
\begin{array}{c|c|c|c|c}
&SO(2N+1)&USp(2N)&O(2N)&USp(2N)'\\ \hline 
\theta_{RR}&1/2&0&0&1/2 \\
\theta_{NS}&0&1/2&0&1/2 \\
\textrm{D3-brane charge}&1/4&1/4&-1/4&1/4 \\
\textrm{orientifold}&\widetilde{\textrm{O3}}^{-}&\textrm{O3}^+&\textrm{O3}^{-}&\widetilde{\textrm{O3}}^+ \\
\textrm{$S$ operation}&\textrm{O3}^+&\widetilde{\textrm{O3}}^{-}&\textrm{O3}^{-}&\widetilde{\textrm{O3}}^+ \\
\end{array}.
\end{align}

When the fundamental string supported on the $09$ directions, 
the half-BPS electric Wilson line in the fundamental representation is introduced in $\mathcal{N}=4$ SYM theory \cite{Maldacena:1998im,Rey:1998ik}. 
When $k$ fundamental strings terminate on the $N$ D3-branes at one end and another D3-brane at the other, 
one obtains the Wilson line transforming in the rank-$k$ symmetric tensor product of the fundamental representation  
\cite{Drukker:2005kx,Gomis:2006sb,Gomis:2006im,Rodriguez-Gomez:2006fmx,Yamaguchi:2007ps}. 
We can introduce a D5-brane supported on the $045678$ directions without further breaking supersymmetry. 
The Wilson line transforming in the rank-$k$ antisymmetric tensor product of the fundamental representation can be realized 
when $k$ fundamental strings end on the $N$ D3-branes and on the D5-brane 
\cite{Yamaguchi:2006tq,Gomis:2006sb,Rodriguez-Gomez:2006fmx,Hartnoll:2006hr}. 
For orthogonal gauge theories there exist the Wilson line operator transforming in the spinor representation. 
There is no counterpart in the unitary or symplectic gauge theories. 
It is realized by a D5-brane which is conserved modulo two due to the topological restrictions from the vanishing $\theta_{NS}$ 
in the presence of O3$^-$- or $\widetilde{\textrm{O3}}^-$-plane \cite{Witten:1998xy}. 

The brane configuration of the Wilson lines in $\mathcal{N}=4$ orthogonal or symplectic SYM theories is summarized as follows: 
\begin{align}
\label{brane_conf}
\begin{array}{c|cccccccccc}
&0&1&2&3&4&5&6&7&8&9 \\ \hline
\textrm{$N$ D3}&\circ&\circ&\circ&&&&\circ&&&\\
\textrm{O3}&\circ&\circ&\circ&&&&\circ&&&\\
\textrm{F1}&\circ&&&&&&&&&\circ\\
\textrm{D5}&\circ&&&\circ&\circ&\circ&&\circ&\circ&\\
\end{array}. 
\end{align}
Here $\circ$ denotes the directions in which branes are extended. 
The configuration preserves $SO(1,2)$ $\times$ $SO(3)$ $\times$ $SO(5)$ global symmetry, 
the bosonic subgroup of the 1d $\mathcal{N}=8$ superconformal group $OSp(4^*|4)$.

Upon the $S$ transformation, 
the brane configuration (\ref{brane_conf}) turns into 
\begin{align}
\label{brane_conf2}
\begin{array}{c|cccccccccc}
&0&1&2&3&4&5&6&7&8&9 \\ \hline
\textrm{$N$ D3}&\circ&\circ&\circ&&&&\circ&&&\\
\textrm{O3}&\circ&\circ&\circ&&&&\circ&&&\\
\textrm{D1}&\circ&&&&&&&&&\circ\\
\textrm{NS5}&\circ&&&\circ&\circ&\circ&&\circ&\circ&\\
\end{array}, 
\end{align}
where the transformation of the O3-planes follow from (\ref{O3_SYM}). 
The S-dual brane configuration (\ref{brane_conf2}) realizes the magnetic 't Hooft line operators in $\mathcal{N}=4$ SYM theories. 
For example, the D1-brane realizes the basic 't Hooft line operator of magnetic charge $(1,0^{N-1})$ in the S-dual gauge theory. 

\subsection{Holographic dual geometries}
The holographically dual geometry is obtained by taking the near-horizon limit of a stack of D3-branes in the background of O3-plane. 
It is argued \cite{Witten:1998xy} that $\mathcal{N}=4$ SYM theories with orthogonal or symplectic gauge groups are 
holographically dual to Type IIB string theory on $AdS_5$ $\times$ $\mathbb{RP}^5$ where $\mathbb{RP}^5$ $\cong$ $S^5/\mathbb{Z}_2$ is the five-dimensional projective plane. 
The radial direction of the $\mathbb{R}^6/\mathbb{Z}_2$ factor labeled by $x^3$, $x^4$, $x^5$, $x^7$, $x^8$ and $x^9$ 
in the configuration (\ref{brane_conf}) becomes one of the $AdS_5$ coordinates and the other five angular directions form $\mathbb{RP}^5$. 
The directions parallel to the D3-branes parameterized by $x^0$, $x^1$, $x^2$ and $x^6$ transform into the other $AdS_5$ coordinates. 

In the absence of O3-plane, the near-horizon limit of the brane configuration (\ref{brane_conf}) or (\ref{brane_conf2}) 
leads to the Type IIB supergravity solutions with the $AdS_2$ $\times$ $S^2$ $\times$ $S^4$ factor. 
D'Hoker, Estes and Gutperle find the supergravity solution with the form \cite{Yamaguchi:2006te,Lunin:2006xr,DHoker:2007mci}
\begin{align}
\label{eqn 17}
AdS_2\times S^2\times S^4\times \Sigma, 
\end{align}
where $\Sigma$ is a two-dimensional surface, over which the $AdS_2$ $\times$ $S^2$ $\times$ $S^4$ part is warped. 
See \cite{DHoker:2007mci} for detailed descriptions. 
Here, we consider the $\mathbb{Z}_{2}$ orientifold of the configurations (\ref{eqn 17}) due to the O3-plane, 
and refer them as the orientifold of bubbling Wilson line geometries. 
There are also discrete torsions $\theta _{RR},\theta _{NS}$ specific to different gauge groups, see the above Table (2.1).
The base space $\Sigma$ for the $AdS_{2}\times S^{2}\times S^{4}$ fibration can be taken as a Riemann surface 
which is parametrized by a $u$-plane \cite{DHoker:2007mci}. 
The boundary of $\Sigma $ can be taken as a horizontal real axis 
with $u=+\infty $ and $u=-\infty $ being identified. 
We have the $\mathbb{Z}_{2}$ identification $\mathrm{Re}(u)\rightarrow -\mathrm{Re}(u)$. 
Consider the $\mathbb{Z}_{2}$ fixed point of the $\mathbb{Z}_{2}$ action on the boundary of Riemann surface to be $u_{\ast }$ $\in \partial \Sigma$. 
We make a choice that $u_{\ast }=0$, as a reference point. 
When the $\mathbb{Z}_{2}$ fixed point $u_{\ast }$ is in the $S^{4}$-shrinking interval, at the fixed point $u_{\ast }$, 
the $S^{4} $ fiber of the covering space, after the quotient, becomes the $S^{4}/\mathbb{Z}_{2}\simeq \mathbb{RP}^{4}$ fiber. 
Together with the overal $AdS_{2}$ fiber, there is $AdS_{2}$ $\times $ $\mathbb{RP}^{4}$. 

From the holographic point of view, 
the Wilson lines transforming in the irreducible representations can be understood as certain configuration of Type IIB string theory on the geometry (\ref{eqn 17}). 
A basic example is the Wilson line transforming in the fundamental representation. 
It is holographically dual to Type IIB string wrapping $AdS_2$ $\subset$ $AdS_5$ \cite{Maldacena:1998im,Rey:1998ik}. 

Furthermore, the rank-$k$ symmetric Wilson line is dual to 
the D3-brane of $k$ units of fundamental string charge wrapping $AdS_2$ $\times$ $S^2$ \cite{Drukker:2005kx}. 
Near the $S^{2}$-degeneration locus on the above base space $\Sigma $, i.e. near the D3 near-horizon region, the $S^{4}$ over a segment form a
topological $5$-cycle threading the $F_{5}$ flux. 
This region corresponds to the Neumann boundary condition, at which the $S^{2}$ shrinks \cite{DHoker:2007mci}. 
Hence locally there is $AdS_{2}$ $\times $ $S^{2}$. 

On the other hand, the rank-$k$ antisymmetric Wilson line is dual to 
the D5-brane of $k$ units of fundamental string charge wrapped on $AdS_2$ $\times$ $S^4$ \cite{Yamaguchi:2006tq}. 
Near the $S^{4}$-degeneration locus on the above base space $\Sigma $, 
i.e. near the D5 near-horizon region, the $S^{2}$ over a segment form a topological $3$-cycle threading the $F_{3}$ flux. 
This region corresponds to the Dirichlet boundary condition, at which the $S^{4}$ shrinks \cite{DHoker:2007mci}. 
In the quotient space, hence locally there is $AdS_{2}\times S^{4}$. 

In the presence of the O3$^{-}$-plane or $\widetilde{\textrm{O3}}^-$-plane, 
there can exist the D5-brane wrapping $AdS_2$ $\times$ $\mathbb{RP}^4$. 
It is holographically dual to the Wilson line in the spinor representation for orthogonal gauge theories \cite{Witten:1998xy}. 
In this case, the D5-brane is wrapped on $AdS_2$ $\times$ $\mathbb{RP}^4$ 
in the near-horizon geometry $AdS_5$ $\times$ $\mathbb{RP}^5$ and it can be viewed as a string wrapping $AdS_2$, 
which is called a fat string \cite{Witten:1998xy}. 
This is consistent to the fact that the tensor product of two spinor representations split into a sum of tensor products of the fundamental representations. 
In this case, the antisymmetric Wilson line that is dual to the D5-brane wrapping $AdS_2$ $\times$ $S^4$ can be thought of as 
the double cover of the D5-brane wrapped on $AdS_2$ $\times$ $\mathbb{RP}^4$ dual to the spinor Wilson line.

More generally, the Wilson line transforming in the irreducible representation 
whose highest weight is labeled by the Young diagram $\lambda$ $=$ $(\lambda_1,\cdots,\lambda_r)$ 
correspond to $r$ coincident D3-branes $(\textrm{D3}_1,\cdots,\textrm{D3}_r)$ wrapping the $AdS_2$ $\times$ $S^2$, 
where D3$_i$ is the $i$-th D3-brane with $\lambda_i$ units of fundamental string charge dissolved in it. 
Alternatively, it can be also described in terms of a configuration with $\lambda_1$ D5-branes 
$(\textrm{D5}_1,\cdots,\textrm{D5}_{\lambda_1})$ where D5$_j$ is the $j$-th D5-brane of $\lambda_j'$ units of fundamental string charge dissolved in it. 
For very large representations, e.g. representations of rectangular Young diagram, one must have backreaction of the geometries. 
The backreaction introduces new cycles in the geometries. 
The D3- or D5-branes are dissolved into fluxes and are described by the backreacted geometries with these new cycles, through which fluxes are threading. 
The Wilson lines in large representations with multiple rows are described by the backreacted geometries of D3 on these $AdS_{2}$ $\times $ $S^{2}$. 
The D3 flux $F_{5}$ contains the component through the above $5$-cycle. 
The Wilson lines in large representations with multiple columns are described by the backreacted geometries of D5 on these $AdS_{2}$ $\times $ $S^{4}$. 
The D5 flux $F_{3}$ contains the component through the above $3$-cycle. 

\section{Line defect correlators}
\label{sec_lineindex}

\subsection{Wilson line defect correlators}
\label{sec_Wline}
The Wilson line operator inserts a point-like electrically charged particle 
whose worldline is the curve on which the Wilson line is supported. 
It is characterized by a representations of gauge group $G$. 
It can be described by the trace in the representation of the gauge group 
of the holonomy matrix associated with the parallel transport along the curve. 

To describe the line defect indices, we introduce
\begin{align}
\label{qpoch_def}
(a;q)_{0}&:=1,\qquad
(a;q)_{n}:=\prod_{k=0}^{n-1}(1-aq^{k}),\qquad 
(q)_{n}:=\prod_{k=1}^{n}(1-q^{k}),
\nonumber \\
(a;q)_{\infty}&:=\prod_{k=0}^{\infty}(1-aq^{k}),\qquad 
(q)_{\infty}:=\prod_{k=1}^{\infty} (1-q^k), 
\end{align}
with $a, q \in \mathbb{C}$ and $|q|<1$. 
Also we use the notation defined by 
\begin{align}
(ax^{\pm};q)_{n}:=(ax;q)_{n}(ax^{-1};q)_{n}. 
\end{align}

The line defect indices for the Wilson line operators $W_{\mathcal{R}_i}$ transforming in the representation $\mathcal{R}_i$, $i=1,\cdots, k$, 
in $\mathcal{N}=4$ SYM theory of gauge group $G$ can be evaluated by the matrix integral of the form \cite{Gang:2012yr}
\begin{align}
\label{Wilson_intgral1}
&
\langle W_{\mathcal{R}_1}\cdots W_{\mathcal{R}_k}\rangle^{G}(t;q)
\nonumber\\
&=\frac{1}{|\textrm{Weyl}(G)|}
\frac{(q)_{\infty}^{2\textrm{rank}(G)}}
{(q^{\frac12}t^{\pm2};q)_{\infty}^{\textrm{rank}(G)}}
\oint \prod_{\alpha\in \textrm{root}(G)}ds 
\frac{(s^{\alpha};q)_{\infty}(qs^{\alpha};q)_{\infty}}
{(q^{\frac12}t^2s^{\alpha};q)_{\infty}(q^{\frac12}t^{-2}s^{\alpha};q)_{\infty}}
\prod_{i=1}^{k}\chi^{\mathfrak{g}}_{\mathcal{R}_i},
\end{align}
where 
$|\textrm{Weyl}(G)|$ is the order of Weyl group for gauge group $G$ 
and $\chi_{\mathcal{R}_i}^{\mathfrak{g}}$ is the character of the representation $\mathcal{R}_i$ of gauge algebra $\mathfrak{g}$. 
Without any insertion of the character, it is nothing but the flavored Schur index of $\mathcal{N}=4$ SYM theory defined by 
\begin{align}
\mathcal{I}^{G}(t;q)&:=\Tr (-1)^{F}q^{J+\frac{H+C}{4}} t^{H-C}, 
\end{align}
where the trace is taken over the Hilbert space on the $S^3$ as a cohomology of the chosen supercharges. 
$F$ is the Fermion number, $J$ the generator of the rotation, $C$ and $H$ the Cartan generators of the $SU(2)_C$ and $SU(2)_H$ 
where $SU(2)_C\times SU(2)_H$ is the subgroup of the $SU(4)_R$ R-symmetry group. 
The additional degrees of freedom due to the inserted Wilson line operators can be obtained from the normalized correlation function defined by
\begin{align}
\langle \mathcal{W}_{\mathcal{R}_1}\cdots \mathcal{W}_{\mathcal{R}_k}\rangle^{G}(t;q)
&=\frac{\langle W_{\mathcal{R}_1}\cdots W_{\mathcal{R}_k}\rangle^G(t;q)}
{\mathcal{I}^G(t;q)}. 
\end{align}

The line defect indices admit special fugacity limits. 
The unflavored line defect indices are obtained by simply setting $t$ to $1$. 
Another interesting limit is the half-BPS limit where we keep $\mathfrak{q}$ $=$ $q^{1/4}t$ being finite and take $q$ to $0$. 
In the half-BPS limit, the Wilson line defect indices (\ref{Wilson_intgral1}) reduces to 
\begin{align}
\label{halfBPSgeneral}
&
\langle W_{\mathcal{R}_1}\cdots W_{\mathcal{R}_k}\rangle^{G}_{\textrm{$\frac12$BPS}}(\mathfrak{q})
\nonumber\\
&=
\frac{1}{|\textrm{Weyl}(G)|}\frac{1}{(1-\mathfrak{q}^{2})^{\textrm{rank}(G)}}
\oint
\prod_{\alpha \in \textrm{root}(G)} 
ds 
\frac{(1-s^{\alpha})}{(1-\mathfrak{q}^2 s^{\alpha})}
\prod_{i=1}^{k}\chi^{\mathfrak{g}}_{\mathcal{R}_i}. 
\end{align}
The integral (\ref{halfBPSgeneral}) can be viewed as a $\mathfrak{q}$-deformation of the Hall inner product for which 
the Hall-Littlewood functions $P_{\lambda}(s,\mathfrak{t})$ with $\mathfrak{t}$ $=$ $\mathfrak{q}^2$ are orthogonal, 
as argued in \cite{Hatsuda:2022xdv,Hatsuda:2023iwi,Hatsuda:2023imp}. 
For unitary gauge theories we can evaluate it by making use of the following formula: 
\begin{align}
\frac{1}{N!}
\oint \prod_{i=1}^N \frac{ds_i}{2\pi is_i}
\frac{\prod_{i\neq j} 1-\frac{s_i}{s_j}}{\prod_{i,j}1-\mathfrak{t}\frac{s_i}{s_j}}
P_{\mu}(s;\mathfrak{t})P_{\lambda}(s^{-1};\mathfrak{t})
&=\frac{\delta_{\mu\lambda}}{(\mathfrak{t};\mathfrak{t})_{N-l(\mu)}\prod_{j\ge 1}(\mathfrak{t};\mathfrak{t})_{m_j(\mu)}},
\end{align}
where $l(\mu)$ is the length of $\mu$ and $m_j(\mu)$ the multiplicity of $\mu$. 

Note that when we take $\mathfrak{q}\rightarrow 0$ in (\ref{halfBPSgeneral}), 
we get the leading coefficient in the $\mathfrak{q}$-series expansion of the correlators. 
The resulting matrix integral of the form 
\footnote{As this is also obtained from (\ref{Wilson_intgral1}) when $q$ is set to zero, the leading coefficients in the $q$-series expansion of the Schur line defect correlators are the same. }
\begin{align}
\frac{1}{|\textrm{Weyl}(G)|}
\oint
\prod_{\alpha \in \textrm{root}(G)} 
ds (1-s^{\alpha})
\prod_{i=1}^{k}\chi^{\mathfrak{g}}_{\mathcal{R}_i}
\end{align}
is studied in the context of random matrix theory and L-function (see e.g. \cite{MR1659828,MR1794267}). 
With the help of Weyl's integration formula, one can compute the integral. 
It gives rise to the multiplicities of the trivial representation in the tensor powers of the representations in which the inserted Wilson lines transform. 

\subsection{'t Hooft line defect correlators}
\label{sec_tHooft}
The 't Hooft line operator describes a probe point-like monopole 
whose worldline is identified with the curve in spacetime that supports the 't Hooft line. 
The half-BPS 't Hooft line operator in $\mathcal{N}=4$ SYM theory can be characterized by magnetic charge $B$ as an element of the cocharacter lattice of gauge group $G$. 
It can be defined by a singular Dirac monopole configuration of the fields. 

The two-point function of the 't Hooft line operators $T_{B}$ with magnetic charge $B$ 
in $\mathcal{N}=4$ SYM theory of gauge group $G$ can be evaluated by the matrix integral of the form \cite{Gang:2012yr}
\begin{align}
\label{tHooft_integral1}
&\langle T_{B} T_{B}\rangle^{G}(t;q)
=
\sum_{v\in \textrm{Rep}(B)}
\frac{1}{|\textrm{Weyl}(B)|}
\frac{(q)_{\infty}^{2\textrm{rank}(G)}}
{(q^{\frac12}t^{\pm2};q)_{\infty}^{\textrm{rank}(G)}}
\oint \prod_{\alpha\in \textrm{root}(G)}ds 
\nonumber\\
&\times 
\frac{(q^{\frac{|\alpha(B)|}{2}}s^{\alpha};q)_{\infty}(q^{1+\frac{|\alpha(B)}{2}|}s^{\alpha};q)_{\infty}}
{(q^{\frac{1+|\alpha(B)|}{2}}t^2s^{\alpha};q)_{\infty}(q^{\frac{1+|\alpha(B)|}{2}}t^{-2}s^{\alpha};q)_{\infty}}
Z_{\textrm{bubb}}^{(B,v)}(t,s;q). 
\end{align}
Here $|\textrm{Weyl}(B)|$ is the order of the Weyl group for unbroken gauge group 
and $\alpha(B)$ $\in$ $\mathbb{Z}$ is defined by $[B,E_{\alpha}]$ $=$ $\alpha(B)E_{\alpha}$ 
where $E_{\alpha}$ are the ladder operators associated to roots $\alpha$ of $\mathfrak{g}$. 
$Z_{\textrm{bubb}}^{(B,v)}(t,s;q)$ is the index that encodes the monopole bubbling effect 
where $v$ is the possible screened asymptotic magnetic charge of the singular monopole. 
It is obtainable from the monopole bubbling index on $S^1\times \mathbb{R}^3$ \cite{Ito:2011ea} (see also \cite{Brennan:2018yuj,Brennan:2018rcn,Hayashi:2019rpw,Hayashi:2020ofu}). 
In this paper, we focus on the minuscule representations for which there is no monopole bubbling effect 
so that there is no contribution from the monopole bubbling index. 

\section{$\mathfrak{so}(2N+1)$}
\label{sec_Btype}
Consider $\mathcal{N}=4$ SYM theory of gauge group associated with the Lie algebra $\mathfrak{g}$ $=$ $\mathfrak{so}(2N+1)$. 
The center of $Spin(2N+1)$ is $\mathbb{Z}_2$ so that the spinor representation carries the $\mathbb{Z}_2$ charge. 
The line operators in the gauge theory carry electric and magnetic charges 
\begin{align}
(z_e,z_m)\in \mathbb{Z}_2\times \mathbb{Z}_2. 
\end{align}
The $Spin(2N+1)$ gauge theory has the line operators with
\begin{align}
(z_e,z_m)&=(0,0), \qquad (z_e,z_m)=(1,0), 
\end{align}
including the electric Wilson line in the spinor representation. 
The $SO(2N+1)_+$ gauge theory has the lines with
\begin{align}
(z_e,z_m)&=(0,0), \qquad (z_e,z_m)=(0,1), 
\end{align}
including the 't Hooft line. 
The $SO(2N+1)_-$ gauge theory has the lines with
\begin{align}
(z_e,z_m)&=(0,0), \qquad (z_e,z_m)=(1,1), 
\end{align}
including the dyonic line. 

The Lie algebra of the Langlands dual gauge group is $\mathfrak{g}^{\vee}$ $=$ $\mathfrak{usp}(2N)$. 
The $USp(2N)$ has a $\mathbb{Z}_2$ center under which the fundamental representation is charged. 
In particular, the Wilson line operator transforming in the spinor representation for $Spin(2N+1)$ gauge theory 
is conjectured to be S-dual to the 't Hooft line operator of $B=(\frac12^N)$ for $USp(2N)/\mathbb{Z}_2$ gauge theory \cite{Aharony:2013hda}. 

The spinor representation of $Spin(2N+1)$ has dimension $2^N$. 
The character of the spinor representation of $\mathfrak{so}(2N+1)$ is given by
\begin{align}
\label{ch_so2N+1_sp}
\chi_{\textrm{sp}}^{\mathfrak{so}(2N+1)}
&=\prod_{i=1}^{N}(s_i^{\frac12}+s_i^{-\frac12}). 
\end{align}

For the fundamental representation of $\mathfrak{so}(2N+1)$ we have the character 
\begin{align}
\label{ch_so2N+1_fund}
\chi_{\tiny \yng(1)}^{\mathfrak{so}(2N+1)}
&=1+\sum_{i=1}^{N}(s_i+s_i^{-1}). 
\end{align}

More generally, the character of the irreducible representation 
with the highest weight labeled by the Young diagram $\lambda$ is given by \cite{MR1153249}
\begin{align}
\label{ch_so2N+1_irrep}
\chi_{\lambda}^{\mathfrak{so}(2N+1)}
&=\frac{\det (s_j^{\lambda_i+N-i+1/2} -s_{j}^{-(\lambda_i+N-i+1/2)})}
{\det(s_j^{N-i+1/2}-s_j^{-(N-i+1/2)})}. 
\end{align}

Let us explain a general method to evaluate the line defect correlators.
We use the character expansion method \cite{Dolan:2007rq} (see also \cite{Sei:2023fjk}).
We note that the $SO(2N+1)$ line defect correlators can be written as
\begin{align}
\label{Wso2N+1}
&\langle W_{\mathcal{R}_1}\cdots W_{\mathcal{R}_k}\rangle^{SO(2N+1)}\notag\\
&=\int \! d\mu^{SO(2N+1)}
\exp \left[ \sum_{n=1}^\infty \frac{1}{n} f_n(q, t) \frac{\overline{P}_n(s)^2-\overline{P}_{2n}(s)}{2} \right]\prod_{i=1}^{k}\chi^{\mathfrak{so}(2N+1)}_{\mathcal{R}_i}(s). 
\end{align}
Here the measure of the maximal torus for $SO(2N+1)$ is given by
\begin{align}
d\mu^{SO(2N+1)}&=\frac{1}{2^N N!}\prod_{i=1}^N \frac{ds_i}{2\pi i s_i}(1-s_i)(1-s_i^{-1}) \notag \\
&\quad \times \prod_{1\leq i<j\leq N} (1-s_is_j)(1-s_i^{-1}s_j^{-1})(1-s_is_j^{-1})(1-s_i^{-1}s_j)
\end{align}
and we have introduced 
\begin{align}
f_n(q,t)=\frac{q^{n/2}(t^{2n}+t^{-2n})-2q^n}{1-q^n},
\label{eq:f_n}
\end{align}
and 
\begin{align}
P_m(s)&:=\sum_{i=1}^N (s_i^m+s_i^{-m}),\label{P_m}\\ 
\overline{P}_m(s)&:=1+P_m(s)=1+\sum_{i=1}^N (s_i^m+s_i^{-m}).
\end{align}
Defining 
\begin{align}
\overline{M}_n(s)=\frac{\overline{P}_n(s)^2-\overline{P}_{2n}(s)}{2}=P_n(s)+\frac{P_n(s)^2-P_{2n}(s)}{2}, 
\end{align} 
the exponential part of (\ref{Wso2N+1}) can be expanded as
\begin{align}
\exp \left( \sum_{n=1}^\infty \frac{1}{n} f_n(q,t)\overline{M}_n(s) \right)
=\sum_{\lambda} \frac{1}{z_\lambda}f_\lambda(q,t)\overline{M}_\lambda(s), 
\end{align}
where for each partition $\lambda=(\lambda_1, \lambda_2, \dots)=(1^{m_1}2^{m_2}\dots )$, $z_\lambda$, $f_\lambda(q,t)$, $\overline{M}_\lambda(s)$ are defined by
\begin{align}
z_\lambda=\prod_{i=1}^\infty i^{m_i} m_i!,\qquad 
f_\lambda(q,t)=\prod_{i=1}^{\ell(\lambda)} f_{\lambda_i}(q,t),\qquad
\overline{M}_\lambda(s)=\prod_{i=1}^{\ell(\lambda)} \overline{M}_{\lambda_i}(s). 
\end{align}
Hence the line defect correlator is evaluated as 
\begin{align}
\langle W_{\mathcal{R}_1}\cdots W_{\mathcal{R}_k}\rangle^{SO(2N+1)}
=\sum_{\lambda} \frac{1}{z_\lambda}f_\lambda(q,t) \int \! d\mu^{SO(2N+1)}
\overline{M}_{\lambda}(s)\prod_{i=1}^{k}\chi^{\mathfrak{so}(2N+1)}_{\mathcal{R}_i}(s).
\end{align}
Accordingly, we should evaluate the integral
\begin{align}
\int \! d\mu^{SO(2N+1)}\overline{M}_{\lambda}(s)\prod_{i=1}^{k}\chi^{\mathfrak{so}(2N+1)}_{\mathcal{R}_i}(s).
\end{align}
As explained in Appendix~\ref{app:character}, the character $\chi^{\mathfrak{so}(2N+1)}_{\mathcal{R}_i}(s)$ can be written in terms of $\overline{P}_m(s)$.
Therefore, the key ingredient of the integral is the following formula:
\begin{align}
\int \! d\mu^{SO(2N+1)} \overline{P}_\mu(s)=\sum_{\nu \in R_{2N+1}(|\mu|)} \chi_\nu^S(\mu)+\sum_{\nu \in W_{2N+1}(|\mu|)} \chi_\nu^S(\mu), 
\end{align}
where $\overline{P}_\mu(s)=\overline{P}_{\mu_1}(s)\overline{P}_{\mu_2}(s)\cdots$, and $\chi_\nu^S(\mu)$ is the character  of the symmetric group for the representation $\nu$ and the conjugacy class $\mu$. We also define subsets of partitions of $p$ as
\begin{align}
R_{n}(p)&=\{ \lambda \vdash p \,|\, \ell(\lambda)\leq n \text{ and } \forall \lambda_i \text{ is even} \}, \\
W_{n}(p)&=\{ \lambda \vdash p \,|\, \ell(\lambda)= n \text{ and } \forall \lambda_i \text{ is odd} \}.
\end{align}
It is well-known that the symmetric group character $\chi_\nu^S(\mu)$ is determined by the Murnaghan-Nakayama rule \cite{MR1507257,MR5729}, or more technically can be read off from the Frobenius formula:
\begin{align}
s_\lambda=\sum_{\mu \vdash \lambda} \frac{\chi_\lambda^S(\mu)}{z_\mu} p_\mu, 
\end{align}
because we can easily express the Schur polynomial $s_\lambda$ in terms of the power sum polynomial $p_\mu$ by combining the Jacobi-Trudi formula with Newton's identities.

Given the following expansion:
\begin{align}
\overline{M}_{\lambda}(s)\prod_{i=1}^{k}\chi^{\mathfrak{so}(2N+1)}_{\mathcal{R}_i}(s)=\sum_\mu a_{\lambda, \mathcal{R}}^\mu \overline{P}_\mu(s),
\end{align}
we can easily perform the integral as
\begin{align}
&\int \! d\mu^{SO(2N+1)}\overline{M}_{\lambda}(s)\prod_{i=1}^{k}\chi^{\mathfrak{so}(2N+1)}_{\mathcal{R}_i}(s)
=\sum_\mu a_{\lambda, \mathcal{R}}^\mu \int \! d\mu^{SO(2N+1)} \overline{P}_\mu(s) \notag \\
&=\sum_\mu a_{\lambda, \mathcal{R}}^\mu \Biggl(\sum_{\nu \in R_{2N+1}(|\mu|)} \chi_\nu^S(\mu)+\sum_{\nu \in W_{2N+1}(|\mu|)} \chi_\nu^S(\mu)\Biggr). 
\end{align}
The correlator is finally given by
\begin{align}
&\langle W_{\mathcal{R}_1}\cdots W_{\mathcal{R}_k}\rangle^{SO(2N+1)}\notag \\
&=\sum_{\lambda} \frac{1}{z_\lambda}f_\lambda(q,t) \sum_\mu a_{\lambda, \mathcal{R}}^\mu \Biggl(\sum_{\nu \in R_{2N+1}(|\mu|)} \chi_\nu^S(\mu)+\sum_{\nu \in W_{2N+1}(|\mu|)} \chi_\nu^S(\mu)\Biggr). 
\end{align}
We do not have the general expression for the coefficients $a_{\lambda, \mathcal{R}}^\mu$. However,
in practical computations, it is rather sufficient to fix $a_{\lambda, \mathcal{R}}^\mu$ for partitions of lower numbers. We can do it by using a symbolic computational system.

We note that in the presence of a pair of the Wilson lines in the spinor representation, 
for which the character is given by (\ref{ch_so2N+1_sp}), 
the associated correlator can be evaluated from the $USp(2N)$ integral. 
Since the square of the character is written as
\begin{align}
(\chi_\text{sp}^{\mathfrak{so}(2N+1)})^2=\prod_{i=1}^N (1+s_i)(1+s_i^{-1}),
\end{align}
we find 
\begin{align}
\label{sp_soM}
d\mu^{SO(2N+1)} (\chi_\text{sp}^{\mathfrak{so}(2N+1)})^2=d\mu^{USp(2N)}. 
\end{align}
where the measure $d\mu^{USp(2N)}$ of the $USp(2N)$ group is given by
\begin{align}
d\mu^{USp(2N)}&=\frac{1}{2^N N!}\prod_{i=1}^N\frac{ds_i}{2\pi i s_i} (1-s_i^2)(1-s_i^{-2}) \notag \\
&\quad \times \prod_{1\leq i<j\leq N} (1-s_is_j)(1-s_i^{-1}s_j^{-1})(1-s_is_j^{-1})(1-s_i^{-1}s_j).
\label{eq:dmu-USp}
\end{align}
By using (\ref{sp_soM}) we get
\begin{align}
&\langle W_\text{sp} W_\text{sp}\rangle^{Spin(2N+1)}
=\int \! d\mu^{USp(2N)}
\exp \left[ \sum_{n=1}^\infty \frac{1}{n} f_n(q, t) \overline{M}_n(s) \right]\notag \\
&=\int \! d\mu^{USp(2N)}
\exp \left[ \sum_{n=1}^\infty \frac{1}{n} f_n(q, t) \biggl(P_n(s)+\frac{P_n(s)^2-P_{2n}(s)}{2} \biggr)\right].
\end{align}
Here we have $Spin(2N+1)$ instead of $SO(2N+1)$ for the gauge group since we are considering the spinor representation. So we have equipped the correlators with the superscript $Spin(2N+1)$. 
This expression is useful to see the large $N$ behavior.

\subsection{$\mathfrak{so}(3)$}
The Schur index of $\mathcal{N}=4$ $SO(3)$ or equivalently $USp(2)$ theory is equal to that of $\mathcal{N}=4$ $SU(2)$ theory. 
According to the formulae of the Schur indices for $\mathcal{N}=4$ theories with unitary gauge group in \cite{Hatsuda:2022xdv}, 
it can be expressed as
\begin{align}
\label{Schur_so3_sp2}
\mathcal{I}^{SO(3)}(t;q)&=\mathcal{I}^{USp(2)}(t;q)
\nonumber\\
&=-\frac{(q^{\frac12}t^{\pm2};q)_{\infty}}{(q;q)_{\infty}^2}
\sum_{\begin{smallmatrix}
p_1, p_2\in \mathbb{Z}\\
p_1<p_2\\
\end{smallmatrix}}
\frac{(q^{\frac12}t^{-2})^{p_1+p_2-2}}
{(1-q^{p_1-\frac12}t^2) (1-q^{p_2-\frac12}t^2)}. 
\end{align}

\subsubsection{Spinor Wilson line}
We begin with $\mathcal{N}=4$ gauge theory based on the Lie algebra $\mathfrak{g}=\mathfrak{so}(3)$. 
The Wilson line in the spinor representation exists for $Spin(3)$ gauge theory. 
When one inserts two or more Wilson lines in the spinor representation for $Spin(3)$ gauge theory, the correlation function is non-trivial. 
The two-point function of a pair of the spinor Wilson lines can be evaluated as
\begin{align}
\label{so3_wsp}
&
\langle W_{\textrm{sp}} W_{\textrm{sp}}\rangle^{Spin(3)}(t;q)
\nonumber\\
&=\frac12 \frac{(q)_{\infty}^2}{(q^{\frac12}t^{\pm2};q)_{\infty}}
\oint \frac{ds}{2\pi is}
\frac{(s^{\pm};q)_{\infty} (qs^{\pm};q)_{\infty}}
{(q^{\frac12}t^{2}s^{\pm};q)_{\infty}(q^{\frac12}t^{-2}s^{\pm};q)_{\infty}}(s^{\frac12}+s^{-\frac12})^2. 
\end{align}
According to the isomorphism $Spin(3)$ $\cong$ $USp(2)$, 
this is equal to the two-point function (\ref{sp2_wfund}) of the Wilson lines in the fundamental representation for $USp(2)$ gauge theory. 

The spinor Wilson line for $\mathcal{N}=4$ $Spin(3)$ gauge theory is S-dual to 
the 't Hooft line with $B=(\frac12)$ for $\mathcal{N}=4$ $USp(2)/\mathbb{Z}_2$ gauge theory. 
The two-point function of these S-dual 't Hooft lines can be computed as 
\begin{align}
\label{sp2_tsp}
&
\langle T_{(\frac12)} T_{(\frac12)}\rangle^{USp(2)/\mathbb{Z}_2}(t;q)
\nonumber\\
&=\frac{(q)_{\infty}^2}{(q^{\frac12}t^{\pm2};q)_{\infty}}
\oint \frac{ds}{2\pi is}
\frac{(q^{\frac12}s^{\pm2};q)_{\infty}(q^{\frac32}s^{\pm2};q)_{\infty}}
{(qt^{2}s^{\pm2};q)_{\infty}(qt^{-2}s^{\pm2};q)_{\infty}}. 
\end{align}
As expected from S-duality, the two expressions (\ref{so3_wsp}) and (\ref{sp2_tsp}) coincide
\begin{align}
\langle W_{\textrm{sp}} W_{\textrm{sp}}\rangle^{Spin(3)}(t;q)
&=\langle T_{(\frac12)} T_{(\frac12)}\rangle^{USp(2)/\mathbb{Z}_2}(t;q). 
\end{align}

We observe that according to the isomorphism $Spin(3)$ $\cong$ $SU(2)$, 
the line defect correlation functions for gauge theories based on the Lie algebras $\mathfrak{so}(3)$ is calculable from the exact results \cite{Hatsuda:2023iwi} for $U(2)$ gauge theory. 
We obtain 
\begin{align}
\label{su2w1_exact}
&\langle W_{\textrm{sp}} W_{\textrm{sp}}\rangle^{Spin(3)}(t;q)
=\langle T_{(\frac12)} T_{(\frac12)}\rangle^{USp(2)/\mathbb{Z}_2}(t;q)
\nonumber\\
&=\langle W_{\tiny \yng(1)} W_{\tiny \yng(1)}\rangle^{SU(2)}(t;q)
=\frac{(q^{\frac12}t^{\pm2};q)_{\infty}}
{(qt^{\pm4};q)_{\infty}}
\sum_{m\in \mathbb{Z}\setminus \{0,n\}}
\frac{t^{2m}-t^{-2m}}{t^2-t^{-2}}
\frac{q^{\frac{m-1}{2}}}{1-q^m}. 
\end{align}

Again the half-BPS limits of the two-point functions (\ref{so3_wsp}) and (\ref{sp2_tsp}) are simply obtained from the results in \cite{Hatsuda:2023iwi}. 
We get 
\begin{align}
\label{halfBPSso3_wsp}
&
\langle W_{\textrm{sp}} W_{\textrm{sp}}\rangle^{Spin(3)}_{\textrm{$\frac12$BPS}}(\mathfrak{q})
=\langle T_{(\frac12)} T_{(\frac12)}\rangle^{USp(2)/\mathbb{Z}_2}_{\textrm{$\frac12$BPS}}(\mathfrak{q})
\nonumber\\
&=\langle W_{\tiny \yng(1)} W_{\tiny \yng(1)}\rangle^{SU(2)}_{\textrm{$\frac12$BPS}}(\mathfrak{q})
=\frac{1+\mathfrak{q}^2}{1-\mathfrak{q}^4}
=\frac{1}{1-\mathfrak{q}^2}. 
\end{align}

More generally, the correlators with an even number of spinor Wilson lines are non-trivial whereas those with an odd number vanish. 
We have
\begin{align}
\label{so3_wspmult}
\langle \underbrace{W_{\textrm{sp}} W_{\textrm{sp}} \cdots W_{\textrm{sp}}}_{2k} \rangle^{Spin(3)}(t;q)
&=\langle \underbrace{W_{\tiny \yng(1)} W_{\tiny \yng(1)}\cdots W_{\tiny \yng(1)}}_{2k}\rangle^{SU(2)}(t;q). 
\end{align}
These can be also computed from the exact results from the results in \cite{Hatsuda:2023iwi} for $U(2)$ gauge theory. 
For example, in the half-BPS limit we have
\begin{align}
\langle W_{\textrm{sp}} W_{\textrm{sp}} W_{\textrm{sp}} W_{\textrm{sp}} \rangle^{Spin(3)}_{\textrm{$\frac12$BPS}}(\mathfrak{q})
&=\frac{2+3\mathfrak{q}^2+\mathfrak{q}^4}{1-\mathfrak{q}^4}, \\
\langle W_{\textrm{sp}} W_{\textrm{sp}} W_{\textrm{sp}} W_{\textrm{sp}} W_{\textrm{sp}} W_{\textrm{sp}} \rangle^{Spin(3)}_{\textrm{$\frac12$BPS}}(\mathfrak{q})
&=\frac{5+9\mathfrak{q}^2+5\mathfrak{q}^4+\mathfrak{q}^6}{1-\mathfrak{q}^4}, \\
\langle W_{\textrm{sp}} W_{\textrm{sp}} W_{\textrm{sp}}W_{\textrm{sp}} W_{\textrm{sp}} W_{\textrm{sp}} W_{\textrm{sp}} W_{\textrm{sp}} \rangle^{Spin(3)}_{\textrm{$\frac12$BPS}}(\mathfrak{q})
&=\frac{14+28\mathfrak{q}^2+20\mathfrak{q}^4+7\mathfrak{q}^6+\mathfrak{q}^8}{1-\mathfrak{q}^4}, \\
\langle W_{\textrm{sp}} W_{\textrm{sp}} W_{\textrm{sp}} W_{\textrm{sp}} W_{\textrm{sp}} W_{\textrm{sp}} W_{\textrm{sp}} W_{\textrm{sp}} W_{\textrm{sp}} W_{\textrm{sp}} \rangle^{Spin(3)}_{\textrm{$\frac12$BPS}}(\mathfrak{q})
&=\frac{42+90\mathfrak{q}^2+75\mathfrak{q}^4+35\mathfrak{q}^6+9\mathfrak{q}^8+\mathfrak{q}^{10}}{1-\mathfrak{q}^4}. 
\end{align}
Here we observe that 
the expansion coefficients of the numerator in the half-BPS limit of the $2k$-point function of the spinor Wilson lines 
have an interesting combinatorial interpretation. 
We find that 
\begin{align}
\label{halfBPSso3_wspkpt}
\langle \underbrace{W_{\textrm{sp}} \cdots W_{\textrm{sp}}}_{2k} \rangle^{Spin(3)}_{\textrm{$\frac12$BPS}}(\mathfrak{q})
&=\mathcal{I}^{SO(3)}_{\textrm{$\frac12$BPS}}(\mathfrak{q})\sum_{i=0}^{k}a_{k\ \textrm{sp}}^{\mathfrak{so}(3)}(i)\mathfrak{q}^{2i}\notag \\
&=\frac{1}{1-\mathfrak{q}^4}\sum_{i=0}^{k}a_{k\ \textrm{sp}}^{\mathfrak{so}(3)}(i)\mathfrak{q}^{2i}, 
\end{align}
where 
\begin{align}
a_{k\ \textrm{sp}}^{\mathfrak{so}(3)}(i)&=(2i+1)\frac{(2k)!}{(k-i)!(k+i+1)!}
\nonumber\\
&=C_{k+i+1,2i+1}, 
\end{align}
where 
\begin{align}
C_{n,m}&=\frac{m}{n}\left(
\begin{matrix}
2n-m-1\\
n-1\\
\end{matrix}
\right)
\end{align}
is the transposed Catalan triangle. 
In particular, $a_{k\ \textrm{sp}}^{\mathfrak{so}(3)}(0)$ is the $k$-th Catalan number
\begin{align}
C_k&=\frac{1}{k+1}\left(\begin{matrix} 2k\\ k\\ \end{matrix}\right)
\nonumber\\
&=\prod_{1\le i\le j\le k-1}\frac{i+j+2}{i+j}. 
\end{align} 
The Catalan number is known as the multiplicity of the trivial representation, i.e. one-dimensional representation 
in the $2k$-th tensor power of the spin representation of $Spin(3)$ or equivalently that of the fundamental representation of $SU(2)$ \cite{MR106711}. 
Furthermore, the coefficients $a_{k\ \textrm{sp}}^{\mathfrak{so}(3)}(i)$ 
the multiplicity of the $(2i+1)$-dimensional irreducible representation $V_{2i}$, 
i.e. the rank-$i$ symmetric representation, in the $2k$-th tensor power of the spinor representation $V_1$ of $Spin(3)$ 
or that of the fundamental representation of $SU(2)$ \cite{MR2761939}. 
As we will see, the formula (\ref{halfBPSso3_wspkpt}) can be used to compute other correlators. 

The Catalan number is obtained from the generating function
\begin{align}
\frac{1-\sqrt{1-4x}}{2x}&=\sum_{k=0}^{\infty}a_{k\ \textrm{sp}}^{\mathfrak{so}(3)}(0) x^k, 
\end{align}
with $a_{0\ \textrm{sp}}^{\mathfrak{so}(3)}(0)$ $\equiv$ $1$. 
Furthermore, the sequences of the subsequent coefficients $a_{k\ \textrm{sp}}^{\mathfrak{so}(3)}(i)$ can be 
viewed as the $(2i+2)$-th  convolutions of the Catalan number in that they are generated by
\begin{align}
\frac{1}{x^{i+1}}
\left(\frac{1-\sqrt{1-4x}}{2}\right)^{2i+1}
&=\sum_{k=0}^{\infty}a_{k\ \textrm{sp}}^{\mathfrak{so}(3)}(i) x^k. 
\end{align}
Therefore the generating function for the $2k$-point function of the spinor Wilson lines is given by
\begin{align}
\sum_{k=0}^\infty x^k \langle \underbrace{W_{\textrm{sp}} \cdots W_{\textrm{sp}}}_{2k} \rangle^{Spin(3)}_{\textrm{$\frac12$BPS}}(\mathfrak{q}) 
&=\frac{1}{1-\mathfrak{q}^4}\cdot\frac{2(1-\sqrt{1-4x})}{4x-(1-\sqrt{1-4x})^2 \mathfrak{q}^2}. 
\end{align}

\subsubsection{Fundamental Wilson line}
The Wilson line operator in the fundamental (vector) representation exists in $SO(3)$ gauge theories as a basic electric line. 
The two-point function of the fundamental Wilson lines is 
\begin{align}
\label{so3_wfund}
&
\langle W_{\tiny \yng(1)} W_{\tiny \yng(1)}\rangle^{SO(3)}(t;q)
\nonumber\\
&=\frac12 \frac{(q)_{\infty}^2}{(q^{\frac12}t^{\pm2};q)_{\infty}}
\oint \frac{ds}{2\pi is}
\frac{(s^{\pm};q)_{\infty} (qs^{\pm};q)_{\infty}}
{(q^{\frac12}t^{2}s^{\pm};q)_{\infty}(q^{\frac12}t^{-2}s^{\pm};q)_{\infty}}(1+s+s^{-1})^2. 
\end{align}
This is equivalent to the two-point function of the Wilson lines in the rank-$2$ symmetric representation for $SU(2)$ gauge theory. 
It follows from \cite{Hatsuda:2023iwi} that 
\begin{align}
\label{so3_wfund_exact}
&
\langle W_{\tiny \yng(1)} W_{\tiny \yng(1)}\rangle^{SO(3)}(t;q)
=\langle W_{\tiny \yng(2)} W_{\tiny \yng(2)}\rangle^{SU(2)}(t;q)
\nonumber\\
&=
\frac{(q^{\frac12}t^{\pm2};q)_{\infty}}
{(qt^{\pm4};q)_{\infty}}
\left[
\frac32\sum_{m\in \mathbb{Z}\setminus \{0\}}
\left(
\frac{t^{2m}-t^{-2m}}{t-t^{-2}}
\frac{q^{\frac{m-1}{2}}}{1-q^m}
\right)
-\frac{2}{1-q}
-\frac{q^{\frac12}(t^2+t^{-2})}{1-q^2}
\right]. 
\end{align}

In the half-BPS limit, the two-point function of the Wilson lines in the fundamental representation for $SO(3)$ gauge theory is given by
\begin{align}
\label{halfBPSso3_wfund}
\langle W_{\tiny \yng(1)} W_{\tiny \yng(1)}\rangle^{SO(3)}_{\textrm{$\frac12$BPS}}(\mathfrak{q})
&=\langle W_{\tiny \yng(2)} W_{\tiny \yng(2)}\rangle^{SU(2)}_{\textrm{$\frac12$BPS}}(\mathfrak{q})
\nonumber\\
&=\frac{1+\mathfrak{q}^2+\mathfrak{q}^4}{1-\mathfrak{q}^4}
\nonumber\\
&=\frac{1-\mathfrak{q}^6}{(1-\mathfrak{q}^2)(1-\mathfrak{q}^4)}. 
\end{align}

Note that unlike the spinor Wilson line, the one-point function of the fundamental Wilson line is non-trivial. 
Although the $U(2)$ gauge theory has no non-vanishing one-point function, 
such correlation functions of the fundamental Wilson lines can be also expressed by making use of the results \cite{Hatsuda:2023iwi} for $U(2)$ gauge theory 
according to the decomposition into the spinor representations
\begin{align}
\langle \underbrace{W_{\tiny \yng(1)}\cdots W_{\tiny \yng(1)}}_{k} \rangle^{SO(3)}(t;q)&=
\sum_{i=0}^{k}\left(\begin{matrix}k\\i\\ \end{matrix}\right)
(-1)^i \langle \underbrace{W_{\textrm{sp}}\cdots W_{\textrm{sp}}}_{2(k-i)} \rangle^{SO(3)}(t;q), 
\end{align}
where we have used the relation $\chi_{\tiny \yng(1)}^{\mathfrak{so}(3)}$ $=$ ${\chi_{\textrm{sp}}^{\mathfrak{so}(3)}}^2-1$. 
In particular, the half-BPS limits of the correlation functions of the fundamental Wilson lines can be exactly computed using the formula (\ref{halfBPSso3_wspkpt}). 
For example, the one-point function is computed as 
\begin{align}
\langle W_{\tiny \yng(1)}\rangle^{SO(3)}(t;q)
&=\langle W_{\textrm{sp}} W_{\textrm{sp}}\rangle^{SO(3)}(t;q)-\mathcal{I}^{SO(3)}(t;q)
\nonumber\\
&=-\frac{(q^{\frac12}t^{\pm2};q)_{\infty}}{(q;q)_{\infty}^2}
\sum_{\begin{smallmatrix}
p_1, p_2\in \mathbb{Z}\\
p_1+1<p_2\\
\end{smallmatrix}}
\frac{(q^{\frac12}t^{-2})^{p_1+p_2-2}}
{(1-q^{p_1-\frac12}t^2) (1-q^{p_2-\frac12}t^2)}
\end{align}
and in the half-BPS limit it becomes 
\begin{align}
\label{halfBPSso3_1pt_fund}
\langle W_{\tiny \yng(1)}\rangle^{SO(3)}_{\textrm{$\frac12$BPS}}(\mathfrak{q})
&=\langle W_{\textrm{sp}} W_{\textrm{sp}}\rangle^{SO(3)}_{\textrm{$\frac12$BPS}}(\mathfrak{q})
-\mathcal{I}^{SO(3)}_{\textrm{$\frac12$BPS}}(\mathfrak{q})
\nonumber\\
&=\frac{\mathfrak{q}^2}{1-\mathfrak{q}^4}. 
\end{align}
Similarly, we can compute the higher-point functions although they are more computationally expensive. 
For example, the half-BPS limit of the three-point function can be evaluated as
\begin{align}
\langle W_{\tiny \yng(1)} W_{\tiny \yng(1)} W_{\tiny \yng(1)} \rangle^{SO(3)}_{\textrm{$\frac12$BPS}}(\mathfrak{q})
&=\langle W_{\textrm{sp}} W_{\textrm{sp}} W_{\textrm{sp}} W_{\textrm{sp}} W_{\textrm{sp}} W_{\textrm{sp}} \rangle^{SO(3)}_{\textrm{$\frac12$BPS}}(\mathfrak{q})
-3\langle W_{\textrm{sp}} W_{\textrm{sp}} W_{\textrm{sp}} W_{\textrm{sp}} \rangle^{SO(3)}_{\textrm{$\frac12$BPS}}(\mathfrak{q})
\nonumber\\
&+3\langle W_{\textrm{sp}} W_{\textrm{sp}} \rangle^{SO(3)}_{\textrm{$\frac12$BPS}}(\mathfrak{q})
-\mathcal{I}^{SO(3)}_{\textrm{$\frac12$BPS}}(\mathfrak{q})
\nonumber\\
&=\frac{1+3\mathfrak{q}^2+2\mathfrak{q}^4+\mathfrak{q}^6}{1-\mathfrak{q}^4}.
\end{align}
The four-, five- and six-point functions are computed as
\begin{align}
\langle W_{\tiny \yng(1)}W_{\tiny \yng(1)}W_{\tiny \yng(1)}W_{\tiny \yng(1)} \rangle^{SO(3)}_{\textrm{$\frac12$BPS}}(\mathfrak{q})
&=\frac{3+6\mathfrak{q}^2+6\mathfrak{q}^4+3\mathfrak{q}^6+\mathfrak{q}^8}{1-\mathfrak{q}^4}, \\
\langle W_{\tiny \yng(1)}W_{\tiny \yng(1)}W_{\tiny \yng(1)}W_{\tiny \yng(1)}W_{\tiny \yng(1)} \rangle^{SO(3)}_{\textrm{$\frac12$BPS}}(\mathfrak{q})
&=\frac{6+15\mathfrak{q}^2+15\mathfrak{q}^4+10\mathfrak{q}^6+4\mathfrak{q}^8+\mathfrak{q}^{10}}{1-\mathfrak{q}^4}, \\
\langle W_{\tiny \yng(1)}W_{\tiny \yng(1)}W_{\tiny \yng(1)}W_{\tiny \yng(1)}W_{\tiny \yng(1)}W_{\tiny \yng(1)} \rangle^{SO(3)}_{\textrm{$\frac12$BPS}}(\mathfrak{q})
&=\frac{15+36\mathfrak{q}^2+40\mathfrak{q}^4+29\mathfrak{q}^6+15\mathfrak{q}^8+5\mathfrak{q}^{10}+\mathfrak{q}^{12}}{1-\mathfrak{q}^4}. 
\end{align}
We note that the expansion coefficients of the multi-point functions of the fundamental Wilson lines have a combinatorial interpretation. 
We observe that the half-BPS limit of the $k$-point function of the fundamental Wilson lines can be expanded as
\begin{align}
\label{halfBPS_kpt_so3fund}
\langle \underbrace{W_{\tiny \yng(1)}\cdots W_{\tiny \yng(1)}}_{k} \rangle^{SO(3)}_{\textrm{$\frac12$BPS}}(\mathfrak{q})
&=\frac{\sum_{i=0}^k a_{k\ \tiny\yng(1)}^{\mathfrak{so}(3)}(i) \mathfrak{q}^{2i}}
{1-\mathfrak{q}^4}, 
\end{align}
where 
\begin{align}
\label{so3fund_coeff}
a_{k\ \tiny \yng(1)}^{\mathfrak{so}(3)}(i)&=c_k^{(i)}-c_k^{(i+1)}
\end{align}
and $c_k^{(i)}$ is obtained from the generating function \cite{euler1767observationes}
\begin{align}
(1+x+x^2)^n&=\sum_{i=-k}^{k} c_k^{(i)} x^{k+i}. 
\end{align}
The triangle of the numbers $a_{k\ \tiny \yng(1)}^{\mathfrak{so}(3)}(i)$ is shown as follows: 
\begin{align}
\label{so3fund_coeff_table}
\begin{array}{c|ccccccccc}
k\setminus i&0&1&2&3&4&5&6&7& \\ \hline 
0&1&&&&&&&& \\ 
1&0&1&&&&&&& \\ 
2&1&1&1&&&&&& \\ 
3&1&3&2&1&&&&& \\ 
4&3&6&6&3&1&&&& \\ 
5&6&15&15&10&4&1&&& \\ 
6&15&36&40&29&15&5&1&& \\ 
7&36&91&105&84&49&21&6&1& \\ 
8&91&232&280&238&154&76&28&7&1 \\ 
\end{array}
\end{align}
The numbers (\ref{so3fund_coeff_table}) encode the multiplicities of the rank-$i$ symmetric representations 
in the $k$-th symmetric tensor power of the fundamental representations of $SO(3)$ \cite{MR4023517}. 
In particular, the first two coefficients $a_{k\ \tiny \yng(1)}^{\mathfrak{so}(3)}(0)$ and $a_{k\ \tiny \yng(1)}^{\mathfrak{so}(3)}(1)$ 
appearing in the expansion (\ref{halfBPS_kpt_so3fund}) are identified with the Riordan numbers $R_k$ and $R_{k+1}$. 
The Riordan number is given by
\begin{align}
R_n&=\sum_{i=0}^{n}(-1)^{n-i}\left(
\begin{matrix}
n\\
i\\
\end{matrix}
\right)C_i, 
\end{align}
where $C_i$ is the $i$-th Catalan number. 
The generating function for the $n$-th Riordan number is given by
\begin{align}
\sum_{n=1}^{\infty}R_n x^n=\frac{1}{2x}\left(
1-\frac{\sqrt{1-3x}}{\sqrt{1+x}}
\right). 
\end{align}
The Riordan number is also known to be the multiplicity of the trivial representation in the $k$-th symmetric tensor power  
of the fundamental representation of $SO(3)$ or equivalently that of the adjoint representation of $SU(2)$ \cite{MR1691863}. 

\subsubsection{Symmetric Wilson lines}
Similarly, the two-point function of the Wilson lines in the rank-$k$ symmetric representation for $SO(3)$ gauge theory is equivalent to 
the two-point function of the Wilson lines in the rank-$2k$ symmetric representation in $SU(2)$ gauge theory. 
Making use of the closed-form in \cite{Hatsuda:2023iwi}, 
it can be written as
\begin{align}
\label{so3_wsymk}
\langle W_{(k)} W_{(k)}\rangle^{SO(3)}(t;q)
&=\frac{(q^{\frac12}t^{\pm2};q)_{\infty}}
{(qt^{\pm4};q)_{\infty}}
\Biggl[
\frac{2k+1}{2}\sum_{m\in \mathbb{Z}\setminus \{0\}}
\left(
\frac{t^{2m}-t^{-2m}}{t-t^{-2}}
\frac{q^{\frac{m-1}{2}}}{1-q^m}
\right)
\nonumber\\
&-\sum_{m=1}^{2k}
(2k-m+1)
\left(
\frac{t^{2m}-t^{-2m}}{t-t^{-2}}
\frac{q^{\frac{m-1}{2}}}{1-q^m}
\right)
\Biggr]. 
\end{align}

It follows that the half-BPS limit of the two-point function of the Wilson line in the rank-$k$ symmetric representation is given by
\begin{align}
\label{halfBPSso3_wsymk}
\langle W_{(k)} W_{(k)}\rangle^{SO(3)}_{\textrm{$\frac12$BPS}}(\mathfrak{q})
&=\frac{1+\mathfrak{q}^2+\cdots+\mathfrak{q}^{4k}}{1-\mathfrak{q}^4}
\nonumber\\
&=\frac{1-\mathfrak{q}^{4k+2}}{(1-\mathfrak{q}^2) (1-\mathfrak{q}^4)}. 
\end{align}
In the large representation limit $k\rightarrow \infty$, the expression (\ref{halfBPSso3_wsymk}) becomes 
\begin{align}
\label{halfBPSso3_wsymlarge}
\langle W_{(\infty)} W_{(\infty)}\rangle^{SO(3)}_{\textrm{$\frac12$BPS}}(\mathfrak{q})
&=\frac{1}{(1-\mathfrak{q}^2) (1-\mathfrak{q}^4)}. 
\end{align}

There also exist non-trivial correlation functions of the Wilson lines in the irreducible representations for $SO(3)$ gauge theory, 
whose counterparts vanish in unitary gauge theory. 
Again they can be expressed in terms of the multi-point functions of the spinor Wilson lines. 
For rank-$2$ symmetric Wilson line, we have
\begin{align}
\langle \underbrace{W_{\tiny \yng(2)} \cdots W_{\tiny \yng(2)}}_{k} \rangle^{SO(3)}(t;q)
&=\sum_{k_1+k_2+k_3=k}
\left(
\begin{matrix}
k\\
k_1,k_2,k_3\\
\end{matrix}
\right)
(-3)^{k_2}
\langle \underbrace{W_{\textrm{sp}}\cdots W_{\textrm{sp}}}_{4k_1+2k_2} \rangle^{SO(3)}(t;q)
\end{align}
according to the decomposition 
$\chi_{\tiny \yng(2)}^{\mathfrak{so}(3)}$ $=$ 
${\chi_{\textrm{sp}}^{\mathfrak{so}(3)}}^4-3{\chi_{\textrm{sp}}^{\mathfrak{so}(3)}}^2+1$. 
For example, in the half-BPS limit, we obtain the exact forms 
\begin{align}
\langle W_{\tiny \yng(2)} \rangle^{SO(3)}_{\textrm{$\frac12$BPS}}(\mathfrak{q})
&=\langle W_{\textrm{sp}} W_{\textrm{sp}}W_{\textrm{sp}} W_{\textrm{sp}}  \rangle^{SO(3)}_{\textrm{$\frac12$BPS}}(\mathfrak{q})
-3\langle W_{\textrm{sp}} W_{\textrm{sp}}  \rangle^{SO(3)}_{\textrm{$\frac12$BPS}}(\mathfrak{q})
+\mathcal{I}^{SO(3)}_{\textrm{$\frac12$BPS}}(\mathfrak{q})
\nonumber\\
&=\frac{\mathfrak{q}^4}{1-\mathfrak{q}^4}, \\
\langle W_{\tiny \yng(2)} W_{\tiny \yng(2)} W_{\tiny \yng(2)} \rangle^{SO(3)}_{\textrm{$\frac12$BPS}}(\mathfrak{q})
&=\langle \underbrace{W_{\textrm{sp}} \cdots W_{\textrm{sp}}}_{12} \rangle^{SO(3)}_{\textrm{$\frac12$BPS}}(\mathfrak{q})
-9\langle \underbrace{W_{\textrm{sp}} \cdots W_{\textrm{sp}}}_{10} \rangle^{SO(3)}_{\textrm{$\frac12$BPS}}(\mathfrak{q})
\nonumber\\
&+30\langle \underbrace{W_{\textrm{sp}} \cdots W_{\textrm{sp}}}_{8} \rangle^{SO(3)}_{\textrm{$\frac12$BPS}}(\mathfrak{q})
-45\langle \underbrace{W_{\textrm{sp}} \cdots W_{\textrm{sp}}}_{6} \rangle^{SO(3)}_{\textrm{$\frac12$BPS}}(\mathfrak{q})
\nonumber\\
&+30\langle \underbrace{W_{\textrm{sp}} \cdots W_{\textrm{sp}}}_{4} \rangle^{SO(3)}_{\textrm{$\frac12$BPS}}(\mathfrak{q})
-9\langle W_{\textrm{sp}} W_{\textrm{sp}} \rangle^{SO(3)}_{\textrm{$\frac12$BPS}}(\mathfrak{q})
+\mathcal{I}^{SO(3)}_{\textrm{$\frac12$BPS}}(\mathfrak{q})
\nonumber\\
&=\frac{1+3\mathfrak{q}^2+5\mathfrak{q}^4+4\mathfrak{q}^6+3\mathfrak{q}^8+2\mathfrak{q}^{10}+\mathfrak{q}^{12}}
{1-\mathfrak{q}^4}. 
\end{align}
Similarly, for rank-$3$ symmetric Wilson line we have
\begin{align}
&
\langle \underbrace{W_{\tiny \yng(3)} \cdots W_{\tiny \yng(3)}}_{k} \rangle^{SO(3)}(t;q)
\nonumber\\
&=\sum_{k_1+k_2+k_3+k_4=k}
\left(
\begin{matrix}
k\\
k_1,k_2,k_3,k_4\\
\end{matrix}
\right)
(-1)^{k_2+k_4}
5^{k_2}
6^{k_3}
\langle \underbrace{W_{\textrm{sp}}\cdots W_{\textrm{sp}}}_{6k_2+4k_1+2k_2} \rangle^{SO(3)}(t;q). 
\end{align}
For example, the half-BPS limits of the one- and three-point functions can be computed as
\begin{align}
&
\langle W_{\tiny \yng(3)} \rangle^{SO(3)}_{\textrm{$\frac12$BPS}}(\mathfrak{q})
\nonumber\\
&=\langle \underbrace{W_{\textrm{sp}} \cdots W_{\textrm{sp}}}_{6} \rangle^{SO(3)}_{\textrm{$\frac12$BPS}}(\mathfrak{q})
-5\langle \underbrace{W_{\textrm{sp}} \cdots W_{\textrm{sp}}}_{4} \rangle^{SO(3)}_{\textrm{$\frac12$BPS}}(\mathfrak{q})
+6\langle W_{\textrm{sp}} W_{\textrm{sp}} \rangle^{SO(3)}_{\textrm{$\frac12$BPS}}(\mathfrak{q})
-\mathcal{I}^{SO(3)}_{\textrm{$\frac12$BPS}}(\mathfrak{q})
\nonumber\\
&=\frac{\mathfrak{q}^6}{1-\mathfrak{q}^4}
, \\
&
\langle W_{\tiny \yng(3)} W_{\tiny \yng(3)} W_{\tiny \yng(3)} \rangle^{SO(3)}_{\textrm{$\frac12$BPS}}(\mathfrak{q})
\nonumber\\
&=\langle \underbrace{W_{\textrm{sp}} \cdots W_{\textrm{sp}}}_{18} \rangle^{SO(3)}_{\textrm{$\frac12$BPS}}(\mathfrak{q})
-15\langle \underbrace{W_{\textrm{sp}} \cdots W_{\textrm{sp}}}_{16} \rangle^{SO(3)}_{\textrm{$\frac12$BPS}}(\mathfrak{q})
+93\langle \underbrace{W_{\textrm{sp}} \cdots W_{\textrm{sp}}}_{14} \rangle^{SO(3)}_{\textrm{$\frac12$BPS}}(\mathfrak{q})
\nonumber\\
&-308\langle \underbrace{W_{\textrm{sp}} \cdots W_{\textrm{sp}}}_{12} \rangle^{SO(3)}_{\textrm{$\frac12$BPS}}(\mathfrak{q})
+588\langle \underbrace{W_{\textrm{sp}} \cdots W_{\textrm{sp}}}_{10} \rangle^{SO(3)}_{\textrm{$\frac12$BPS}}(\mathfrak{q})
-651\langle \underbrace{W_{\textrm{sp}} \cdots W_{\textrm{sp}}}_{8} \rangle^{SO(3)}_{\textrm{$\frac12$BPS}}(\mathfrak{q})
\nonumber\\
&+399\langle \underbrace{W_{\textrm{sp}} \cdots W_{\textrm{sp}}}_{6} \rangle^{SO(3)}_{\textrm{$\frac12$BPS}}(\mathfrak{q})
-123\langle \underbrace{W_{\textrm{sp}} \cdots W_{\textrm{sp}}}_{4} \rangle^{SO(3)}_{\textrm{$\frac12$BPS}}(\mathfrak{q})
+18\langle W_{\textrm{sp}} W_{\textrm{sp}} \rangle^{SO(3)}_{\textrm{$\frac12$BPS}}(\mathfrak{q})
-\mathcal{I}^{SO(3)}_{\textrm{$\frac12$BPS}}(\mathfrak{q})
\nonumber \\
&=\frac{1+3\mathfrak{q}^2+5\mathfrak{q}^4+7\mathfrak{q}^6+6\mathfrak{q}^8+5\mathfrak{q}^{10}+4\mathfrak{q}^{12}+3\mathfrak{q}^{14}+2\mathfrak{q}^{16}+\mathfrak{q}^{18}}
{1-\mathfrak{q}^4}. 
\end{align}

More generally, we can compute the correlators of higher-rank symmetric Wilson lines by making use of the decomposition
\begin{align}
\chi_{(k)}^{\mathfrak{so}(3)}&=\sum_{n=0}^{k}(-1)^n \left(\begin{matrix}
2k-n\\
n\\
\end{matrix}\right)
{\chi_{\textrm{sp}}^{\mathfrak{so}(3)}}^{2k-2n}. 
\end{align}
Here we remark several exact closed-form expressions for the half-BPS limits of the correlators of the symmetric Wilson lines with general ranks. 
We find that the half-BPS limit of the one-point function is given by
\begin{align}
\label{halfBPSso3_1pt_symk}
\langle W_{(k)}\rangle^{SO(3)}_{\textrm{$\frac12$BPS}}(\mathfrak{q})
&=\frac{\mathfrak{q}^{2k}}{1-\mathfrak{q}^4}. 
\end{align}
While we have found the two-point function (\ref{halfBPSso3_wsymk}) of the rank-$k$ symmetric Wilson lines, 
the two-point functions of a pair of the Wilson lines transforming in the representations of different ranks is also non-trivial. 
For $k\le l$, we obtain the half-BPS limit of the two-point function of the form
\begin{align}
\label{halfBPSso3_2pt_symksyml}
\langle W_{(k)}W_{(l)}\rangle^{SO(3)}_{\textrm{$\frac12$BPS}}(\mathfrak{q})
&=\frac{\mathfrak{q}^{2(l-k)} (1-\mathfrak{q}^{4k+2})}{(1-\mathfrak{q}^2)(1-\mathfrak{q}^4)}. 
\end{align}
Note that the two-point function (\ref{halfBPSso3_2pt_symksyml}) vanishes in the large representation limit except for $k=l$. 
The half-BPS limit of the three-point function is 
\begin{align}
\langle W_{(k)}W_{(k)}W_{(k)} \rangle^{SO(3)}_{\textrm{$\frac12$BPS}}(\mathfrak{q})
&=\frac{1+\mathfrak{q}^2-3\mathfrak{q}^{2k+2}+\mathfrak{q}^{6k+4}}{(1-\mathfrak{q}^2)^2(1-\mathfrak{q}^4)}. 
\end{align}
In the large representation limit, it reduces to 
\begin{align}
\langle W_{(\infty)} W_{(\infty)} W_{(\infty)} \rangle^{SO(3)}_{\textrm{$\frac12$BPS}}(\mathfrak{q})
&=\frac{1}{(1-\mathfrak{q}^2)^3}. 
\end{align}
For $k$-point function of the symmetric Wilson line operators of equal rank with $k>3$, the expression diverges in the large representation limit. 
For example, the four-point function is evaluated as
\begin{align}
\langle W_{(k)}W_{(k)}W_{(k)}W_{(k)} \rangle^{SO(3)}_{\textrm{$\frac12$BPS}}(\mathfrak{q})
&=\frac{2k+1-3\mathfrak{q}^2-(2k+1)\mathfrak{q}^4+4\mathfrak{q}^{4k+4}-\mathfrak{q}^{8k+6}}
{(1-\mathfrak{q}^2)^3(1-\mathfrak{q}^4)}. 
\end{align}

\subsection{$\mathfrak{so}(5)$}

\subsubsection{Spinor Wilson line}
For $\mathcal{N}=4$ SYM theory associated with the $\mathfrak{so}(5)$ gauge algebra, 
the Wilson line operator in the spinor representation is allowed for $Spin(5)$ gauge theory. 
There is no non-trivial one-point function of the spinor Wilson line. 
The two-point function of the Wilson lines in the spinor representation takes the form 
\begin{align}
\label{so5_wsp}
&
\langle W_{\textrm{sp}} W_{\textrm{sp}}\rangle^{Spin(5)}(t;q)
\nonumber\\
&=\frac18 \frac{(q)_{\infty}^4}{(q^{\frac12}t^{\pm};q)_{\infty}^2}
\oint \prod_{i=1}^2 \frac{ds_i}{2\pi is_i}
\frac{(s_i^{\pm};q)_{\infty}(qs_i^{\pm};q)_{\infty}}
{(q^{\frac12}t^2 s_i^{\pm};q)_{\infty}(q^{\frac12}t^{-2}s_i^{\pm};q)_{\infty}}
\nonumber\\
&\times 
\frac{(s_1^{\pm}s_2^{\mp};q)_{\infty}(s_1^{\pm}s_2^{\pm};q)_{\infty}
(qs_1^{\pm}s_2^{\mp};q)_{\infty}(qs_1^{\pm}s_2^{\pm};q)_{\infty}}
{(q^{\frac12}t^2s_1^{\pm}s_2^{\mp};q)_{\infty}(q^{\frac12}t^2s_1^{\pm}s_2^{\pm};q)_{\infty}
(q^{\frac12}t^{-2}s_1^{\pm}s_2^{\mp};q)_{\infty}(q^{\frac12}t^{-2}s_1^{\pm}s_2^{\pm};q)_{\infty}}
\prod_{i=1}^2(s_i^{\frac12}+s_i^{-\frac12})^2. 
\end{align}
The exact closed-form expressions for the two-point functions are not simply derived from the results for unitary gauge theories, 
however, as we have the isomorphism $Spin(5)$ $\cong$ $USp(4)$, 
this agrees with the two-point function (\ref{sp4_wfund}) of the Wilson lines in the fundamental representation for $USp(4)$ gauge theory. 

The spinor Wilson line for $Spin(5)$ is conjecturally S-dual to the 't Hooft lines of $B=(\frac12,\frac12)$ in $USp(4)/\mathbb{Z}_2$ gauge theory. 
The two-point function of the dual 't Hooft lines can be evaluated as 
\begin{align}
\label{sp4_tsp}
&
\langle T_{(\frac12,\frac12)} T_{(\frac12,\frac12)}\rangle^{USp(4)/\mathbb{Z}_2}(t;q)
\nonumber\\
&=\frac12 \frac{(q)_{\infty}^4}{(q^{\frac12}t^{\pm2};q)_{\infty}^2}
\oint \prod_{i=1}^2 \frac{ds_i}{2\pi is_i}
\frac{(q^{\frac12}s_i^{\pm2};q)_{\infty}(q^{\frac32}s_i^{\pm2};q)_{\infty}}
{(qt^2s_i^{\pm2};q)_{\infty}(qt^{-2}s_i^{\pm2};q)_{\infty}}
\nonumber\\
&\times 
\frac{(s_1^{\pm}s_2^{\mp};q)_{\infty}(q^{\frac12}s_1^{\pm}s_2^{\pm};q)_{\infty}
(qs_1^{\pm}s_2^{\mp};q)_{\infty}(q^{\frac32}s_1^{\pm}s_2^{\pm};q)_{\infty}}
{(q^{\frac12}t^2s_1^{\pm}s_2^{\mp};q)_{\infty}(qt^2s_1^{\pm}s_2^{\pm};q)_{\infty}
(q^{\frac12}t^{-2}s_1^{\pm}s_2^{\mp};q)_{\infty}(qt^{-2}s_1^{\pm}s_2^{\pm};q)_{\infty}}. 
\end{align}
The two-point functions (\ref{so5_wsp}) and (\ref{sp4_tsp}) agree with each other
\begin{align}
\langle W_{\textrm{sp}} W_{\textrm{sp}}\rangle^{Spin(5)}(t;q)
&=\langle T_{(\frac12,\frac12)} T_{(\frac12,\frac12)}\rangle^{USp(4)/\mathbb{Z}_2}(t;q). 
\end{align}

The expression (\ref{sp4_tsp}) reduces to the half-BPS index of $\mathcal{N}=4$ $U(2)$ gauge theory in the half-BPS limit. 
Accordingly, we find that the half-BPS limits of the two-point functions (\ref{so5_wsp}) and (\ref{sp4_tsp}) are given by
\begin{align}
\label{halfBPSso5_wsp}
\langle W_{\textrm{sp}} W_{\textrm{sp}}\rangle^{Spin(5)}_{\textrm{$\frac12$BPS}}(\mathfrak{q})
&=\langle T_{(\frac12,\frac12)} T_{(\frac12,\frac12)}\rangle^{USp(4)/\mathbb{Z}_2}_{\textrm{$\frac12$BPS}}(\mathfrak{q})
\nonumber\\
&=\frac{1+\mathfrak{q}^2+\mathfrak{q}^4+\mathfrak{q}^6}{(1-\mathfrak{q}^4)(1-\mathfrak{q}^8)}
\nonumber\\
&=\frac{1}{(1-\mathfrak{q}^2)(1-\mathfrak{q}^4)}. 
\end{align}

More generally, even-point correlation functions of the spinor Wilson lines are non-trivial. 
For four-, six-, eight- and ten-point functions, we find the following exact form expressions in the half-BPS limits: 
\begin{align}
\langle \underbrace{W_{\textrm{sp}} \cdots W_{\textrm{sp}}}_{4} \rangle^{Spin(5)}_{\textrm{$\frac12$BPS}}(\mathfrak{q})
&=\frac{3+6\mathfrak{q}^2+8\mathfrak{q}^4+9\mathfrak{q}^6+6\mathfrak{q}^8+3\mathfrak{q}^{10}+\mathfrak{q}^{12}}
{(1-\mathfrak{q}^4)(1-\mathfrak{q}^8)}, \\
\langle \underbrace{W_{\textrm{sp}} \cdots W_{\textrm{sp}}}_{6} \rangle^{Spin(5)}_{\textrm{$\frac12$BPS}}(\mathfrak{q})
&=\frac{1}
{(1-\mathfrak{q}^4)(1-\mathfrak{q}^8)}
(14+40\mathfrak{q}^2+66\mathfrak{q}^{4}+85\mathfrak{q}^6
\nonumber\\
&+81\mathfrak{q}^{8}+59\mathfrak{q}^{10}+34\mathfrak{q}^{12}+15\mathfrak{q}^{14}+5\mathfrak{q}^{16}+\mathfrak{q}^{18}), \\
\langle \underbrace{W_{\textrm{sp}} \cdots W_{\textrm{sp}}}_{8} \rangle^{Spin(5)}_{\textrm{$\frac12$BPS}}(\mathfrak{q})
&=\frac{1}{(1-\mathfrak{q}^4)(1-\mathfrak{q}^8)}
(84+300\mathfrak{q}^2+581\mathfrak{q}^4+840\mathfrak{q}^6+945\mathfrak{q}^8+842\mathfrak{q}^{10}
\nonumber\\
&+616\mathfrak{q}^{12}+378\mathfrak{q}^{14}+195\mathfrak{q}^{16}+83\mathfrak{q}^{18}+28\mathfrak{q}^{20}+7\mathfrak{q}^{22}+\mathfrak{q}^{24}). 
\end{align}
They can be written as
\begin{align}
\langle \underbrace{W_{\textrm{sp}} \cdots W_{\textrm{sp}}}_{2k} \rangle^{Spin(5)}_{\textrm{$\frac12$BPS}}(\mathfrak{q})
&=\frac{\sum_{i=0}^{3k} a_{k\ \textrm{sp}}^{\mathfrak{so}(5)}(i) \mathfrak{q}^{2i}}
{(1-\mathfrak{q}^4)(1-\mathfrak{q}^8)}. 
\end{align}
We see that the coefficient $a_{k\ \textrm{sp}}^{\mathfrak{so}(5)}(0)$ is given by
\begin{align}
\label{spin5_spmult1}
a_{k\ \textrm{sp}}^{\mathfrak{so}(5)}(0) &=C_{k}C_{k+2}-C_{k+1}^2
\nonumber\\
&=\frac{24 (2k+1)! (2k-1)!}{(k-1)!k! (k+2)!(k+3)!}
\nonumber\\
&=\prod_{1\le i\le j\le k-1}\frac{i+j+4}{i+j}, 
\end{align}
where $C_k$ is the $k$-th Catalan number. 
The number (\ref{spin5_spmult1}) is known to be the multiplicity of the trivial representation 
in the $2k$-th tensor power of the spinor representation of $Spin(5)$. 
It is generated from
\begin{align}
{_{3}F_{2}}\left(1,\frac12,\frac32;3,4;16x\right)
&=\sum_{k=0}^{\infty} a_{k\ \textrm{sp}}^{\mathfrak{so}(5)}(0) x^k, 
\end{align}
where 
\begin{align}
{_{p}F_{q}}\left(a_1,\cdots,a_p;b_1,\cdots,b_q;z\right)
&=\sum_{k=0}^{\infty} \frac{(a_1)_k (a_2)_k\cdots (a_p)_k}
{(b_1)_k (b_2)_k\cdots (b_q)_k}\frac{z^k}{k!}
\end{align}
is the generalized hypergeometric function. 
Also we observe that the coefficient $a_{k\ \textrm{sp}}^{\mathfrak{so}(5)}(1)$ agrees with
\begin{align}
a_{k\ \textrm{sp}}^{\mathfrak{so}(5)}(1) &=\frac{60(2k)!(2k+2)!}{(k-1)!k! (k+3)!(k+4)!}. 
\end{align}

\subsubsection{Fundamental Wilson line}
The Wilson line operator transforming in the fundamental representation 
appears as a minimal electrically charged line in $SO(5)$ gauge theory. 

For such Wilson lines, the one-point functions are non-trivial. 
We find that the half-BPS limits of the $k$-point correlation functions of the fundamental Wilson lines for $SO(5)$ SYM theory take the from 
\begin{align}
\langle \underbrace{W_{\tiny \yng(1)}\cdots W_{\tiny \yng(1)}}_{k} \rangle^{SO(5)}_{\textrm{$\frac12$BPS}}(\mathfrak{q})
&=\frac{\sum_{i=0}^{2k} a_{k\ \tiny \yng(1)}^{\mathfrak{so}(5)}(i) \mathfrak{q}^{2i}}{(1-\mathfrak{q}^4)(1-\mathfrak{q}^8)}, 
\end{align}
where $a_{k\ \tiny \yng(1)}^{\mathfrak{so}(5)}(i)$ is some non-negative integer. 
In other words, the normalized correlator is given by a polynomial in $\mathfrak{q}^2$ with positive integer coefficients. 
For example, for $k=1,\cdots, 5$, we obtain 
\begin{align}
\label{halfBPSso5_1pt_fund}
\langle W_{\tiny \yng(1)} \rangle^{SO(5)}_{\textrm{$\frac12$BPS}}(\mathfrak{q})
&=\frac{\mathfrak{q}^4}{(1-\mathfrak{q}^4)(1-\mathfrak{q}^8)}, \\
\label{halfBPSso5_2pt_fund}
\langle W_{\tiny \yng(1)} W_{\tiny \yng(1)}\rangle^{SO(5)}_{\textrm{$\frac12$BPS}}(\mathfrak{q})
&=\frac{1+\mathfrak{q}^2+\mathfrak{q}^4+\mathfrak{q}^6+\mathfrak{q}^8}{(1-\mathfrak{q}^4)(1-\mathfrak{q}^8)}
\nonumber\\
&=\frac{1-\mathfrak{q}^{10}}{(1-\mathfrak{q}^2)(1-\mathfrak{q}^4)(1-\mathfrak{q}^8)}, \\
\label{halfBPSso5_3pt_fund}
\langle W_{\tiny \yng(1)} W_{\tiny \yng(1)} W_{\tiny \yng(1)}\rangle^{SO(5)}_{\textrm{$\frac12$BPS}}(\mathfrak{q})
&=\frac{\mathfrak{q}^2+3\mathfrak{q}^4+3\mathfrak{q}^6+3\mathfrak{q}^8+2\mathfrak{q}^{10}+\mathfrak{q}^{12}}
{(1-\mathfrak{q}^4)(1-\mathfrak{q}^8)}, \\
\label{halfBPSso5_4pt_fund}
\langle W_{\tiny \yng(1)} W_{\tiny \yng(1)} W_{\tiny \yng(1)} W_{\tiny \yng(1)}\rangle^{SO(5)}_{\textrm{$\frac12$BPS}}(\mathfrak{q})
&=\frac{1}{(1-\mathfrak{q}^4)(1-\mathfrak{q}^8)}
(3+3\mathfrak{q}^2+9\mathfrak{q}^4+15\mathfrak{q}^6+12\mathfrak{q}^8
\nonumber\\
&+12\mathfrak{q}^{10}+6\mathfrak{q}^{12}+\mathfrak{q}^{16}), \\
\langle W_{\tiny \yng(1)}W_{\tiny \yng(1)} W_{\tiny \yng(1)} W_{\tiny \yng(1)} W_{\tiny \yng(1)}\rangle^{SO(5)}_{\textrm{$\frac12$BPS}}(\mathfrak{q})
&=\frac{1}{(1-\mathfrak{q}^4)(1-\mathfrak{q}^8)}
(1+10\mathfrak{q}^2+24\mathfrak{q}^4+36\mathfrak{q}^6+44\mathfrak{q}^8
\nonumber\\
&+41\mathfrak{q}^{10}+31\mathfrak{q}^{12}+19\mathfrak{q}^{14}+10\mathfrak{q}^{16}
+4\mathfrak{q}^{18}+\mathfrak{q}^{20}). 
\end{align}
It follows that the coefficient $a_{k\ \tiny \yng(1)}^{\mathfrak{so}(5)}(0) $ is given by
\begin{align}
\label{so5_fundmult0}
a_{k\ \tiny \yng(1)}^{\mathfrak{so}(5)}(0) 
&=\sum_{i=0}^{\lfloor \frac{k}{2} \rfloor}C_{i}C_{i+1}
\left(
\begin{matrix}
k\\
2i\\
\end{matrix}
\right)
-\sum_{i=0}^{\lfloor \frac{k+1}{2}\rfloor}
C_i^2 
\left(
\begin{matrix}
k\\
2i-1\\
\end{matrix}
\right)
\nonumber\\
&=-k { _{3}F_{2}}\left(\frac32,\frac12-\frac{k}{2};3,3;16\right)
+ { _{3}F_{2}}\left(\frac32,\frac12-\frac{k}{2};2,3;16\right). 
\end{align}
The number (\ref{so5_fundmult0}) is known to be equal to the multiplicity of the trivial representation 
in the $k$-th tensor power of the fundamental representation of $SO(5)$ \cite{MR2697358}. 

\subsubsection{Adjoint Wilson line}
Next consider the adjoint Wilson line that transforms in the rank-$2$ antisymmetric representation. 
We find that 
the half-BPS limits of the correlation functions of the Wilson lines in the rank-$2$ antisymmetric representation take the form
\begin{align}
\langle \underbrace{W_{\tiny \yng(1,1)}\cdots W_{\tiny \yng(1,1)}}_{k} \rangle^{SO(5)}_{\textrm{$\frac12$BPS}}(\mathfrak{q})
&=\frac{\sum_{i=0}^{3k} a_{k\ \tiny \yng(1,1)}^{\mathfrak{so}(5)}(i) \mathfrak{q}^{2i}}
{(1-\mathfrak{q}^4)(1-\mathfrak{q}^8)}, 
\end{align}
where $a_{k\ \tiny \yng(1,1)}^{\mathfrak{so}(5)}(i)$ are some positive definite integers. 
For example, for $k=1,\cdots, 4$ we have 
\begin{align}
\label{halfBPSso5_wasym2_1pt}
\langle W_{\tiny \yng(1,1)} \rangle^{SO(5)}_{\textrm{$\frac12$BPS}}(\mathfrak{q})
&=\frac{\mathfrak{q}^2+\mathfrak{q}^6}{(1-\mathfrak{q}^4)(1-\mathfrak{q}^8)}
\nonumber\\
&=\frac{\mathfrak{q}^2}{(1-\mathfrak{q}^4)^2}, \\
\label{halfBPSso5_wasym2}
\langle W_{\tiny \yng(1,1)} W_{\tiny \yng(1,1)}\rangle^{SO(5)}_{\textrm{$\frac12$BPS}}(\mathfrak{q})
&=\frac{1+\mathfrak{q}^2+3\mathfrak{q}^4+2\mathfrak{q}^6+3\mathfrak{q}^8+\mathfrak{q}^{10}+\mathfrak{q}^{12}}
{(1-\mathfrak{q}^4)(1-\mathfrak{q}^8)}
\nonumber\\
&=\frac{(1-\mathfrak{q}^6)(1-\mathfrak{q}^8)}{(1-\mathfrak{q}^2)(1-\mathfrak{q}^4)^3}, \\
\langle W_{\tiny \yng(1,1)} W_{\tiny \yng(1,1)} W_{\tiny \yng(1,1)}\rangle^{SO(5)}_{\textrm{$\frac12$BPS}}(\mathfrak{q})
&=\frac{1}{(1-\mathfrak{q}^4)(1-\mathfrak{q}^8)}
(
1+6\mathfrak{q}^2+9\mathfrak{q}^4+16\mathfrak{q}^6+15\mathfrak{q}^{8}
\nonumber\\
&+15\mathfrak{q}^{10}+9\mathfrak{q}^{12}
+6\mathfrak{q}^{14}+2\mathfrak{q}^{16}+\mathfrak{q}^{18}
), \\
\langle W_{\tiny \yng(1,1)} W_{\tiny \yng(1,1)}W_{\tiny \yng(1,1)} W_{\tiny \yng(1,1)}\rangle^{SO(5)}_{\textrm{$\frac12$BPS}}(\mathfrak{q})
&=\frac{1}{(1-\mathfrak{q}^4)(1-\mathfrak{q}^8)}
(
6+22\mathfrak{q}^2+54\mathfrak{q}^4+82\mathfrak{q}^6+15\mathfrak{q}^8
\nonumber\\
&+15\mathfrak{q}^{10}+9\mathfrak{q}^{12}+6\mathfrak{q}^{14}+2\mathfrak{q}^{16}+\mathfrak{q}^{18}
). 
\end{align}

\subsubsection{Symmetric Wilson lines}
While the rank of the antisymmetric representation is not greater than $2$, 
there is no restriction on the rank for the symmetric representation of $SO(5)$. 
Again it follows that the half-BPS limits of the correlation functions of the Wilson lines in the rank-$2$ symmetric representations can be written as
\begin{align}
\langle \underbrace{W_{\tiny (2)} \cdots W_{\tiny (2)}}_{k} \rangle^{SO(5)}_{\textrm{$\frac12$BPS}}(\mathfrak{q})
&=\frac{\sum_{i=0}^{4k}a_{k\ \tiny \yng(2)}^{\mathfrak{so}(5)}(i) \mathfrak{q}^{2i}
}{(1-\mathfrak{q}^4)(1-\mathfrak{q}^8)}, 
\end{align}
with $a_{k\ \tiny \yng(2)}^{\mathfrak{so}(5)}(i)$ being some positive definite integers. 
For example, we find
\begin{align}
\langle W_{\tiny \yng(2)} \rangle^{SO(5)}_{\textrm{$\frac12$BPS}}(\mathfrak{q})
&=\frac{\mathfrak{q}^4+\mathfrak{q}^8}{(1-\mathfrak{q}^4)(1-\mathfrak{q}^8)}
\nonumber\\
&=\frac{\mathfrak{q}^4}{(1-\mathfrak{q}^4)^2}, \\
\langle W_{\tiny \yng(2)} W_{\tiny \yng(2)}\rangle^{SO(5)}_{\textrm{$\frac12$BPS}}(\mathfrak{q})
&=\frac{1+\mathfrak{q}^2+2\mathfrak{q}^4+2\mathfrak{q}^6+3\mathfrak{q}^8+2\mathfrak{q}^{10}+3\mathfrak{q}^{12}+\mathfrak{q}^{14}+\mathfrak{q}^{16}}
{(1-\mathfrak{q}^4)(1-\mathfrak{q}^8)}, \\
\langle W_{\tiny \yng(2)} W_{\tiny \yng(2)}W_{\tiny \yng(2)} \rangle^{SO(5)}_{\textrm{$\frac12$BPS}}(\mathfrak{q})
&=
\frac{1}{(1-\mathfrak{q}^4)(1-\mathfrak{q}^8)}
\nonumber\\
&\times 
(
1+3\mathfrak{q}^2+9\mathfrak{q}^4+13\mathfrak{q}^6+20\mathfrak{q}^{8}+21\mathfrak{q}^{10}
\nonumber\\
&+22\mathfrak{q}^{12}+18\mathfrak{q}^{14}+15\mathfrak{q}^{16}+9\mathfrak{q}^{18}
+6\mathfrak{q}^{20}+2\mathfrak{q}^{22}+\mathfrak{q}^{24}
). 
\end{align}

Let us consider the Wilson lines transforming in the symmetric representation of general rank $k$. 
The half-BPS limit of the one-point function is given by 
\begin{align}
\label{halfBPS_so5wsymk}
\langle W_{(k)} \rangle^{SO(5)}_{\textrm{$\frac12$BPS}}(\mathfrak{q})
&=\frac{\mathfrak{q}^{4k-4}+\mathfrak{q}^{4k}}{(1-\mathfrak{q}^4)(1-\mathfrak{q}^8)}
\nonumber\\
&=\frac{\mathfrak{q}^{4k-4}}{(1-\mathfrak{q}^4)^2}. 
\end{align}
The half-BPS limit of the normalized two-point function is given by a polynomial with positive definite integer coefficients of degree $8k$. 
In other words, it can be expressed as
\begin{align}
\langle W_{(k)} W_{(k)} \rangle^{SO(5)}_{\textrm{$\frac12$BPS}}(\mathfrak{q})
&=\frac{\sum_{i=0}^{8k} a_{2\ (k)}^{\mathfrak{so}(5)}(i) \mathfrak{q}^{2i} }{(1-\mathfrak{q}^4)(1-\mathfrak{q}^8)}, 
\end{align}
where $a_{2\ (k)}^{\mathfrak{so}(5)}(i)$ are some positive definite integers. 
For example, for $k=3, 4$ we obtain
\begin{align}
\langle W_{\tiny \yng(3)} W_{\tiny \yng(3)} \rangle^{SO(5)}_{\textrm{$\frac12$BPS}}(\mathfrak{q})
&=\frac{1}{(1-\mathfrak{q}^4)(1-\mathfrak{q}^8)}
(1+\mathfrak{q}^2+2\mathfrak{q}^4+3\mathfrak{q}^6+4\mathfrak{q}^8+4\mathfrak{q}^{10}
\nonumber\\
&+6\mathfrak{q}^{12}+5\mathfrak{q}^{14}+5\mathfrak{q}^{16}+4\mathfrak{q}^{18}+3\mathfrak{q}^{20}+\mathfrak{q}^{22}+\mathfrak{q}^{24}), \\
\langle W_{\tiny \yng(4)} W_{\tiny \yng(4)} \rangle^{SO(5)}_{\textrm{$\frac12$BPS}}(\mathfrak{q})
&=\frac{1}{(1-\mathfrak{q}^4)(1-\mathfrak{q}^8)}
(1+\mathfrak{q}^2+2\mathfrak{q}^4+3\mathfrak{q}^6+5\mathfrak{q}^8+5\mathfrak{q}^{10}
\nonumber\\
&+8\mathfrak{q}^{12}+8\mathfrak{q}^{14}+10\mathfrak{q}^{16}+9\mathfrak{q}^{18}+10\mathfrak{q}^{20}
+7\mathfrak{q}^{22}+7\mathfrak{q}^{24}
\nonumber\\
&+4\mathfrak{q}^{26}+3\mathfrak{q}^{28}+\mathfrak{q}^{30}+\mathfrak{q}^{32}). 
\end{align}

As the rank $k$ of symmetric representation increases, 
the lower order terms in the series expansion of the two-point function of the rank-$k$ symmetric Wilson lines are stabilized. 
The finite $k$ correction appears from the term with $\mathfrak{q}^{k+1}$. 
We find that in the large representation limit $k\rightarrow \infty$, 
the half-BPS limit of the two-point function of the Wilson lines in the rank-$k$ symmetric representation has the expansion
\begin{align}
&
\langle W_{(\infty)} W_{(\infty)}\rangle^{SO(5)}_{\textrm{$\frac12$BPS}}(\mathfrak{q})
\nonumber\\
&=1+\mathfrak{q}^2+3\mathfrak{q}^4+4\mathfrak{q}^6+9\mathfrak{q}^8+11\mathfrak{q}^{10}
+21\mathfrak{q}^{12}+26\mathfrak{q}^{14}+44\mathfrak{q}^{16}+54\mathfrak{q}^{18}+84\mathfrak{q}^{20}+\cdots. 
\end{align}
We find that it agrees with the following expression: 
\begin{align}
\label{halfBPSso5_wsymlarge}
\langle W_{(\infty)} W_{(\infty)}\rangle^{SO(5)}_{\textrm{$\frac12$BPS}}(\mathfrak{q})
&=\frac{1-\mathfrak{q}^{24}}{(1-\mathfrak{q}^2)(1-\mathfrak{q}^4)^2(1-\mathfrak{q}^6)(1-\mathfrak{q}^8)^2(1-\mathfrak{q}^{12})}. 
\end{align}

\subsection{$\mathfrak{so}(7)$}

\subsubsection{Spinor Wilson line}
For SYM theories based on the $\mathfrak{so}(7)$ gauge algebra, the spinor Wilson line exists for $Spin(7)$ gauge theory. 
Again the odd-point correlation function of the spinor Wilson line vanishes. 
The two-point function of the spinor Wilson lines for $Spin(7)$ SYM theory is given by
\begin{align}
\label{so7_wsp}
&
\langle W_{\textrm{sp}} W_{\textrm{sp}}\rangle^{Spin(7)}(t;q)
\nonumber\\
&=\frac{1}{48} \frac{(q)_{\infty}^6}{(q^{\frac12}t^{\pm};q)_{\infty}^3}
\oint \prod_{i=1}^3 \frac{ds_i}{2\pi is_i}
\frac{(s_i^{\pm};q)_{\infty}(qs_i^{\pm};q)_{\infty}}
{(q^{\frac12}t^2 s_i^{\pm};q)_{\infty}(q^{\frac12}t^{-2}s_i^{\pm};q)_{\infty}}
\nonumber\\
&\times 
\prod_{i<j}\frac{(s_i^{\pm}s_j^{\mp};q)_{\infty}(s_i^{\pm}s_j^{\pm};q)_{\infty}(qs_i^{\pm}s_j^{\mp};q)_{\infty}(qs_i^{\pm}s_j^{\pm};q)_{\infty}}
{(q^{\frac12}t^2s_i^{\pm}s_j^{\mp};q)_{\infty}(q^{\frac12}t^2s_i^{\pm}s_j^{\pm};q)_{\infty}
(q^{\frac12}t^{-2}s_i^{\pm}s_j^{\mp};q)_{\infty}(q^{\frac12}t^{-2}s_i^{\pm}s_j^{\pm};q)_{\infty}}
\prod_{i=1}^3(s_i^{\frac12}+s_i^{-\frac12})^2. 
\end{align}

The Wilson line in the spinor representation for $Spin(7)$ gauge theory is S-dual to 
the 't Hooft lines of $B=(\frac12,\frac12,\frac12)$ in $USp(6)/\mathbb{Z}_2$. 
The two-point function of the dual 't Hooft lines is 
\begin{align}
\label{sp6_tsp}
&
\langle T_{(\frac12,\frac12,\frac12)} T_{(\frac12,\frac12,\frac12)}\rangle^{USp(6)/\mathbb{Z}_2}(t;q)
\nonumber\\
&=\frac16 \frac{(q)_{\infty}^6}{(q^{\frac12}t^{\pm2};q)_{\infty}^3}
\oint \prod_{i=1}^3 \frac{ds_i}{2\pi is_i}
\frac{(q^{\frac12}s_i^{\pm2};q)_{\infty}(q^{\frac32}s_i^{\pm2};q)_{\infty}}
{(qt^2s_i^{\pm2};q)_{\infty}(qt^{-2}s_i^{\pm2};q)_{\infty}}
\nonumber\\
&\times 
\prod_{i<j}
\frac{(s_i^{\pm}s_j^{\mp};q)_{\infty}(q^{\frac12}s_i^{\pm}s_j^{\pm};q)_{\infty}
(qs_i^{\pm}s_j^{\mp};q)_{\infty}(q^{\frac32}s_i^{\pm}s_j^{\pm};q)_{\infty}}
{(q^{\frac12}t^2s_i^{\pm}s_j^{\mp};q)_{\infty}(qt^2s_i^{\pm}s_j^{\pm};q)_{\infty}
(q^{\frac12}t^{-2}s_i^{\pm}s_j^{\mp};q)_{\infty}(qt^{-2}s_i^{\pm}s_j^{\pm};q)_{\infty}}. 
\end{align}
This agrees with the expression (\ref{so7_wsp}). 

The matrix integral (\ref{sp6_tsp}) becomes the half-BPS index of $\mathcal{N}=4$ $U(3)$ SYM theory in the half-BPS limit. 
Hence the half-BPS limits of the two-point functions (\ref{so7_wsp}) and (\ref{sp6_tsp}) are given by
\begin{align}
\langle W_{\textrm{sp}} W_{\textrm{sp}}\rangle^{Spin(7)}_{\textrm{$\frac12$BPS}}(\mathfrak{q})
&=\langle T_{(\frac12,\frac12,\frac12)} T_{(\frac12,\frac12,\frac12)}\rangle^{USp(6)/\mathbb{Z}_2}_{\textrm{$\frac12$BPS}}(\mathfrak{q})
\nonumber\\
&=\frac{1+\mathfrak{q}^2+\mathfrak{q}^4+2\mathfrak{q}^6+\mathfrak{q}^8+\mathfrak{q}^{10}+\mathfrak{q}^{12}}
{(1-\mathfrak{q}^4)(1-\mathfrak{q}^8)(1-\mathfrak{q}^{12})}
\nonumber\\
&=\frac{1}{(1-\mathfrak{q}^2)(1-\mathfrak{q}^4)(1-\mathfrak{q}^6)}. 
\end{align}

More generally, there exist non-trivial even-point correlation functions of the spinor Wilson lines. 
It follows that the half-BPS limits of these correlators take the form
\begin{align}
\langle \underbrace{W_{\textrm{sp}} \cdots W_{\textrm{sp}}}_{2k} \rangle^{Spin(7)}_{\textrm{$\frac12$BPS}}(\mathfrak{q})
&=
\frac{\sum_{i=0}^{6k} a_{k\ \textrm{sp}}^{\mathfrak{so}(7)}(i) \mathfrak{q}^{2i}}{(1-\mathfrak{q}^4)(1-\mathfrak{q}^8)(1-\mathfrak{q}^{12})}. 
\end{align}
For example, $k=2, 3$ one finds
\begin{align}
\langle W_{\textrm{sp}} W_{\textrm{sp}} W_{\textrm{sp}} W_{\textrm{sp}} \rangle^{Spin(7)}_{\textrm{$\frac12$BPS}}(\mathfrak{q})
&=\frac{1}{(1-\mathfrak{q}^4)(1-\mathfrak{q}^8)(1-\mathfrak{q}^{12})}
(4+9\mathfrak{q}^2+15\mathfrak{q}^4
\nonumber\\
&+25\mathfrak{q}^6+29\mathfrak{q}^8+32\mathfrak{q}^{10}+33\mathfrak{q}^{12}+26\mathfrak{q}^{14}+20\mathfrak{q}^{16}
\nonumber\\
&+13\mathfrak{q}^{18}+6\mathfrak{q}^{20}+3\mathfrak{q}^{22}+\mathfrak{q}^{24}), \\
\langle W_{\textrm{sp}} W_{\textrm{sp}} W_{\textrm{sp}} W_{\textrm{sp}} W_{\textrm{sp}} W_{\textrm{sp}} \rangle^{Spin(7)}_{\textrm{$\frac12$BPS}}(\mathfrak{q})
&=\frac{1}{(1-\mathfrak{q}^4)(1-\mathfrak{q}^8)(1-\mathfrak{q}^{12})}
(30+105\mathfrak{q}^2+235\mathfrak{q}^4
\nonumber\\
&+435\mathfrak{q}^6+650\mathfrak{q}^8+855\mathfrak{q}^{10}+1010\mathfrak{q}^{12}+1055\mathfrak{q}^{14}
\nonumber\\
&+1006\mathfrak{q}^{16}+865\mathfrak{q}^{18}+665\mathfrak{q}^{20}+470\mathfrak{q}^{22}+299\mathfrak{q}^{24}
\nonumber\\
&+170\mathfrak{q}^{26}+89\mathfrak{q}^{28}+40\mathfrak{q}^{30}+15\mathfrak{q}^{32}+5\mathfrak{q}^{34}+\mathfrak{q}^{36}
). 
\end{align}
We observe that the coefficient $a_{k\ \textrm{sp}}^{\mathfrak{so}(7)}(0)$ agrees with
\begin{align}
a_{k\ \textrm{sp}}^{\mathfrak{so}(7)}(0)
&=\prod_{1\le i\le j\le k-1}\frac{i+j+6}{i+j}. 
\end{align}
This is the multiplicity of the trivial representation in the $2k$-th tensor power of the spinor representation of $Spin(7)$, 
which is obtained from the Littelman paths as the number of triples of non-crossing Dyck paths \cite{MR927758,MR2388243}. 
It can be found from the generating function
\begin{align}
{_{4}F_{3}}\left(1,\frac12,\frac52,\frac32;4,5,6;64x\right)
&=\sum_{k=0}^{\infty} a_{k\ \textrm{sp}}^{\mathfrak{so}(7)}(0) x^k. 
\end{align}

\subsubsection{Fundamental Wilson line}
The Wilson line in the fundamental representation is allowed for $SO(7)$ gauge theories as a minimal electrically charged line. 
We find that the half-BPS limits of the correlation functions of the Wilson lines in the fundamental representation for $SO(7)$ SYM theory can be expressed as
\begin{align}
\langle \underbrace{W_{\tiny \yng(1)} \cdots W_{\tiny \yng(1)} }_{k} \rangle^{SO(7)}_{\textrm{$\frac12$BPS}}(\mathfrak{q})
&=\frac{\sum_{i=0}^{3k} a_{k\ \tiny \yng(1)}^{\mathfrak{so}(7)}(i) \mathfrak{q}^{2i}}{(1-\mathfrak{q}^4)(1-\mathfrak{q}^{8})(1-\mathfrak{q}^{12})}, 
\end{align}
where $a_{k\ \tiny \yng(1)}^{\mathfrak{so}(7)}(i)$ are some positive definite integers. 
For example, the one-, two- and three-point functions are given by
\begin{align}
\langle W_{\tiny \yng(1)}\rangle^{SO(7)}_{\textrm{$\frac12$BPS}}(\mathfrak{q})
&=\frac{\mathfrak{q}^6}{(1-\mathfrak{q}^4)(1-\mathfrak{q}^8)(1-\mathfrak{q}^{12})}, \\
\langle W_{\tiny \yng(1)} W_{\tiny \yng(1)}\rangle^{SO(7)}_{\textrm{$\frac12$BPS}}(\mathfrak{q})
&=\frac{1+\mathfrak{q}^2+\mathfrak{q}^4+\mathfrak{q}^6+\mathfrak{q}^{8}+\mathfrak{q}^{10}+\mathfrak{q}^{12}}
{(1-\mathfrak{q}^4)(1-\mathfrak{q}^8)(1-\mathfrak{q}^{12})}
\nonumber\\
&=\frac{1-\mathfrak{q}^{14}}
{(1-\mathfrak{q}^2)(1-\mathfrak{q}^4)(1-\mathfrak{q}^8)(1-\mathfrak{q}^{12})}, \\
\langle W_{\tiny \yng(1)} W_{\tiny \yng(1)} W_{\tiny \yng(1)}  \rangle^{SO(7)}_{\textrm{$\frac12$BPS}}(\mathfrak{q})
&=\frac{\mathfrak{q}^4+3\mathfrak{q}^6+3\mathfrak{q}^8+3\mathfrak{q}^{10}+3\mathfrak{q}^{12}+3\mathfrak{q}^{14}+2\mathfrak{q}^{16}+\mathfrak{q}^{18}}
{(1-\mathfrak{q}^4)(1-\mathfrak{q}^8)(1-\mathfrak{q}^{12})}. 
\end{align}

\subsubsection{Antisymmetric Wilson lines}
For $SO(7)$ gauge theories, there exist rank-$2$ and rank-$3$ antisymmetric Wilson lines. 
It follows that the normalized correlators of these antisymmetric Wilson lines are given by a polynomial with positive definite integer coefficients in the half-BPS limit 
so that they can be written as
\begin{align}
\langle \underbrace{W_{\tiny \yng(1,1)}\cdots W_{\tiny \yng(1,1)}}_{k} \rangle^{SO(7)}_{\textrm{$\frac12$BPS}}(\mathfrak{q})
&=\frac{\sum_{i=0}^{5k} a_{k\ \tiny \yng(1,1)}^{\mathfrak{so}(7)}(i) \mathfrak{q}^{2i}}{(1-\mathfrak{q}^4)(1-\mathfrak{q}^8)(1-\mathfrak{q}^{12})}, \\
\langle \underbrace{W_{\tiny \yng(1,1,1)}\cdots W_{\tiny \yng(1,1,1)}}_{k} \rangle^{SO(7)}_{\textrm{$\frac12$BPS}}(\mathfrak{q})
&=\frac{\sum_{i=0}^{6k} a_{k\ \tiny \yng(1,1,1)}^{\mathfrak{so}(7)}(i) \mathfrak{q}^{2i}}{(1-\mathfrak{q}^4)(1-\mathfrak{q}^8)(1-\mathfrak{q}^{12})}. 
\end{align}
For example, for the Wilson lines in the rank-$2$ antisymmetric representation, i.e. adjoint Wilson lines, 
the one- and two-point functions are given by
\begin{align}
\label{halfBPS_so7wasym_1pt}
\langle W_{\tiny \yng(1,1)} \rangle^{SO(7)}_{\textrm{$\frac12$BPS}}(\mathfrak{q})
&=\frac{\mathfrak{q}^2+\mathfrak{q}^6+\mathfrak{q}^{10}}
{(1-\mathfrak{q}^4)(1-\mathfrak{q}^8)(1-\mathfrak{q}^{12})}, \\
\label{halfBPS_so7wasym_2pt}
\langle W_{\tiny \yng(1,1)}W_{\tiny \yng(1,1)} \rangle^{SO(7)}_{\textrm{$\frac12$BPS}}(\mathfrak{q})
&=\frac{1}{(1-\mathfrak{q}^4)(1-\mathfrak{q}^{8})(1-\mathfrak{q}^{12})}
\nonumber\\
&\times 
(1+\mathfrak{q}^2+3\mathfrak{q}^4+2\mathfrak{q}^6+5\mathfrak{q}^8
+3\mathfrak{q}^{10}
\nonumber\\
&+5\mathfrak{q}^{12}+2\mathfrak{q}^{14}+3\mathfrak{q}^{16}+\mathfrak{q}^{18}+\mathfrak{q}^{20}). 
\end{align}
Similarly, the correlators of the rank-$3$ antisymmetric Wilson lines are
\begin{align}
\langle W_{\tiny \yng(1,1,1)} \rangle^{SO(7)}_{\textrm{$\frac12$BPS}}(\mathfrak{q})
&=\frac{\mathfrak{q}^4+\mathfrak{q}^8+\mathfrak{q}^{12}}
{(1-\mathfrak{q}^4)(1-\mathfrak{q}^8)(1-\mathfrak{q}^{12})}, \\
\langle W_{\tiny \yng(1,1,1)}W_{\tiny \yng(1,1,1)} \rangle^{SO(7)}_{\textrm{$\frac12$BPS}}(\mathfrak{q})
&=\frac{1}
{(1-\mathfrak{q}^4)(1-\mathfrak{q}^8)(1-\mathfrak{q}^{12})}
\nonumber\\
&\times 
(1+\mathfrak{q}^2+3\mathfrak{q}^4+4\mathfrak{q}^6+7\mathfrak{q}^8+6\mathfrak{q}^{10}+9\mathfrak{q}^{12}
\nonumber\\
&+6\mathfrak{q}^{14}+7\mathfrak{q}^{16}+4\mathfrak{q}^{18}
+3\mathfrak{q}^{20}+\mathfrak{q}^{22}+\mathfrak{q}^{24}
). 
\end{align}

\subsubsection{Symmetric Wilson line}
Also the normalized correlators of the symmetric Wilson lines are given by a polynomial with non-negative integer coefficients in the half-BPS limit. 
For the rank-$l$ symmetric Wilson lines they take the form
\begin{align}
\langle \underbrace{W_{(l)}\cdots W_{(l)}}_{k} \rangle^{SO(7)}_{\textrm{$\frac12$BPS}}(\mathfrak{q})
&=\frac{\sum_{i=0}^{3lk} a_{k\ (l)}^{\mathfrak{so}(7)}(i) \mathfrak{q}^{2i}}{(1-\mathfrak{q}^4)(1-\mathfrak{q}^8)(1-\mathfrak{q}^{12})}. 
\end{align}
For example, the one- and two-point functions of rank-$2$ symmetric Wilson lines are evaluated as
\begin{align}
\langle W_{\tiny \yng(2)} \rangle^{SO(7)}_{\textrm{$\frac12$BPS}}(\mathfrak{q})
&=\frac{\mathfrak{q}^4+\mathfrak{q}^8+\mathfrak{q}^{12}}
{(1-\mathfrak{q}^4)(1-\mathfrak{q}^8)(1-\mathfrak{q}^{12})}, \\
\langle W_{\tiny \yng(2)} W_{\tiny \yng(2)} \rangle^{SO(7)}_{\textrm{$\frac12$BPS}}(\mathfrak{q})
&=\frac{1}{(1-\mathfrak{q}^4)(1-\mathfrak{q}^8)(1-\mathfrak{q}^{12})}
\nonumber\\
&\times 
(1+\mathfrak{q}^2+2\mathfrak{q}^4+2\mathfrak{q}^6+4\mathfrak{q}^8+3\mathfrak{q}^{10}
+5\mathfrak{q}^{12}
\nonumber\\
&+3\mathfrak{q}^{14}+5\mathfrak{q}^{16}+2\mathfrak{q}^{18}+3\mathfrak{q}^{20}+\mathfrak{q}^{22}+\mathfrak{q}^{24}). 
\end{align}

\subsection{$\mathfrak{so}(2N+1)$}
Now we would like to generalize the results for the line defect indices in SYM theories with $\mathfrak{so}(2N+1)$ gauge algebra. 

\subsubsection{Spinor Wilson line}
The Wilson line in the spinor representation is available for $Spin(2N+1)$ gauge theory. 
While the one-point function of the Wilson line in the spinor representation is trivial, 
the two-point function of the Wilson line in the spinor representation for $Spin(2N+1)$ gauge theory can be evaluated as
\begin{align}
\label{so2N+1_wsp}
&
\langle W_{\textrm{sp}} W_{\textrm{sp}}\rangle^{Spin(2N+1)}(t;q)
\nonumber\\
&=\frac{1}{2^N N!} \frac{(q)_{\infty}^{2N}}{(q^{\frac12}t^{\pm};q)_{\infty}^N}
\oint \prod_{i=1}^N \frac{ds_i}{2\pi is_i}
\frac{(s_i^{\pm};q)_{\infty}(qs_i^{\pm};q)_{\infty}}
{(q^{\frac12}t^2 s_i^{\pm};q)_{\infty}(q^{\frac12}t^{-2}s_i^{\pm};q)_{\infty}}
\nonumber\\
&\times 
\prod_{i<j}\frac{(s_i^{\pm}s_j^{\mp};q)_{\infty}(s_i^{\pm}s_j^{\pm};q)_{\infty}(qs_i^{\pm}s_j^{\mp};q)_{\infty}(qs_i^{\pm}s_j^{\pm};q)_{\infty}}
{(q^{\frac12}t^2s_i^{\pm}s_j^{\mp};q)_{\infty}(q^{\frac12}t^2s_i^{\pm}s_j^{\pm};q)_{\infty}
(q^{\frac12}t^{-2}s_i^{\pm}s_j^{\mp};q)_{\infty}(q^{\frac12}t^{-2}s_i^{\pm}s_j^{\pm};q)_{\infty}}
\prod_{i=1}^N(s_i^{\frac12}+s_i^{-\frac12})^2. 
\end{align}

On the other hand, the two-point function of the 't Hooft lines of $B=(\frac12^N)$ for $USp(2N)/\mathbb{Z}_2$ gauge theory can be written as 
\begin{align}
\label{sp2N_tsp}
&
\langle T_{(\frac12^N)} T_{(\frac12^N)}\rangle^{USp(2N)/\mathbb{Z}_2}(t;q)
\nonumber\\
&=\frac{1}{N!} \frac{(q)_{\infty}^{2N}}{(q^{\frac12}t^{\pm2};q)_{\infty}^N}
\oint \prod_{i=1}^N \frac{ds_i}{2\pi is_i}
\frac{(q^{\frac12}s_i^{\pm2};q)_{\infty}(q^{\frac32}s_i^{\pm2};q)_{\infty}}
{(qt^2s_i^{\pm2};q)_{\infty}(qt^{-2}s_i^{\pm2};q)_{\infty}}
\nonumber\\
&\times 
\prod_{i<j}
\frac{(s_i^{\pm}s_j^{\mp};q)_{\infty}(q^{\frac12}s_i^{\pm}s_j^{\pm};q)_{\infty}
(qs_i^{\pm}s_j^{\mp};q)_{\infty}(q^{\frac32}s_i^{\pm}s_j^{\pm};q)_{\infty}}
{(q^{\frac12}t^2s_i^{\pm}s_j^{\mp};q)_{\infty}(qt^2s_i^{\pm}s_j^{\pm};q)_{\infty}
(q^{\frac12}t^{-2}s_i^{\pm}s_j^{\mp};q)_{\infty}(qt^{-2}s_i^{\pm}s_j^{\pm};q)_{\infty}}. 
\end{align}
According to S-duality between 
the Wilson line in the spinor representation for $Spin(2N+1)$ gauge theory 
and the 't Hooft line of $B=(\frac12^N)$ for $USp(2N)/\mathbb{Z}_2$ gauge theory, 
the two expressions (\ref{so2N+1_wsp}) and (\ref{sp2N_tsp}) will be identical. 

In the half-BPS limit, 
the expression (\ref{sp2N_tsp}) reduces to the matrix integral of the half-BPS index for $U(N)$ SYM theory. 
Therefore, the half-BPS limits of the two-point functions (\ref{so2N+1_wsp}) and (\ref{sp2N_tsp}) are evaluated as
\begin{align}
\langle W_{\textrm{sp}} W_{\textrm{sp}}\rangle^{Spin(2N+1)}_{\textrm{$\frac12$BPS}}(\mathfrak{q})
&=\langle T_{(\frac12^N)} T_{(\frac12^N)}\rangle^{USp(2N)/\mathbb{Z}_2}_{\textrm{$\frac12$BPS}}(\mathfrak{q})
\nonumber\\
&=\prod_{n=1}^{N}\frac{1}{(1-\mathfrak{q}^{2n})}. 
\end{align}
Since the the half-BPS index of the $Spin(2N+1)$ gauge theory is given by
\begin{align}
\label{halfBPS_so2N+1}
\mathcal{I}^{Spin(2N+1)}_{\textrm{$\frac12$BPS}}(\mathfrak{q})&=\prod_{n=1}^{N}\frac{1}{1-\mathfrak{q}^{4n}}, 
\end{align}
the normalized two-point function is given by
\begin{align}
\label{nhalfBPS_so2N+1sp}
\langle \mathcal{W}_{\textrm{sp}} \mathcal{W}_{\textrm{sp}}\rangle^{Spin(2N+1)}_{\textrm{$\frac12$BPS}}(\mathfrak{q})
&=\prod_{n=1}^{N}(1+\mathfrak{q}^{2n}). 
\end{align}

While the odd-point correlation functions vanish, the even-point correlation functions are non-trivial. 
In the half-BPS limit they can be expressed as
\begin{align}
\langle \underbrace{W_{\textrm{sp}} \cdots W_{\textrm{sp}}}_{2k} \rangle^{Spin(2N+1)}_{\textrm{$\frac12$BPS}}(\mathfrak{q})
&=\frac{\sum_{i=0}^{\frac{N(N+1)k}{2}} a_{k\ \textrm{sp}}^{\mathfrak{so}(2N+1)}(i) \mathfrak{q}^{2i}}{\prod_{n=1}^{N}(1-\mathfrak{q}^{4n})}, 
\end{align}
with non-negative coefficients $ a_{k\ \textrm{sp}}^{\mathfrak{so}(2N+1)}(i)$. 
First, we propose that the first coefficients $a_{k\ \textrm{sp}}^{\mathfrak{so}(2N+1)}(0)$ can be expressed as 
the determinant of the $N$ $\times$ $N$ Hankel matrix
\begin{align}
\label{B_tensor_sp_0}
a_{k\ \textrm{sp}}^{\mathfrak{so}(2N+1)}(0)&=\det (C_{2N-i-j+k})
\nonumber\\
&=\prod_{1\le i\le j\le k-1}\frac{i+j+2N}{i+j}, 
\end{align}
where $C_k$ is the $k$-th Catalan number. 
The number (\ref{B_tensor_sp_0}) is the multiplicity of the trivial representation in the $2k$-th power of the spinor representation of $Spin(2N+1)$ \cite{MR3959676}. 

\subsubsection{Fundamental Wilson line}
Let us generalize the results for more general line defect indices which is allowed in $SO(2N+1)$ gauge theories. 
The half-BPS limits of the correlation functions of the fundamental Wilson lines take the form
\begin{align}
\langle \underbrace{W_{\tiny \yng(1)} \cdots W_{\tiny \yng(1)}}_{k} \rangle^{SO(2N+1)}_{\textrm{$\frac12$BPS}}(\mathfrak{q})
&=\frac{\sum_{i=0}^{Nk} a_{k\ \tiny\yng(1)}^{\mathfrak{so}(2N+1)}(i)\mathfrak{q}^{2i} }{\prod_{n=1}^{N}(1-\mathfrak{q}^{4n})},
\end{align}
with $a_{k\ \tiny\yng(1)}^{\mathfrak{so}(2N+1)}(i)$ being some positive definite integers. 
In particular the one- and two-point functions are given by
\begin{align}
\label{halfBPSso2N+1_wfund}
\langle W_{\tiny \yng(1)} \rangle^{SO(2N+1)}_{\textrm{$\frac12$BPS}}(\mathfrak{q})
&=\frac{\mathfrak{q}^{2N}}{\prod_{n=1}^{N}(1-\mathfrak{q}^{4n})}, \\
\label{halfBPSso2N+1_wfund}
\langle W_{\tiny \yng(1)} W_{\tiny \yng(1)}\rangle^{SO(2N+1)}_{\textrm{$\frac12$BPS}}(\mathfrak{q})
&=\frac{1+\mathfrak{q}^2+\mathfrak{q}^4+\cdots+\mathfrak{q}^{4N}}{\prod_{n=1}^{N}(1-\mathfrak{q}^{4n})}
\nonumber\\
&=\frac{1-\mathfrak{q}^{4N+2}}{(1-\mathfrak{q}^2)\prod_{n=1}^{N}(1-\mathfrak{q}^{4n})}. 
\end{align}
The normalized two-point functions take the forms
\begin{align}
\langle \mathcal{W}_{\tiny \yng(1)} \rangle^{SO(2N+1)}_{\textrm{$\frac12$BPS}}(\mathfrak{q})&=\mathfrak{q}^{2N}, \\
\langle \mathcal{W}_{\tiny \yng(1)} \mathcal{W}_{\tiny \yng(1)}\rangle^{SO(2N+1)}_{\textrm{$\frac12$BPS}}(\mathfrak{q})&=\frac{1-\mathfrak{q}^{4N+2}}{1-\mathfrak{q}^2}. 
\end{align}

Accordingly, we find the expression for the connected normalized two-point function of the fundamental Wilson lines
\begin{align}
\label{nhalfBPS_so2N+1_wfund}
\langle \mathcal{W}_{\tiny \yng(1)} \mathcal{W}_{\tiny \yng(1)}\rangle^{SO(2N+1)}_{\textrm{$\frac12$BPS},c}(\mathfrak{q})&=
\langle \mathcal{W}_{\tiny \yng(1)} \mathcal{W}_{\tiny \yng(1)}\rangle^{SO(2N+1)}_{\textrm{$\frac12$BPS}}(\mathfrak{q})
-{\langle \mathcal{W}_{\tiny \yng(1)} \rangle^{SO(2N+1)}_{\textrm{$\frac12$BPS}}(\mathfrak{q})}^2
\nonumber\\
&=\frac{1-\mathfrak{q}^{4N}}{1-\mathfrak{q}^2}. 
\end{align}

\subsubsection{Adjoint Wilson line}
For the adjoint Wilson lines, the correlators take the form
\begin{align}
\langle \underbrace{W_{\tiny \yng(1,1)} \cdots W_{\tiny \yng(1,1)}}_{k} \rangle^{SO(2N+1)}_{\textrm{$\frac12$BPS}}(\mathfrak{q})
&=\frac{\sum_{i=0}^{(2N-1)k} a_{k\ \tiny\yng(1)}^{\mathfrak{so}(2N+1)}(i)\mathfrak{q}^{2i} }{\prod_{n=1}^{N}(1-\mathfrak{q}^{4n})},
\end{align}
where $ a_{k\ \tiny\yng(1)}^{\mathfrak{so}(2N+1)}(i)$ are positive definite integers. 
The one-point function of the adjoint Wilson line is given by
\begin{align}
\label{halfBPS_so2N+1wadj_1pt}
\langle W_{\tiny \yng(1,1)} \rangle^{SO(2N+1)}_{\textrm{$\frac12$BPS}}(\mathfrak{q})
&=\frac{\mathfrak{q}^2+\mathfrak{q}^6+\cdots+\mathfrak{q}^{4N-2}}
{\prod_{n=1}^{N}(1-\mathfrak{q}^{4n})}
\nonumber\\
&=\frac{\mathfrak{q}^2(1-\mathfrak{q}^{4N})}{(1-\mathfrak{q}^4)\prod_{n=1}^{N}(1-\mathfrak{q}^{4n})}. 
\end{align}
The normalized two-point function takes the forms
\begin{align}
\langle \mathcal{W}_{\tiny \yng(1,1)} \rangle^{SO(2N+1)}_{\textrm{$\frac12$BPS}}(\mathfrak{q})&=\frac{\mathfrak{q}^2(1-\mathfrak{q}^{4N})}{1-\mathfrak{q}^4}. 
\end{align}

\subsubsection{Symmetric Wilson line}
For the rank-$l$ symmetric Wilson lines, the correlation functions can be written as
\begin{align}
\langle \underbrace{W_{(l)} \cdots W_{(l)}}_{k} \rangle^{SO(2N+1)}_{\textrm{$\frac12$BPS}}(\mathfrak{q})
&=\frac{\sum_{i=0}^{Nlk} a_{k\ (l)}^{\mathfrak{so}(2N+1)}(i)\mathfrak{q}^{2i} }{\prod_{n=1}^{N}(1-\mathfrak{q}^{4n})},
\end{align}
with $ a_{k\ (l)}^{\mathfrak{so}(2N+1)}(i)$ are positive definite integers. 

\section{$\mathfrak{usp}(2N)$}
\label{sec_Ctype}
Let us consider $\mathcal{N}=4$ SYM theories with gauge group associated with the Lie algebra $\mathfrak{g}$ $=$ $\mathfrak{usp}(2N)$. 
The center of $USp(2N)$ is $\mathbb{Z}_2$ under which the fundamental representation carries the $\mathbb{Z}_2$ charge. 
The line operators are labeled by electric and magnetic charges 
\begin{align}
(z_e,z_m)\in \mathbb{Z}_2\times \mathbb{Z}_2. 
\end{align}
The $USp(2N)$ gauge theory has the line operators with 
\begin{align}
(z_e,z_m)&=(0,0), \qquad (z_e,z_m)=(1,0), 
\end{align}
including the electric Wilson line in the fundamental representation. 
The $(USp(2N)/\mathbb{Z}_2)_+$ gauge theory contains the lines with 
\begin{align}
(z_e,z_m)&=(0,0), \qquad (z_e,z_m)=(0,1), 
\end{align}
including the magnetic 't Hooft line. 
The $(USp(2N)/\mathbb{Z}_2)_-$ gauge theory admits the lines with 
\begin{align}
(z_e,z_m)&=(0,0), \qquad (z_e,z_m)=(1,1), 
\end{align}
including the dyonic line. 
These conventions are consistent with those introduced for $SO(2N+1)_{\pm}$ 
as $USp(2)/\mathbb{Z}_2$ $\cong$ $SO(3)$ and $USp(4)/\mathbb{Z}_2$ $\cong$ $SO(5)$. 

The Lie algebra of the Langlands gauge group is $\mathfrak{g}^{\vee}$ $=$ $\mathfrak{so}(2N+1)$. 
The $USp(2N)$ gauge theory has the Wilson line operator transforming in the fundamental representation 
which is S-dual to the 't Hooft line operator with $B=(1,0)$ in $SO(2N+1)$ gauge theory. 

The character of the fundamental representation for $\mathfrak{usp}(2N)$ is 
\begin{align}
\label{ch_usp2N_fund}
\chi_{\tiny \yng(1)}^{\mathfrak{usp}(2N)}
&=\sum_{i=1}^{N}(s_i+s_i^{-1}). 
\end{align}

The character of the irreducible representation with highest weight $\lambda$ can be written as \cite{MR1153249}
\begin{align}
\label{ch_usp2N_irrep}
\chi_{\tiny \yng(1)}^{\mathfrak{usp}(2N)}
&=\frac{\det(s_j^{\lambda_i+N-i+1}-s_j^{-\lambda_i-N+i-1})}{\det(s_j^{N-i+1}-s_j^{-N+i-1})}. 
\end{align}

The line defect correlators can be written as
\begin{align}
&\langle W_{\mathcal{R}_1}\cdots W_{\mathcal{R}_k}\rangle^{USp(2N)}\notag\\
&=\int \! d\mu^{USp(2N)}
\exp \left( \sum_{n=1}^\infty \frac{1}{n}f_n(q,t) L_n(s) \right)\prod_{i=1}^{k}\chi^{\mathfrak{usp}(2N)}_{\mathcal{R}_i}(s), 
\end{align}
where $d\mu^{USp(2N)}$ and $f_n(q,t)$ are given by \eqref{eq:dmu-USp} and \eqref{eq:f_n}, respectively.
Also we have introduced the notation
\begin{equation}
\begin{aligned}
L_n(s)=\frac{P_n(s)^2+P_{2n}(s)}{2},
\end{aligned}
\end{equation}
where the power sum function $P_m(s)$ is defined by \eqref{P_m}.
We can evaluate them by using the character expansion method.
As in the $SO(2N+1)$ case, the exponential part has the following expansion:
\begin{align}
\exp \left( \sum_{n=1}^\infty \frac{1}{n} f_n(q,t)L_n(s) \right)
=\sum_{\lambda} \frac{1}{z_\lambda}f_\lambda(q,t)L_\lambda(s).
\end{align}
Therefore the correlator is written as 
\begin{align}
\langle W_{\mathcal{R}_1}\cdots W_{\mathcal{R}_k}\rangle^{USp(2N)}
=\sum_{\lambda} \frac{1}{z_\lambda}f_\lambda(q,t) \int \! d\mu^{USp(2N)}
L_{\lambda}(s)\prod_{i=1}^{k}\chi^{\mathfrak{usp}(2N)}_{\mathcal{R}_i}(s). 
\end{align}
We expand the integrand in terms of $P_\mu(s)$ as
\begin{align}
L_{\lambda}(s)\prod_{i=1}^{k}\chi^{\mathfrak{usp}(2N)}_{\mathcal{R}_i}(s)=\sum_\mu b_{\lambda, \mathcal{R}}^\mu P_\mu(s),
\end{align}
and use the integral formula \cite{Sei:2023fjk}:
\begin{align}
\int \! d\mu^{USp(2N)} P_\mu(s)=\sum_{\nu \in R_{2N}^c(|\mu|)} \chi_\nu^S(\mu), 
\end{align}
where  the subset $R_{n}^c(p)$ of partitions of $p$ is defined by
\begin{align}
R_{n}^c(p)=\{ \lambda \vdash p \,|\, \ell(\lambda)\leq n \text{ and } \forall \lambda_i' \text{ is even} \} . 
\end{align}
Then, we can easily perform the integral as
\begin{align}
\int \! d\mu^{USp(2N)}L_{\lambda}(s)\prod_{i=1}^{k}\chi^{\mathfrak{usp}(2N)}_{\mathcal{R}_i}(s)
&=\sum_\mu b_{\lambda, \mathcal{R}}^\mu \int \! d\mu^{USp(2N)} P_\mu(s) \notag \\
&=\sum_\mu b_{\lambda, \mathcal{R}}^\mu \sum_{\nu \in R_{2N}^c(|\mu|)} \chi_\nu^S(\mu). 
\end{align}
The correlator is finally given by
\begin{align}
\langle W_{\mathcal{R}_1}\cdots W_{\mathcal{R}_k}\rangle^{USp(2N)}
=\sum_{\lambda} \frac{1}{z_\lambda}f_\lambda(q,t) \sum_\mu b_{\lambda, \mathcal{R}}^\mu \sum_{\nu \in R_{2N}^c(|\mu|)} \chi_\nu^S(\mu). 
\end{align}

\subsection{$\mathfrak{usp}(2)$}

\subsubsection{Fundamental Wilson line}
The $USp(2)$ gauge theory admits the Wilson line in the fundamental representation. 
The one-point function of the fundamental Wilson line vanishes. 
The two-point function of the Wilson lines in the fundamental representation for $USp(2)$ gauge theory reads
\begin{align}
\label{sp2_wfund}
&
\langle W_{\tiny \yng(1)} W_{\tiny \yng(1)}\rangle^{USp(2)}(t;q)
\nonumber\\
&=\frac12 \frac{(q)_{\infty}^2}{(q^{\frac12}t^{\pm2};q)_{\infty}}
\oint \frac{ds}{2\pi is}
\frac{(s^{\pm2};q)_{\infty}(qs^{\pm2};q)_{\infty}}
{(q^{\frac12}t^{2}s^{\pm2};q)_{\infty}(q^{\frac12}t^{-2}s^{\pm2};q)_{\infty}}
(s+s^{-1})^2. 
\end{align}

The fundamental Wilson line for $USp(2)$ gauge theory is S-dual to the 't Hooft line of $B=(1)$ for $SO(3)_+$ gauge theory. 
The two-point function of the dual 't Hooft lines is given by
\begin{align}
\label{so3_t1}
&
\langle T_{(1)} T_{(1)}\rangle^{SO(3)}(t;q)
\nonumber\\
&=\frac{(q)_{\infty}^2}{(q^{\frac12}t^{\pm2};q)_{\infty}}
\oint \frac{ds}{2\pi is}
\frac{(q^{\frac12}s^{\pm};q)_{\infty}(q^{\frac32}s^{\pm};q)_{\infty}}
{(qt^{2}s^{\pm};q)_{\infty}(qt^{-2}s^{\pm};q)_{\infty}}. 
\end{align}
The expressions (\ref{sp2_wfund}) and (\ref{so3_t1}) are equivalent as a consequence of S-duality. 
Besides, as we have the isomorphism $USp(2)$ $\cong$ $Spin(3)$ so that the fundamental Wilson line for $USp(2)$ is identified with the spinor Wilson line for $Spin(3)$, 
they agree with the two-point functions (\ref{so3_wsp}) and (\ref{sp2_tsp}). 
Thus we have 
\begin{align}
\langle W_{\tiny \yng(1)} W_{\tiny \yng(1)}\rangle^{USp(2)}(t;q)
&=\langle W_{\textrm{sp}} W_{\textrm{sp}}\rangle^{Spin(3)}(t;q)
\nonumber\\
=\langle T_{(1)} T_{(1)}\rangle^{SO(3)}(t;q)
&=\langle T_{(\frac12)} T_{(\frac12)}\rangle^{USp(2)/\mathbb{Z}_2}(t;q). 
\end{align}
In fact, in the half-BPS limit, the two-point function (\ref{so3_t1}) simply becomes
\begin{align}
\label{halfBPSusp2_wfund}
\langle W_{\tiny \yng(1)} W_{\tiny \yng(1)}\rangle^{USp(2)}_{\textrm{$\frac12$BPS}}(\mathfrak{q})
&=\langle T_{(1)} T_{(1)}\rangle^{SO(3)}_{\textrm{$\frac12$BPS}}(\mathfrak{q})
\nonumber\\
&=\frac{1+\mathfrak{q}^2}{1-\mathfrak{q}^4}
\nonumber\\
&=\frac{1}{1-\mathfrak{q}^2}, 
\end{align}
which is consistent with (\ref{halfBPSso3_wsp}). 

Similarly, the even-point correlation function of the fundamental Wilson lines are non-trivial 
and they are equal to the correlation function (\ref{so3_wspmult}) of the $Spin(3)$ spinor Wilson lines. 

\subsubsection{Adjoint Wilson line}
The two-point function of the adjoint Wilson line transforming in the rank-$2$ symmetric representation for $USp(2)$ gauge theory with the form 
\begin{align}
\label{sp2_wsym2}
&
\langle W_{\tiny \yng(2)} W_{\tiny \yng(2)}\rangle^{USp(2)}(t;q)
\nonumber\\
&=\frac12 \frac{(q)_{\infty}^2}{(q^{\frac12}t^{\pm2};q)_{\infty}}
\oint \frac{ds}{2\pi is}
\frac{(s^{\pm 2};q)_{\infty} (qs^{\pm 2};q)_{\infty}}
{(q^{\frac12}t^{2}s^{\pm 2};q)_{\infty}(q^{\frac12}t^{-2}s^{\pm 2};q)_{\infty}}(1+s^2+s^{-2})^2
\end{align}
is equivalent to the two-point function (\ref{so3_wfund}) of the Wilson line in the fundamental representation for $SO(3)$ gauge theory. 
Also the multi-point functions are equal to those of the $SO(3)$ fundamental Wilson lines. 

\subsubsection{Symmetric Wilson line}
There exist non-trivial one-point functions in $USp(2)$ gauge theory. 
According to the isomorphism $USp(2)$ $\cong$ $Spin(3)$, it follows that 
one-point function of odd rank symmetric Wilson line vanishes 
and that of even rank symmetric Wilson line is non-trivial. 
We have 
\begin{align}
\langle W_{(2k)} \rangle^{USp(2)}(t;q)&=\langle W_{(k)} \rangle^{SO(3)}(t;q). 
\end{align}
The half-BPS limit of the one-point function is equal to (\ref{halfBPSso3_1pt_symk})
\begin{align}
\langle W_{(2k)}\rangle^{USp(2)}_{\textrm{$\frac12$BPS}}(\mathfrak{q})
&=\frac{\mathfrak{q}^{2k}}{(1-\mathfrak{q}^4)}. 
\end{align}

More generally, the two-point function of the Wilson lines in the rank-$k$ symmetric representation for $USp(2)$ gauge theory is 
identical to the two-point function of the Wilson lines transforming in the rank-$k$ symmetric product of the spinor representation in $Spin(3)$ gauge theory. 
The half-BPS limit of the two-point function of the rank-$k$ symmetric Wilson lines is evaluated as
\begin{align}
\label{halfBPSusp2_wsymk}
\langle W_{(k)} W_{(k)}\rangle^{USp(2)}_{\textrm{$\frac12$BPS}}(\mathfrak{q})
&=\frac{1-\mathfrak{q}^{2k+2}}{(1-\mathfrak{q}^{2})(1-\mathfrak{q}^4)}. 
\end{align}
When $k=2l$, it agrees with the two-point function (\ref{halfBPSso3_wsymk}) of the rank-$l$ symmetric Wilson lines for $SO(3)$ gauge theory. 

In the large representation limit $k\rightarrow \infty$, 
the two-point function (\ref{halfBPSusp2_wsymk}) becomes 
\begin{align}
\langle W_{(\infty)} W_{(\infty)}\rangle^{USp(2)}_{\textrm{$\frac12$BPS}}(\mathfrak{q})
&=\frac{1}{(1-\mathfrak{q}^2)(1-\mathfrak{q}^4)}. 
\end{align}

\subsection{$\mathfrak{usp}(4)$}

\subsubsection{Fundamental Wilson line}
The Wilson line in the fundamental representation is allowed for $USp(4)$ gauge theory. 
While the one-point function of the fundamental Wilson line is trivial, the two-point function of the Wilson lines in the fundamental representation is non-trivial. 
It can be evaluated as
\begin{align}
\label{sp4_wfund}
&
\langle W_{\tiny \yng(1)} W_{\tiny \yng(1)}\rangle^{USp(4)}(t;q)
\nonumber\\
&=\frac{1}{8}\frac{(q)_{\infty}^4}{(q^{\frac12}t^{\pm2};q)_{\infty}^2}
\oint \prod_{i=1}^{2}\frac{ds_i}{2\pi is_i}
\frac{(s_i^{\pm2};q)_{\infty} (qs_i^{\pm2};q)_{\infty}}
{(q^{\frac12}t^{2}s_i^{\pm2};q)_{\infty}(q^{\frac12}t^{-2}s_i^{\pm2};q)_{\infty}}
\nonumber\\
&\times 
\frac{(s_1^{\pm}s_2^{\mp};q)_{\infty}(s_1^{\pm}s_2^{\pm};q)_{\infty}
(qs_1^{\pm}s_2^{\mp};q)_{\infty}(qs_1^{\pm}s_2^{\pm};q)_{\infty}}
{(q^{\frac12}t^2s_1^{\pm}s_2^{\mp};q)_{\infty}(q^{\frac12}t^2s_1^{\pm}s_2^{\pm};q)_{\infty}
(q^{\frac12}t^{-2}s_1^{\pm}s_2^{\mp};q)_{\infty}(q^{\frac12}t^{-2}s_1^{\pm}s_2^{\pm};q)_{\infty}}
\left[\sum_{i=1}^2(s_i+s_i^{-1})\right]^2, 
\end{align}

The fundamental Wilson line in $USp(4)$ SYM theory is S-dual to the 't Hooft line of $B=(1,0)$ for $SO(5)_+$ SYM theory. 
The two-point function of the S-dual  't Hooft lines reads
\begin{align}
\label{so5_t1}
&
\langle T_{(1,0)} T_{(1,0)}\rangle^{SO(5)}(t;q)
\nonumber\\
&=\frac12 \frac{(q)_{\infty}^4}{(q^{\frac12}t^{\pm2};q)_{\infty}^2}
\oint \prod_{i=1}^2 \frac{ds_i}{2\pi is_i}
\frac{(q^{\frac12}s_1^{\pm};q)_{\infty}(s_2^{\pm};q)_{\infty}(q^{\frac32}s_1^{\pm};q)_{\infty}(qs_2^{\pm};q)_{\infty}}
{(qt^2s_1^{\pm};q)_{\infty}(q^{\frac12}t^2s_2^{\pm};q)_{\infty}(qt^{-2}s_1^{\pm};q)_{\infty}(q^{\frac12}t^{-2}s_2^{\pm};q)_{\infty}}
\nonumber\\
&\times 
\frac{(q^{\frac12}s_1^{\pm}s_2^{\mp};q)_{\infty}(q^{\frac12}s_1^{\pm}s_2^{\pm};q)_{\infty}
(q^{\frac32}s_1^{\pm}s_2^{\mp};q)_{\infty}(q^{\frac32}s_1^{\pm}s_2^{\pm};q)_{\infty}}
{(qt^2s_1^{\pm}s_2^{\mp};q)_{\infty}(qt^2s_1^{\pm}s_2^{\pm};q)_{\infty}
(qt^{-2}s_1^{\pm}s_2^{\mp};q)_{\infty}(qt^{-2}s_1^{\pm}s_2^{\pm};q)_{\infty}}. 
\end{align}
This precisely agrees with the expression (\ref{sp4_wfund}). 
In addition, according to the isomorphism $USp(4)$ $\cong$ $Spin(5)$, 
these two-point functions agree with the expressions (\ref{so5_wsp}), (\ref{sp4_tsp}). 
Therefore, we have 
\begin{align}
\langle W_{\tiny \yng(1)} W_{\tiny \yng(1)}\rangle^{USp(4)}(t;q)
&=\langle W_{\textrm{sp}} W_{\textrm{sp}}\rangle^{Spin(5)}(t;q)
\nonumber\\
=\langle T_{(1,0)} T_{(1,0)}\rangle^{SO(5)}(t;q)
&=\langle T_{(\frac12,\frac12)} T_{(\frac12,\frac12)}\rangle^{USp(4)/\mathbb{Z}_2}(t;q). 
\end{align}
In fact, the half-BPS limit of the two-point function (\ref{so5_t1}) is shown to take the same form as 
the half-BPS index of $\mathcal{N}=4$ $U(2)$ SYM theory. 
In other words, we have 
\begin{align}
\langle W_{\tiny \yng(1)} W_{\tiny \yng(1)}\rangle^{USp(4)}_{\textrm{$\frac12$BPS}}(\mathfrak{q})
&=\langle T_{(1,0)} T_{(1,0)}\rangle^{SO(5)}_{\textrm{$\frac12$BPS}}(\mathfrak{q})
\nonumber\\
&=\frac{1}{(1-\mathfrak{q}^2)(1-\mathfrak{q}^4)}, 
\end{align}
which is consistent with the expressions (\ref{halfBPSso5_wsp}). 

Furthermore, the even-point correlation functions of the fundamental Wilson lines are non-trivial. 
They are equal to those of the $Spin(5)$ spinor Wilson lines. 

\subsubsection{Antisymmetric Wilson line}
For the rank-$2$ antisymmetric representation of $USp(4)$, the one-point function is non-trivial. 
Since $USp(4)$ $\cong$ $Spin(5)$, we have
\begin{align}
\langle W_{\tiny \yng(1,1)} \rangle^{USp(4)}(t;q)
&=\langle W_{\tiny \yng(1)} \rangle^{SO(5)}(t;q). 
\end{align}
Analogously, for the multi-point functions one finds
\begin{align}
\langle \underbrace{W_{\tiny \yng(1,1)} \cdots W_{\tiny \yng(1,1)}}_{k} \rangle^{USp(4)}(t;q)
&=\langle \underbrace{W_{\tiny \yng(1)} \cdots W_{\tiny \yng(1)}}_{k} \rangle^{SO(5)}(t;q). 
\end{align}

\subsubsection{Symmetric Wilson line}
For the adjoint Wilson lines transforming in the rank-$2$ symmetric representation for $USp(4)$ SYM theory, 
the correlators agree with those of the adjoint Wilson lines in the rank-$2$ antisymmetric representation for $SO(5)$ SYM theory. 
It follows that 
\begin{align}
\langle \underbrace{W_{\tiny \yng(2)} \cdots W_{\tiny \yng(2)}}_{k} \rangle^{USp(4)}(t;q)
&=\langle \underbrace{W_{\tiny \yng(1,1)} \cdots W_{\tiny \yng(1,1)}}_{k} \rangle^{SO(5)}(t;q). 
\end{align}

For the Wilson line in the odd-rank symmetric representation, 
the even-point correlation functions are non-trivial, while the one-point correlation function vanishes. 
For rank-$2l$ symmetric Wilson line 
the correlation functions are given by those of the Wilson lines associated with the rectangular Young diagram $(l^2)$ in $SO(5)$ SYM theory
\begin{align}
\langle \underbrace{W_{(2l)} \cdots W_{(2l)}}_{k} \rangle^{USp(4)}(t;q)
&=\langle \underbrace{W_{(l^2)} \cdots W_{(l^2)}}_{k} \rangle^{SO(5)}(t;q). 
\end{align}
In the half-BPS limit, they can be written as
\begin{align}
\langle \underbrace{W_{(2l)} \cdots W_{(2l)}}_{k} \rangle^{USp(4)}_{\textrm{$\frac12$BPS}}(\mathfrak{q})
&=\frac{\sum_{i=0}^{3lk} a_{k\ (2l)}^{\mathfrak{usp}(4)}(i) \mathfrak{q}^{2i}}{(1-\mathfrak{q}^4)(1-\mathfrak{q}^8)}, 
\end{align}
with $a_{k\ (2l)}^{\mathfrak{usp}(4)}(i)$ being non-negative integers. 

When the rank $l$ of symmetric representation grows, 
the lower order terms in the series expansion of the two-point function are stabilized. 
In the large representation limit $l\rightarrow \infty$, 
the half-BPS limit of the two-point function of the rank-$l$ symmetric Wilson lines in $USp(4)$ gauge theory 
or equivalently that of the rectangular Wilson lines in $SO(5)$ gauge theory can be expanded as
\begin{align}
&
\langle W_{(\infty)} W_{(\infty)}\rangle^{USp(4)}_{\textrm{$\frac12$BPS}}(\mathfrak{q})
=\langle W_{(\infty^2)} W_{(\infty^2)}\rangle^{SO(5)}_{\textrm{$\frac12$BPS}}(\mathfrak{q})
\nonumber\\
&=1+\mathfrak{q}^2+4\mathfrak{q}^4+5\mathfrak{q}^6+13\mathfrak{q}^{8}+16\mathfrak{q}^{10}
+33\mathfrak{q}^{12}+41\mathfrak{q}^{14}+73\mathfrak{q}^{16}+90\mathfrak{q}^{18}+145\mathfrak{q}^{20}+\cdots. 
\end{align}
We find that it is given by
\begin{align}
\langle W_{(\infty)} W_{(\infty)}\rangle^{USp(4)}_{\textrm{$\frac12$BPS}}(\mathfrak{q})
&=\langle W_{(\infty^2)} W_{(\infty^2)}\rangle^{SO(5)}_{\textrm{$\frac12$BPS}}(\mathfrak{q})
\nonumber\\
&=
\frac{1-\mathfrak{q}^{16}}{(1-\mathfrak{q}^2)(1-\mathfrak{q}^4)^3(1-\mathfrak{q}^6)(1-\mathfrak{q}^{8})^2}. 
\end{align}

\subsubsection{Rectangular Wilson line}
In addition, the correlation function of the Wilson lines associated with the rectangular diagram $(l^2)$ in $USp(4)$ gauge theory 
is identical to that of the symmetric Wilson lines in $SO(5)$ gauge theory
\begin{align}
\langle \underbrace{W_{(l^2)}\cdots W_{(l^2)}}_{k} \rangle^{USp(4)}(t;q)&=
\langle \underbrace{W_{(l)}\cdots W_{(l)}}_{k} \rangle^{SO(5)}(t;q). 
\end{align}

\subsection{$\mathfrak{usp}(6)$}

\subsubsection{Fundamental Wilson line}
The $USp(6)$ gauge theory has the Wilson line in the fundamental representation. 
The one-point function of the Wilson line in the fundamental representation in $USp(6)$ SYM theory vanishes. 
The two-point function of the fundamental Wilson lines in $USp(6)$ SYM theory reads 
\begin{align}
\label{sp6_wfund}
&
\langle W_{\tiny \yng(1)} W_{\tiny \yng(1)}\rangle^{USp(6)}(t;q)
\nonumber\\
&=\frac{1}{48}\frac{(q)_{\infty}^{6}}{(q^{\frac12}t^{\pm2};q)_{\infty}^3}
\oint \prod_{i=1}^{3}\frac{ds_i}{2\pi is_i}
\frac{(s_i^{\pm2};q)_{\infty} (qs_i^{\pm2};q)_{\infty}}
{(q^{\frac{1}{2}}t^{2}s_i^{\pm2};q)_{\infty}(q^{\frac12}t^{-2}s_i^{\pm2};q)_{\infty}}
\nonumber\\
&\times 
\prod_{i<j}\frac{(s_i^{\pm}s_j^{\mp};q)_{\infty}(s_i^{\pm}s_j^{\pm};q)_{\infty}(qs_i^{\pm}s_j^{\mp};q)_{\infty}(qs_i^{\pm}s_j^{\pm};q)_{\infty}}
{(q^{\frac12}t^2s_i^{\pm}s_j^{\mp};q)_{\infty}(q^{\frac12}t^2s_i^{\pm}s_j^{\pm};q)_{\infty}
(q^{\frac12}t^{-2}s_i^{\pm}s_j^{\mp};q)_{\infty}(q^{\frac12}t^{-2}s_i^{\pm}s_j^{\pm};q)_{\infty}}
\left[\sum_{i=1}^3(s_i+s_i^{-1})\right]^2. 
\end{align}
The two-point function of the S-dual 't Hooft lines with $B=(1,0,0)$ for $SO(7)$ SYM theory can be written as
\begin{align}
\label{so7_t1}
&
\langle T_{(1,0,0)} T_{(1,0,0)}\rangle^{SO(7)}(t;q)
\nonumber\\
&=\frac{1}{8} \frac{(q)_{\infty}^{6}}{(q^{\frac12}t^{\pm2};q)_{\infty}^3}
\oint \prod_{i=1}^{3} \frac{ds_i}{2\pi is_i}
\frac{(q^{\frac12\delta_{i,1}}s_i^{\pm};q)_{\infty}(q^{1+\frac12\delta_{i,1}}s_i^{\pm};q)_{\infty}}
{(q^{\frac{1+\delta_{i,1}}{2}}t^2s_i^{\pm};q)_{\infty}(q^{\frac{1+\delta_{i,1}}{2}}t^{-2}s_i^{\pm};q)_{\infty}}
\nonumber\\
&\times 
\prod_{i<j}
\frac{(q^{\frac12\delta_{i+j,1}}s_i^{\pm}s_j^{\mp};q)_{\infty}(q^{\frac12\delta_{i+j,1}}s_i^{\pm}s_j^{\mp};q)_{\infty}
(q^{1+\frac{1}{2}\delta_{i+j,1}}s_i^{\pm}s_j^{\mp};q)_{\infty}(q^{1+\frac{1}{2}\delta_{i+j,1}}s_i^{\pm}s_j^{\pm};q)_{\infty}}
{(q^{\frac{1+\delta_{i+j,1}}{2}}t^2s_i^{\pm}s_j^{\mp};q)_{\infty}(q^{\frac{1+\delta_{i+j,1}}{2}}t^2s_i^{\pm}s_j^{\pm};q)_{\infty}
(q^{\frac{1+\delta_{i+j,1}}{2}}t^{-2}s_i^{\pm}s_j^{\mp};q)_{\infty}(q^{\frac{1+\delta_{i+j,1}}{2}}t^{-2}s_i^{\pm}s_j^{\pm};q)_{\infty}}. 
\end{align}
The two expressions (\ref{sp6_wfund}) and (\ref{so7_t1}) agree with each other. 

In the half-BPS limit, the two-point function (\ref{sp6_wfund}) of the Wilson lines in the fundamental representation for $USp(6)$ gauge theory
or equivalently the two-point function (\ref{so7_t1}) of the 't Hooft lines of $B=(1,0,0)$ for $SO(7)$ gauge theory reduces to
\begin{align}
\langle W_{\tiny \yng(1)} W_{\tiny \yng(1)}\rangle^{USp(6)}_{\textrm{$\frac12$BPS}}(\mathfrak{q})
&=\langle T_{(1,0,0)} T_{(1,0,0)}\rangle^{SO(7)}_{\textrm{$\frac12$BPS}}(\mathfrak{q})
\nonumber\\
&=\frac{1+\mathfrak{q}^2+\mathfrak{q}^4+\mathfrak{q}^6+\mathfrak{q}^8+\mathfrak{q}^{10}}{(1-\mathfrak{q}^4)(1-\mathfrak{q}^8(1-\mathfrak{q}^{12}))}
\nonumber\\
&=\frac{1}{(1-\mathfrak{q}^2)(1-\mathfrak{q}^4)(1-\mathfrak{q}^8)}. 
\end{align}

Let us consider the multi-point functions. 
While odd-point correlation functions of the fundamental Wilson lines vanish in $USp(6)$ SYM theory, 
there exist non-trivial even-point correlation functions. 
In the half-BPS limit, the $2k$-point function takes the form 
\begin{align}
\langle \underbrace{W_{\tiny \yng(1)} \cdots W_{\tiny \yng(1)}}_{2k} \rangle^{USp(6)}_{\textrm{$\frac12$BPS}}(\mathfrak{q})
&=\frac{\sum_{i=0}^{5k} a_{k\ \tiny\yng(1)}^{\mathfrak{usp}(6)}(i)\mathfrak{q}^{2i}}{(1-\mathfrak{q}^4)(1-\mathfrak{q}^8)(1-\mathfrak{q}^{12})}, 
\end{align}
where $a_{k\ \tiny\yng(1)}^{\mathfrak{usp}(6)}(i)$ are some positive definite integers. 
For example, the four-point function is evaluated as
\begin{align}
\langle W_{\tiny \yng(1)} W_{\tiny \yng(1)} W_{\tiny \yng(1)} W_{\tiny \yng(1)} \rangle^{USp(6)}_{\textrm{$\frac12$BPS}}(\mathfrak{q})
&=\frac{1}{(1-\mathfrak{q}^{4})(1-\mathfrak{q}^{8})(1-\mathfrak{q}^{12})}
(3+6\mathfrak{q}^2+9\mathfrak{q}^4+12\mathfrak{q}^6+14\mathfrak{q}^8
\nonumber\\
&+15\mathfrak{q}^{10}+12\mathfrak{q}^{12}+9\mathfrak{q}^{14}+6\mathfrak{q}^{16}+3\mathfrak{q}^{18}+\mathfrak{q}^{20}). 
\end{align}
The sequence of the first coefficients $a_{k\ \tiny\yng(1)}^{\mathfrak{usp}(6)}(0)$ is in accordance with 
the multiplicity of the trivial representation of the $2k$-th tensor power of the fundamental representation of $USp(6)$ \cite{MR1235279}. 
It is also known to be the sequence of the $2k$-th moment of the trace of random matrix in $USp(6)$ \cite{MR1659828,MR2555991,MR3502944}. 
It has a generating function
\begin{align}
\det \left( 
\begin{matrix}
F_0&F_1&F_2\\
F_1&F_2&F_3\\
F_2&F_3&F_4\\
\end{matrix}
 \right)
&=
\sum_{k=0}^{\infty} \frac{a_{k\ \tiny\yng(1)}^{\mathfrak{usp}(6)}(0)}{(2k)!} x^{2k}, 
\end{align}
where 
\begin{align}
\label{F_gene_usp}
F_{m}(x)&:=\sum_{j=0}^{m}\left(
\begin{matrix}
m\\
j\\
\end{matrix}
\right) \left(
I_{2j-m}(2x)-I_{2j-m+2}(2x)
\right)
\end{align}
and
\begin{align}
\label{modBessel1st}
I_{k}(2x)&:=\sum_{n=0}^{\infty}\frac{x^{2n+k}}{n! (n+k)!}
\end{align}
is the modified Bessel function of the first kind of order $k$. 

\subsubsection{Antisymmetric Wilson line}
Let us turn to the correlation functions of the Wilson lines transforming in the antisymmetric representations. 
For the Wilson line transforming in the irreducible representation labeled by the Young diagram with even number of boxes, 
one finds non-vanishing odd-point correlation functions. 

For example, for the rank-$2$ antisymmetric Wilson line, 
the half-BPS limits of the one- and two-point functions are are given by
\begin{align}
\langle W_{\tiny \yng(1,1)} \rangle^{USp(6)}_{\textrm{$\frac12$BPS}}(\mathfrak{q})
&=\frac{\mathfrak{q}^4+\mathfrak{q}^8}{(1-\mathfrak{q}^4)(1-\mathfrak{q}^8)(1-\mathfrak{q}^{12})}, \\
\langle W_{\tiny \yng(1,1)} W_{\tiny \yng(1,1)}\rangle^{USp(6)}_{\textrm{$\frac12$BPS}}(\mathfrak{q})
&=\frac{1+\mathfrak{q}^2+2\mathfrak{q}^4+2\mathfrak{q}^6+3\mathfrak{q}^8+2\mathfrak{q}^{10}+3\mathfrak{q}^{12}+\mathfrak{q}^{14}+\mathfrak{q}^{16}}
{(1-\mathfrak{q}^4)(1-\mathfrak{q}^8)(1-\mathfrak{q}^{12})}. 
\end{align}

For the rank-$3$ antisymmetric Wilson line, the one-point function vanishes. 
The half-BPS limit of the two-point functions of the Wilson lines in the antisymmetric representation are evaluated as
\begin{align}
\langle W_{\tiny \yng(1,1,1)} W_{\tiny \yng(1,1,1)}\rangle^{USp(6)}_{\textrm{$\frac12$BPS}}(\mathfrak{q})
&=\frac{1+\mathfrak{q}^2+\mathfrak{q}^4+2\mathfrak{q}^6+2\mathfrak{q}^8+2\mathfrak{q}^{10}+2\mathfrak{q}^{12}+\mathfrak{q}^{14}+\mathfrak{q}^{16}+\mathfrak{q}^{18}}
{(1-\mathfrak{q}^4)(1-\mathfrak{q}^8)(1-\mathfrak{q}^{12})}. 
\end{align}

\subsubsection{Symmetric Wilson line}
For the Wilson lines transforming in the adjoint representation, i.e. the rank-$2$ symmetric representations, 
we find that the correlation functions agree with those of the adjoint Wilson line for $SO(7)$ gauge theory 
\begin{align}
\langle \underbrace{W_{\tiny \yng(2)} \cdots W_{\tiny \yng(2)}}_{k} \rangle^{USp(6)}(t;q)
&=\langle \underbrace{W_{\tiny \yng(1,1)} \cdots W_{\tiny \yng(1,1)}}_{k} \rangle^{SO(7)}(t;q). 
\end{align}
For example, the half-BPS limits of the Willson lines are given by
\begin{align}
\langle W_{\tiny \yng(2)} \rangle^{USp(6)}_{\textrm{$\frac12$BPS}}(\mathfrak{q})
&=\frac{\mathfrak{q}^2+\mathfrak{q}^6+\mathfrak{q}^{10}}{(1-\mathfrak{q}^4)(1-\mathfrak{q}^8)(1-\mathfrak{q}^{12})}, \\
\langle W_{\tiny \yng(2)} W_{\tiny \yng(2)}\rangle^{USp(6)}_{\textrm{$\frac12$BPS}}(\mathfrak{q})
&=\frac{1}{(1-\mathfrak{q}^4)(1-\mathfrak{q}^8)(1-\mathfrak{q}^{12})}
\nonumber\\
&\times 
(1+\mathfrak{q}^2+3\mathfrak{q}^4+2\mathfrak{q}^6+5\mathfrak{q}^8+3\mathfrak{q}^{10}
\nonumber\\
&+5\mathfrak{q}^{12}+2\mathfrak{q}^{14}+3\mathfrak{q}^{16}+\mathfrak{q}^{18}+\mathfrak{q}^{20}
), 
\end{align}
which agree with (\ref{halfBPS_so7wasym_1pt}) and (\ref{halfBPS_so7wasym_2pt}) respectively. 

More generally, for the rank-$2l$ symmetric Wilson lines, 
the correlation functions take the form
\begin{align}
\langle \underbrace{W_{(2l)}\cdots W_{(2l)}}_{k} \rangle^{USp(6)}_{\textrm{$\frac12$BPS}}(\mathfrak{q})
&=\frac{\sum_{i=0}^{5lk} a_{k\ (2l)}^{\mathfrak{usp}(6)}(i) \mathfrak{q}^{2i} }{(1-\mathfrak{q}^4)(1-\mathfrak{q}^8)(1-\mathfrak{q}^{12})}
\end{align}
where $a_{k\ (2l)}^{\mathfrak{usp}(6)}(i)$ are some positive definite integers. 
For example for $k=1$, $l=2$, we have 
\begin{align}
\langle W_{\tiny \yng(4)} \rangle^{USp(6)}_{\textrm{$\frac12$BPS}}(\mathfrak{q})
&=\frac{\mathfrak{q}^4+\mathfrak{q}^8+2\mathfrak{q}^{12}+\mathfrak{q}^{16}+\mathfrak{q}^{20}}
{(1-\mathfrak{q}^4)(1-\mathfrak{q}^8)(1-\mathfrak{q}^{12})}. 
\end{align}

\subsection{$\mathfrak{usp}(2N)$}
Let us consider $\mathcal{N}=4$ SYM theory based on the gauge algebra $\mathfrak{usp}(2N)$ of general rank. 

\subsubsection{Fundamental Wilson line}
The $USp(2N)$ SYM theory admits the Wilson line in the fundamental representation. 
The one-point function of the fundamental Wilson line vanishes. 
The two-point function of the fundamental Wilson lines is given by
\begin{align}
\label{sp2N_wfund}
&
\langle W_{\tiny \yng(1)} W_{\tiny \yng(1)}\rangle^{USp(2N)}(t;q)
\nonumber\\
&=\frac{1}{2^N N!}\frac{(q)_{\infty}^{2N}}{(q^{\frac12}t^{\pm2};q)_{\infty}^N}
\oint \prod_{i=1}^{N}\frac{ds_i}{2\pi is_i}
\frac{(s_i^{\pm2};q)_{\infty} (qs_i^{\pm2};q)_{\infty}}
{(q^{\frac{1}{2}}t^{2}s_i^{\pm2};q)_{\infty}(q^{\frac12}t^{-2}s_i^{\pm2};q)_{\infty}}
\nonumber\\
&\times 
\prod_{i<j}\frac{(s_i^{\pm}s_j^{\mp};q)_{\infty}(s_i^{\pm}s_j^{\pm};q)_{\infty}(qs_i^{\pm}s_j^{\mp};q)_{\infty}(qs_i^{\pm}s_j^{\pm};q)_{\infty}}
{(q^{\frac12}t^2s_i^{\pm}s_j^{\mp};q)_{\infty}(q^{\frac12}t^2s_i^{\pm}s_j^{\pm};q)_{\infty}
(q^{\frac12}t^{-2}s_i^{\pm}s_j^{\mp};q)_{\infty}(q^{\frac12}t^{-2}s_i^{\pm}s_j^{\pm};q)_{\infty}}
\left[\sum_{i=1}^N(s_i+s_i^{-1})\right]^2. 
\end{align}
It agrees with the two-point function of the 't Hooft line of $B=(1,0^{N-1})$ in $SO(2N+1)$ SYM theory 
\begin{align}
\label{so2N+1_t1}
&
\langle T_{(1,0^{N-1})} T_{(1,0^{N-1})}\rangle^{SO(2N+1)}(t;q)
\nonumber\\
&=\frac{1}{2^{N-1}(N-1)!} \frac{(q)_{\infty}^{2N}}{(q^{\frac12}t^{\pm2};q)_{\infty}^N}
\oint \prod_{i=1}^{N} \frac{ds_i}{2\pi is_i}
\frac{(q^{\frac12\delta_{i,1}}s_i^{\pm};q)_{\infty}(q^{1+\frac12\delta_{i,1}}s_i^{\pm};q)_{\infty}}
{(q^{\frac{1+\delta_{i,1}}{2}}t^2s_i^{\pm};q)_{\infty}(q^{\frac{1+\delta_{i,1}}{2}}t^{-2}s_i^{\pm};q)_{\infty}}
\nonumber\\
&\times 
\prod_{i<j}
\frac{(q^{\frac12\delta_{i+j,1}}s_i^{\pm}s_j^{\mp};q)_{\infty}(q^{\frac12\delta_{i+j,1}}s_i^{\pm}s_j^{\mp};q)_{\infty}
(q^{1+\frac{1}{2}\delta_{i+j,1}}s_i^{\pm}s_j^{\mp};q)_{\infty}(q^{1+\frac{1}{2}\delta_{i+j,1}}s_i^{\pm}s_j^{\pm};q)_{\infty}}
{(q^{\frac{1+\delta_{i+j,1}}{2}}t^2s_i^{\pm}s_j^{\mp};q)_{\infty}(q^{\frac{1+\delta_{i+j,1}}{2}}t^2s_i^{\pm}s_j^{\pm};q)_{\infty}
(q^{\frac{1+\delta_{i+j,1}}{2}}t^{-2}s_i^{\pm}s_j^{\mp};q)_{\infty}(q^{\frac{1+\delta_{i+j,1}}{2}}t^{-2}s_i^{\pm}s_j^{\pm};q)_{\infty}}. 
\end{align} 

The closed-form expression for the half-BPS limit can be derived from the matrix integral (\ref{so2N+1_t1}). 
We find that the half-BPS limit of the two-point function of the Wilson lines in the fundamental representation for $USp(2N)$ SYM theory 
or equivalently that of the 't Hooft line of $B=(1,0^{N-1})$ in $SO(2N+1)$ SYM theory is given by
\begin{align}
\langle W_{\tiny \yng(1)} W_{\tiny \yng(1)}\rangle^{USp(2N)}_{\textrm{$\frac12$BPS}}(\mathfrak{q})
&=\langle T_{(1,0^{N-1})} T_{(1,0^{N-1})}\rangle^{SO(2N+1)}_{\textrm{$\frac12$BPS}}(\mathfrak{q})
\nonumber\\
&=\frac{1}{(1-\mathfrak{q}^2)\prod_{n=1}^{N-1}(1-\mathfrak{q}^{4n})}. 
\end{align}
Since the half-BPS index of $USp(2N)$ gauge theory is given by
\begin{align}
\mathcal{I}^{USp(2N)}_{\textrm{$\frac12$BPS}}(\mathfrak{q})
&=\prod_{n=1}^{N}\frac{1}{1-\mathfrak{q}^{4n}}, 
\end{align}
We obtain the expression for the normalized two-point function 
\begin{align}
\label{nhalfBPS_sp2N_wfund}
\langle \mathcal{W}_{\tiny \yng(1)} \mathcal{W}_{\tiny \yng(1)}\rangle^{USp(2N)}_{\textrm{$\frac12$BPS}}(\mathfrak{q})
&=\frac{1-\mathfrak{q}^{4N}}{1-\mathfrak{q}^2},
\end{align}
which agrees with the normalized two-point function (\ref{nhalfBPS_so2N+1_wfund}) of the fundamental Wilson lines in $SO(2N+1)$ gauge theory. 

While the odd-point functions of the fundamental Wilson lines vanish, the even-point correlation functions are non-trivial. 
They can be written as
\begin{align}
\langle \underbrace{
W_{\tiny \yng(1)}\cdots W_{\tiny \yng(1)}}_{2k} \rangle^{USp(2N)}_{\textrm{$\frac12$BPS}}(\mathfrak{q})
&=\frac{\sum_{i=0}^{(2N-1)k} a_{k\ \tiny\yng(1)}^{\mathfrak{usp}(2N)}(i) \mathfrak{q}^{2i}}
{\prod_{n=1}^N (1-\mathfrak{q}^{4n})}, 
\end{align}
with $a_{k\ \tiny\yng(1)}^{\mathfrak{usp}(2N)}(i)$ being non-negative integers. 
It follows that the sequence of the first coefficients $a_{k\ \tiny\yng(1)}^{\mathfrak{usp}(2N)}(0)$ is 
equal to the multiplicity of the trivial representation of the $2k$-th tensor power of the fundamental representation of $USp(2N)$ \cite{MR1235279}. 
This is also known as the sequence of the $2k$-th moment of the trace of random matrix in $USp(2N)$ \cite{MR1659828,MR2555991,MR3502944}. 
They can be obtained from the generating function 
\begin{align}
\det (F_{i+j-2}(x))
&=\sum_{k=0}^{\infty}\frac{a_{k\ \tiny\yng(1)}^{\mathfrak{usp}(2N)}(0)}{(2k)!}x^{2k}
\end{align}
with $i,j$ $=$ $1,2,\cdots, N$. 
Here $F_m(x)$ is defined by (\ref{F_gene_usp}) 
and $I_k(2x)$ is the modified Bessel function (\ref{modBessel1st}) of the first kind of order $k$. 

\subsubsection{Antisymmetric Wilson line}
Also $USp(2N)$ gauge theory has non-vanishing one-point function of the Wilson line 
transforming in the irreducible representation labeled by the Young diagram with even number of boxes. 

In the half-BPS limits, the correlators of the rank-$2$ antisymmetric Wilson lines take the form
\begin{align}
\langle \underbrace{W_{\tiny \yng(1,1)} \cdots W_{\tiny \yng(1,1)}}_{k} \rangle^{USp(2N)}_{\textrm{$\frac12$BPS}}(\mathfrak{q})
&=\frac{\sum_{i=0}^{2(N-1)k} a_{k\ \tiny\yng(1,1)}^{\mathfrak{usp}(2N)}(i) \mathfrak{q}^{2i}}{\prod_{n=1}^N(1-\mathfrak{q}^{4n})},
\end{align}
with $a_{k\ \tiny\yng(1,1)}^{\mathfrak{usp}(2N)}(i)$ being non-negative integers. 

\subsubsection{Symmetric Wilson line}
Next consider the rank-$l$ symmetric Wilson lines. 
For $l=2$, it is the adjoint Wilson line. 
The correlation functions of the adjoint Wilson line in $USp(2N)$ gauge theory agree with those in $SO(2N+1)$ gauge theory
\begin{align}
\langle \underbrace{W_{\tiny \yng(2)} \cdots W_{\tiny \yng(2)}}_{k} \rangle^{USp(2N)}(t;q)
&=\langle \underbrace{W_{\tiny \yng(1,1)} \cdots W_{\tiny \yng(1,1)}}_{k} \rangle^{SO(2N+1)}(t;q). 
\end{align}
In particular, the one-point function in the half-BPS limit is equal to (\ref{halfBPS_so2N+1wadj_1pt})
\begin{align}
\langle W_{\tiny \yng(2)} \rangle^{USp(2N)}_{\textrm{$\frac12$BPS}}(\mathfrak{q})
&=\frac{\mathfrak{q}^2+\mathfrak{q}^6+\cdots+\mathfrak{q}^{4N-2}}
{\prod_{n=1}^N (1-\mathfrak{q}^{4n})}
\nonumber\\
&=\frac{\mathfrak{q}^2(1-\mathfrak{q}^{4N})}{(1-\mathfrak{q}^4)
\prod_{n=1}^{N}(1-\mathfrak{q}^{4n})}. 
\end{align}

While the odd-point correlation functions are trivial for the rank-$(2l+1)$ symmetric Wilson lines, 
they are non-trivial for the rank-$2l$ symmetric Wilson lines. 
In the half-BPS limit they take the form
\begin{align}
\langle \underbrace{W_{(2l)} \cdots W_{(2l)}}_{k} \rangle^{USp(2N)}_{\textrm{$\frac12$BPS}}(\mathfrak{q})
&=\frac{\sum_{i=0}^{(2N-1)lk} a_{k\ (2l)}^{\mathfrak{usp}(2N)}(i) \mathfrak{q}^{2i}}{\prod_{n=1}^N(1-\mathfrak{q}^{4n})},
\end{align}
with $a_{k\ (2l)}^{\mathfrak{usp}(2N)}(i)$ being non-negative integers. 

\section{$\mathfrak{so}(2N)$}
\label{sec_Dtype}
For $\mathcal{N}=4$ gauge theory of gauge group based on the Lie algebra $\mathfrak{so}(2N)$, 
the center of $Spin(2N)$ is $\mathbb{Z}_2\times \mathbb{Z}_2$ for $N$ even and $\mathbb{Z}_4$ for $N$ odd. 
In this case, the possible gauge groups depend on whether $N$ is even or odd. 

When $N$ is even, the center of $Spin(2N)$ is $\mathbb{Z}_2^{S}\times \mathbb{Z}_2^{C}$. 
The $Spin(2N)$ gauge theory has two distinct spinor representations, which are not complex conjugates 
in such a way that $\mathbb{Z}_2^{S}$ acts on one spinor representation by $-1$ and on the other by $+1$ 
while the $\mathbb{Z}_2^{C}$ acts in an opposite manner. 
The line operators have charges 
\begin{align}
(z_{e,S},z_{e,C};z_{m,S},z_{m,C})
&\in (\mathbb{Z}_2\times \mathbb{Z}_2)
\times (\mathbb{Z}_2\times \mathbb{Z}_2). 
\end{align}
One finds five kinds of possible gauge groups by taking the quotients by $\mathbb{Z}_2^S$, $\mathbb{Z}_2^{C}$ and 
$\mathbb{Z}_2^{V}$, the diagonal subgroup of $\mathbb{Z}_2^S$ $\times$ $\mathbb{Z}_2^C$ 
\footnote{
While the $Ss(2N)$ and $Sc(2N)$ theories are related with one another by the $\mathbb{Z}_2$ outer-automorphism of $Spin(2N)$, 
it is convenient to deal them separately when one considers S-duality \cite{Aharony:2013hda}. 
}
\cite{Aharony:2013hda}
\begin{align}
&Spin(2N), \nonumber\\
&SO(2N)=Spin(2N)/\mathbb{Z}_2^V, \nonumber\\
&Ss(2N)=Spin(2N)/\mathbb{Z}_2^S, \nonumber\\
&Sc(2N)=Spin(2N)/\mathbb{Z}_2^C, \nonumber\\
&SO(2N)/\mathbb{Z}_2=Spin(2N)/(\mathbb{Z}_2^S\times \mathbb{Z}_2^C). 
\end{align}
The line operators in $Spin(2N)$ and $SO(2N)$ gauge theories carry the charges 
which are the same as those for $Spin(2N+1)$ and $SO(2N+1)$ gauge theories. 
For $N=4d+2$, there are two types of $Ss(2N)$ gauge theory has the line operators with
\begin{align}
Ss(2N)_{+}: 
(z_{e,S},z_{e,C};z_{m,S},z_{m,C})&=(0,0;0,0), \nonumber \\
(z_{e,S},z_{e,C};z_{m,S},z_{m,C})&=(1,0;0,0), \nonumber \\
(z_{e,S},z_{e,C};z_{m,S},z_{m,C})&=(0,0;0,1), \\
Ss(2N)_{-}: 
(z_{e,S},z_{e,C};z_{m,S},z_{m,C})&=(0,0;0,0), \nonumber \\
(z_{e,S},z_{e,C};z_{m,S},z_{m,C})&=(1,0;0,0), \nonumber \\
(z_{e,S},z_{e,C};z_{m,S},z_{m,C})&=(0,1;0,1). 
\end{align}
For $N=4d$, we have
\begin{align}
Ss(2N)_{+}: 
(z_{e,S},z_{e,C};z_{m,S},z_{m,C})&=(0,0;0,0), \nonumber \\
(z_{e,S},z_{e,C};z_{m,S},z_{m,C})&=(1,0;0,0), \nonumber \\
(z_{e,S},z_{e,C};z_{m,S},z_{m,C})&=(0,0;1,0), \\
Ss(2N)_{-}: 
(z_{e,S},z_{e,C};z_{m,S},z_{m,C})&=(0,0;0,0), \nonumber \\
(z_{e,S},z_{e,C};z_{m,S},z_{m,C})&=(1,0;0,0), \nonumber \\
(z_{e,S},z_{e,C};z_{m,S},z_{m,C})&=(0,1;1,0). 
\end{align}
The $SO(2N)/\mathbb{Z}_2$ gauge theory admits the lines with the charges of the form 
\begin{align}
(z_{e,S},z_{e,C};z_{m,S},z_{m,C})&=(0,0;0,0), \nonumber \\
(z_{e,S},z_{e,C};z_{m,S},z_{m,C})&=(n_{SS},n_{SC};1,0), \nonumber \\
(z_{e,S},z_{e,C};z_{m,S},z_{m,C})&=(n_{CS},n_{CC};0,1), 
\end{align}
where $n_{SS}$, $n_{SC}$, $n_{CS}$ and $n_{CC}$ are $0$ or $1$. 
When $N=4d+2$ (resp. $N=4d$), $n_{SC}=n_{CS}$ (resp. $n_{SS}=n_{CC}$). 
Hence there are eight types of $SO(2N)/\mathbb{Z}_2$ theories, which is denoted by 
$(SO(2N)/\mathbb{Z}_2)_{\begin{smallmatrix}n_{SS},&n_{SC}\\n_{CS},&n_{CC}\end{smallmatrix}}$. 

For odd $N$, the center of $Spin(2N)$ is $\mathbb{Z}_4$. 
The two spinor representations are complex conjugates and $\mathbb{Z}_4$ acts as $\pm i$ on the spinors. 
The line operators carry electric and magnetic charges 
\begin{align}
(z_e,z_m)\in \mathbb{Z}_4\times \mathbb{Z}_4. 
\end{align}
There exist three types of gauge groups \cite{Aharony:2013hda}
\begin{align}
&Spin(2N), \nonumber\\
&SO(2N)=Spin(2N)/\mathbb{Z}_2, \nonumber\\
&SO(2N)/\mathbb{Z}_2=Spin(2N)/\mathbb{Z}_4. 
\end{align}
The line operators in the first two theories are the same as those in the $Spin(2N+1)$ and $SO(2N+1)$ gauge theories. 
The $SO(2N)/\mathbb{Z}_2$ theory involves the line operators of charge
\begin{align}
(z_e,z_m)&=(0,0), \qquad (z_e,z_m)=(n,1)
\end{align}
with $n=0,1,2,3$. 
Correspondingly, we have four kinds of theories $(SO(2N)/\mathbb{Z}_2)_{n\mod 4}$. 

Although the Lie algebra of the Langlands dual gauge group is the same, 
gauge theories based on the gauge algebra $\mathfrak{so}(2N)$ map non-trivially under S-duality in the presence of line operators. 
The Wilson line in the spinor representation for $Spin(2N)$ gauge theory 
is S-dual to the 't Hooft line of $B$ $=$ $(\frac12^{N})$ for $SO(2N)/\mathbb{Z}_2$ gauge theory. 
On the other hand, the Wilson line in the fundamental representation for $SO(2N)_+$ 
maps to the 't Hooft line of $B$ $=$ $(1,0^{N-1})$ for $SO(2N)_+$ gauge theory. 

Furthermore, we can consider the gauge theory with disconnected gauge group $O(2N)$ 
by gauging the global discrete $\mathbb{Z}_2$ symmetry of $SO(2N)$ gauge theory. 

For the chiral spinor representation we have the character 
\begin{align}
\chi_{\textrm{sp}}^{\mathfrak{so}(2N)}
&=\frac12 \left[
\prod_{i=1}^N (s_i^{\frac12}+s_i^{-\frac12})
+ \prod_{i=1}^N (s_i^{\frac12}-s_i^{-\frac12})
\right], 
\end{align}
and for the antichiral spinor representations we have
\begin{align}
\chi_{\overline{\textrm{sp}}}^{\mathfrak{so}(2N)}
&=\frac12 \left[
\prod_{i=1}^N (s_i^{\frac12}+s_i^{-\frac12})
-\prod_{i=1}^N (s_i^{\frac12}-s_i^{-\frac12})
\right]. 
\end{align}

The character of the fundamental representation for $\mathfrak{so}(2N)$ is 
\begin{align}
\chi_{\tiny \yng(1)}^{\mathfrak{so}(2N)}
&=\sum_{i=1}^{N}(s_i+s_i^{-1}). 
\end{align}

The character of the irreducible representation of $\mathfrak{so}(2N)$ with highest weight labeled by $\lambda=(\lambda_1,\dots, \lambda_N)$ ($\lambda_1 \geq \dots \geq \lambda_{N-1} \geq |\lambda_N| \geq 0$) is \cite{MR1153249}
\begin{align}
\label{ch_so2N_irrep}
\chi_{\lambda}^{\mathfrak{so}(2N)}
&=\frac{\det (s_j^{\lambda_i+N-i}+s_j^{-\lambda_i-N+i})+\det(s_j^{\lambda_i+N-i}-s_j^{-\lambda_i-N+i})
}{\det(s_j^{N-i}+s_j^{-N+i})}. 
\end{align}
Note that the chiral and antichiral spinor representations correspond to $\lambda=\text{sp}=(1/2,\dots,1/2, 1/2)$ and $\lambda=\overline{\text{sp}}=(1/2,\dots,1/2, -1/2)$, respectively. 
For the disconnected component of $\mathfrak{so}(2N)$ gauge theory, 
we instead employ the character (\ref{ch_usp2N_irrep}) of the $\mathfrak{usp}(2N-2)$. 

The line defect correlators can be written as
\begin{align}
&\langle W_{\mathcal{R}_1}\cdots W_{\mathcal{R}_k}\rangle^{SO(2N)}\notag\\
&=\int \! d\mu^{SO(2N)}
\exp \left( \sum_{n=1}^\infty \frac{1}{n}f_n(q,t) M_n(s) \right)\prod_{i=1}^{k}\chi^{\mathfrak{so}(2N)}_{\mathcal{R}_i}(s), 
\end{align}
where $f_n(q,t)$ is given by \eqref{eq:f_n}. The measure is given by
\begin{equation}
\begin{aligned}
d\mu^{SO(2N)}=\frac{1}{2^{N-1}N!} \prod_{i=1}^{N} \frac{ds_i}{2\pi i s_i} \prod_{1\leq i<j\leq N} (1-s_is_j)(1-s_i^{-1}s_j^{-1})(1-s_is_j^{-1})(1-s_i^{-1}s_j),
\end{aligned}
\end{equation}
and we have introduced the notation
\begin{equation}
\begin{aligned}
M_n(s)=\frac{P_n(s)^2+P_{2n}(s)}{2},
\end{aligned}
\end{equation}
where the power sum function $P_m(s)$ is defined by \eqref{P_m}.

In this case, the application of the character expansion method to line defect correlators is more subtle. For a representation $\lambda$ with length $l(\lambda)<N$, we can apply the character expansion method straightforwardly. However, if $\ell(\lambda)=N$, we have two representations $\lambda=(\lambda_1,\dots, \lambda_{N-1}, \lambda_N)$ and $\overline{\lambda}=(\lambda_1,\dots, \lambda_{N-1}, -\lambda_N)$. In this case, the characters $\chi_{\lambda}^{\mathfrak{so}(2N)}$ and $\chi_{\overline{\lambda}}^{\mathfrak{so}(2N)}$ are not expressed by the power sum function $P_{m}$. We have to use the Littlewood-Richardson rule in order to rewrite the integrand in terms of the characters.

\subsection{$\mathfrak{so}(4)$}
The $SO(4)$ index is equal to the square of the $SU(2)$ index. 
Using the formula in \cite{Hatsuda:2022xdv}, we can write 
\begin{align}
\mathcal{I}^{SO(4)}(t;q)
&=\mathcal{I}^{SU(2)}(t;q) \times \mathcal{I}^{SU(2)}(t;q)
\nonumber\\
&=\frac{(q^{\frac12}t^{\pm2};q)_{\infty}^2}
{(q;q)_{\infty}^4}
\left(
\sum_{\begin{smallmatrix}
p_1, p_2\in \mathbb{Z}\\
p_1<p_2\\
\end{smallmatrix}}
\frac{(q^{\frac12}t^{-2})^{p_1+p_2-2}}
{(1-q^{p_1-\frac12}t^2) (1-q^{p_2-\frac12}t^2)}
\right)^2. 
\end{align}

\subsubsection{Spinor Wilson line}
The $Spin(4)$ gauge theory has the Wilson line transforming in the (anti)chiral spinor representation. 
The chiral representation and the antichiral representation are not complex conjugates. 
The two-point function of the chiral spinor Wilson line and the antichiral spinor Wilson line vanishes. 
However, the two-point function of a pair of the Wilson lines in the chiral spinor representation is non-trivial. 
It can be written as
\begin{align}
\label{so4_wsp}
&
\langle W_{\textrm{sp}} W_{\textrm{sp}}\rangle^{Spin(4)}(t;q)
=\frac{1}{4}\frac{(q)_{\infty}^4}{(q^{\frac12}t^{\pm2};q)_{\infty}^2}
\oint \prod_{i=1}^2 \frac{ds_i}{2\pi is_i}
\nonumber\\
&\times 
\frac{(s_1^{\pm}s_2^{\mp};q)_{\infty}(s_1^{\pm}s_2^{\pm};q)_{\infty}
(qs_1^{\pm}s_2^{\mp};q)_{\infty}(qs_1^{\pm}s_2^{\pm};q)_{\infty}}
{(q^{\frac12}t^2s_1^{\pm}s_2^{\mp};q)_{\infty}(q^{\frac12}t^2s_1^{\pm}s_2^{\pm};q)_{\infty}
(q^{\frac12}t^{-2}s_1^{\pm}s_2^{\mp};q)_{\infty}(q^{\frac12}t^{-2}s_1^{\pm}s_2^{\pm};q)_{\infty}}
\nonumber\\
&\times (s_1^{\frac12}s_2^{\frac12}+ s_1^{-\frac12}s_2^{-\frac12})^2. 
\end{align}
Similarly, one can obtain 
the two-point function of a pair of the Wilson lines in the antichiral spinor representation by replacing $s_2 \to s_2^{-1}$ in the character. Since the other part in the integrand is invariant under this replacement, we conclude that it leads to the same results as (\ref{so4_wsp}). 

The chiral spinor Wilson line in $Spin(4)$ gauge theory is S-dual 
to the 't Hooft lines with $B=(\frac12,\frac12)$ for $(SO(4)/\mathbb{Z}_2)_{\begin{smallmatrix}+&+\\+&+\\\end{smallmatrix}}$ gauge theory. 
The two-point function of the dual 't Hooft lines is evaluated as 
\begin{align}
\label{so4_tsp}
&
\langle T_{(\frac12,\frac12)} T_{(\frac12,\frac12)}\rangle^{SO(4)/\mathbb{Z}_2}(t;q)
=\frac12 \frac{(q)_{\infty}^4}{(q^{\frac12}t^{\pm2};q)_{\infty}^2}
\oint \prod_{i=1}^2 \frac{ds_i}{2\pi is_i}
\nonumber\\
&\times 
\frac{(q^{\frac12}s_1^{\pm}s_2^{\mp};q)_{\infty}(s_1^{\pm}s_2^{\pm};q)_{\infty}
(q^{\frac32}s_1^{\pm}s_2^{\mp};q)_{\infty}(qs_1^{\pm}s_2^{\pm};q)_{\infty}}
{(qt^2s_1^{\pm}s_2^{\mp};q)_{\infty}(q^{\frac12}t^2s_1^{\pm}s_2^{\pm};q)_{\infty}
(qt^{-2}s_1^{\pm}s_2^{\mp};q)_{\infty}(q^{\frac12}t^{-2}s_1^{\pm}s_2^{\pm};q)_{\infty}}. 
\end{align}
In fact, the expressions (\ref{so4_wsp}) and (\ref{so4_tsp}) agree with each other. 

As we have the isomorphism $Spin(4)$ $\cong$ $SU(2)$ $\times$ $SU(2)$ and $SO(4)$ $\cong$ $(SU(2)\times SU(2))/\mathbb{Z}_2$, 
line defect correlation functions can be obtained from the results for $SU(2)$ gauge theory. 
We find that the two-point functions (\ref{so4_wsp}) and (\ref{so4_tsp}) are given by 
\begin{align}
\langle W_{\textrm{sp}} W_{\textrm{sp}}\rangle^{Spin(4)}(t;q)
&=\langle T_{(\frac12,\frac12)} T_{(\frac12,\frac12)}\rangle^{SO(4)/\mathbb{Z}_2}(t;q)
\nonumber\\
&=
\mathcal{I}^{SU(2)}(t;q)
\langle W_{\tiny \yng(1)} W_{\tiny \yng(1)}\rangle^{SU(2)}(t;q), 
\end{align}
where $\mathcal{I}^{SU(2)}(t;q)$ is the $SU(2)$ Schur index which is identical to (\ref{Schur_so3_sp2}) 
and $\langle W_{\tiny \yng(1)} W_{\tiny \yng(1)}\rangle^{SU(2)}(t;q)$ 
is the two-point function (\ref{su2w1_exact}) of the fundamental Wilson lines in $SU(2)$ gauge theory. 

The half-BPS limits of the two-point functions are given by
\begin{align}
\langle W_{\textrm{sp}} W_{\textrm{sp}}\rangle^{Spin(4)}_{\textrm{$\frac12$BPS}}(\mathfrak{q})
&=\langle T_{(\frac12,\frac12)} T_{(\frac12,\frac12)}\rangle^{SO(4)/\mathbb{Z}_2}_{\textrm{$\frac12$BPS}}(\mathfrak{q})
\nonumber\\
&=\frac{1+\mathfrak{q}^2}{(1-\mathfrak{q}^4)^2}
\nonumber\\
&=\frac{1}{(1-\mathfrak{q}^2)(1-\mathfrak{q}^4)}. 
\end{align}

More generally, the even-point correlation functions of the chiral spinor Wilson lines are non-trivial, whereas the odd-point ones vanish. 
Again the even-point correlation functions can be evaluated from the relation to the observables in $SU(2)$ gauge theory
\begin{align}
\langle \underbrace{W_{\textrm{sp}} \cdots W_{\textrm{sp}}}_{2k} \rangle^{Spin(4)}(t;q)
&=\mathcal{I}^{SU(2)}(t;q) \langle \underbrace{W_{\tiny \yng(1)} \cdots W_{\tiny \yng(1)}}_{2k} \rangle^{SU(2)}(t;q). 
\end{align}
We find the closed-form expression for the half-BPS limits of the even-point correlation functions 
\begin{align}
\langle \underbrace{W_{\textrm{sp}} \cdots W_{\textrm{sp}}}_{2k} \rangle^{Spin(4)}_{\textrm{$\frac12$BPS}}(\mathfrak{q})
&=\mathcal{I}_{\textrm{$\frac12$BPS}}^{SO(4)}(\mathfrak{q})\sum_{i=0}^{k} a_{k\ \textrm{sp}}^{\mathfrak{so}(4)}(i) \mathfrak{q}^{2i}\notag \\
&=\frac{1}{(1-\mathfrak{q}^4)^2}\sum_{i=0}^{k} a_{k\ \textrm{sp}}^{\mathfrak{so}(4)}(i) \mathfrak{q}^{2i}, 
\end{align}
where 
\begin{align}
\label{so4_sp_coeff}
a_{k\ \textrm{sp}}^{\mathfrak{so}(4)}(i)&=\frac{(2i+1) (2k)!}{(k-i)!(k+i+1)!}. 
\end{align}
Interestingly, we find $a_{k\ \textrm{sp}}^{\mathfrak{so}(4)}(i)=a_{k\ \textrm{sp}}^{\mathfrak{so}(3)}(i)$.
For example, we obtain
\begin{align}
\langle W_{\textrm{sp}} W_{\textrm{sp}}W_{\textrm{sp}} W_{\textrm{sp}} \rangle^{Spin(4)}_{\textrm{$\frac12$BPS}}(\mathfrak{q})
&=\frac{2+3\mathfrak{q}^2+\mathfrak{q}^4}{(1-\mathfrak{q}^4)^2}, \\
\langle W_{\textrm{sp}} W_{\textrm{sp}}W_{\textrm{sp}}W_{\textrm{sp}}W_{\textrm{sp}} W_{\textrm{sp}} \rangle^{Spin(4)}_{\textrm{$\frac12$BPS}}(\mathfrak{q})
&=\frac{5+9\mathfrak{q}^2+5\mathfrak{q}^4+\mathfrak{q}^6}{(1-\mathfrak{q}^4)^2}, \\
\langle W_{\textrm{sp}} W_{\textrm{sp}}W_{\textrm{sp}}W_{\textrm{sp}}W_{\textrm{sp}}W_{\textrm{sp}}W_{\textrm{sp}} W_{\textrm{sp}} \rangle^{Spin(4)}_{\textrm{$\frac12$BPS}}(\mathfrak{q})
&=\frac{14+28\mathfrak{q}^2+20\mathfrak{q}^4+7\mathfrak{q}^6+\mathfrak{q}^8}{(1-\mathfrak{q}^4)^2}. 
\end{align}
Note again that the sequence of the first coefficients $a_{k\ \textrm{sp}}^{\mathfrak{so}(4)}(0)$ is the $k$-th Catalan number. 

One can also consider correlation functions including both the spinor and antispinor line operators. After some experiments, we find the following factorization structure:
\begin{align}
\langle W_{\textrm{sp}}^{2k} W_{\overline{\textrm{sp}}}^{2m} \rangle^{Spin(4)}_{\textrm{$\frac12$BPS}}(\mathfrak{q})
&=(\mathcal{I}_{\textrm{$\frac12$BPS}}^{SO(4)}(\mathfrak{q}))^{-1} \langle W_{\textrm{sp}}^{2k}  \rangle^{Spin(4)}_{\textrm{$\frac12$BPS}}(\mathfrak{q})
\langle W_{\overline{\textrm{sp}}}^{2m} \rangle^{Spin(4)}_{\textrm{$\frac12$BPS}}(\mathfrak{q})\notag \\
&=(1-\mathfrak{q}^4)^2\langle W_{\textrm{sp}}^{2k}  \rangle^{Spin(4)}_{\textrm{$\frac12$BPS}}(\mathfrak{q})
\langle W_{\overline{\textrm{sp}}}^{2m} \rangle^{Spin(4)}_{\textrm{$\frac12$BPS}}(\mathfrak{q}).
\end{align}
Since we further have $\langle W_{\overline{\textrm{sp}}}^{2m} \rangle^{Spin(4)}_{\textrm{$\frac12$BPS}}(\mathfrak{q})=\langle W_{\textrm{sp}}^{2m} \rangle^{Spin(4)}_{\textrm{$\frac12$BPS}}(\mathfrak{q})$, we finally obtain
\begin{align}
\langle W_{\textrm{sp}}^{2k} W_{\overline{\textrm{sp}}}^{2m} \rangle^{Spin(4)}_{\textrm{$\frac12$BPS}}(\mathfrak{q})
=\frac{1}{(1-\mathfrak{q}^4)^2}\biggl(\sum_{i=0}^{k} a_{k\ \textrm{sp}}^{\mathfrak{so}(4)}(i) \mathfrak{q}^{2i}\biggr)\biggl(\sum_{j=0}^{m} a_{m\ \textrm{sp}}^{\mathfrak{so}(4)}(j) \mathfrak{q}^{2j}\biggr).
\end{align}

\subsubsection{Fundamental Wilson line}
The Wilson line operator transforming in the fundamental representation exists for $SO(4)$ gauge theories as the minimal electrically charged line operator. 
While the one-point function of the fundamental Wilson line vanishes, the two-point function is non-trivial. 
The two-point function of the fundamental Wilson lines can be evaluated from the matrix integral
\begin{align}
\label{so4_wfund}
&
\langle W_{\tiny \yng(1)} W_{\tiny \yng(1)}\rangle^{SO(4)}(t;q)
=\frac{1}{4}\frac{(q)_{\infty}^4}{(q^{\frac12}t^{\pm2};q)_{\infty}^2}
\oint \prod_{i=1}^2 \frac{ds_i}{2\pi is_i}
\nonumber\\
&\times 
\frac{(s_1^{\pm}s_2^{\mp};q)_{\infty}(s_1^{\pm}s_2^{\pm};q)_{\infty}
(qs_1^{\pm}s_2^{\mp};q)_{\infty}(qs_1^{\pm}s_2^{\pm};q)_{\infty}}
{(q^{\frac12}t^2s_1^{\pm}s_2^{\mp};q)_{\infty}(q^{\frac12}t^2s_1^{\pm}s_2^{\pm};q)_{\infty}
(q^{\frac12}t^{-2}s_1^{\pm}s_2^{\mp};q)_{\infty}(q^{\frac12}t^{-2}s_1^{\pm}s_2^{\pm};q)_{\infty}}
\nonumber\\
&\times (s_1+s_2+s_1^{-1}+s_2^{-1})^2. 
\end{align}
The fundamental Wilson line is S-dual to the 't Hooft line of $B=(1,0)$ for $SO(4)$ SYM theory. 
The two-point function of the dual 't Hooft line is given by 
\begin{align}
\label{so4_t1}
&
\langle T_{(1,0)} T_{(1,0)}\rangle^{SO(4)}(t;q)
=\frac{(q)_{\infty}^4}{(q^{\frac12}t^{\pm2};q)_{\infty}^2}
\oint \prod_{i=1}^2 \frac{ds_i}{2\pi is_i}
\nonumber\\
&\times 
\frac{(q^{\frac12}s_1^{\pm}s_2^{\mp};q)_{\infty}(q^{\frac12}s_1^{\pm}s_2^{\pm};q)_{\infty}
(q^{\frac32}s_1^{\pm}s_2^{\mp};q)_{\infty}(q^{\frac32}s_1^{\pm}s_2^{\pm};q)_{\infty}}
{(qt^2s_1^{\pm}s_2^{\mp};q)_{\infty}(qt^2s_1^{\pm}s_2^{\pm};q)_{\infty}
(qt^{-2}s_1^{\pm}s_2^{\mp};q)_{\infty}(qt^{-2}s_1^{\pm}s_2^{\pm};q)_{\infty}}. 
\end{align}
Again we can obtain the exact closed-form for the two-point functions (\ref{so4_wfund}) and (\ref{so4_t1}) 
by making use of the results for the $SU(2)$ gauge theory. 
One finds that 
\begin{align}
\label{so4W1_T1}
\langle W_{\tiny \yng(1)} W_{\tiny \yng(1)}\rangle^{SO(4)}(t;q)
&=\langle W_{\tiny \yng(1)} W_{\tiny \yng(1)}\rangle^{SU(2)}(t;q)^2, 
\end{align}
where $\langle W_{\tiny \yng(1)} W_{\tiny \yng(1)}\rangle^{SU(2)}(t;q)$ 
is the two-point function (\ref{su2w1_exact})of the fundamental Wilson lines in $SU(2)$ gauge theory. 

In the half-BPS limit, the two-point function of the Wilson line in the fundamental representation for $SO(4)$ SYM theory is 
\begin{align}
\langle W_{\tiny \yng(1)} W_{\tiny \yng(1)}\rangle^{SO(4)}_{\textrm{$\frac12$BPS}}(\mathfrak{q})
&=\frac{1}{(1-\mathfrak{q}^2)^2}. 
\end{align}

Furthermore, the even-point correlation functions of the fundamental Wilson lines are non-trivial. 
Again they can be evaluated as squares of the even-point functions of the $SU(2)$ fundamental Wilson lines
\begin{align}
\langle \underbrace{W_{\tiny \yng(1)} \cdots W_{\tiny \yng(1)}}_{2k} \rangle^{SO(4)}(t;q)
&=\langle \underbrace{W_{\tiny \yng(1)} \cdots W_{\tiny \yng(1)}}_{k} \rangle^{SU(2)}(t;q)^2, 
\end{align}
It follows that in the half-BPS limit they take the form
\begin{align}
\langle \underbrace{W_{\tiny \yng(1)} \cdots W_{\tiny \yng(1)}}_{2k} \rangle^{SO(4)}_{\textrm{$\frac12$BPS}}(\mathfrak{q})
&=\frac{\sum_{i=0}^{2k} a_{k\ \tiny\yng(1)}^{\mathfrak{so}(4)}(i) \mathfrak{q}^{2i}}
{(1-\mathfrak{q}^4)^2}, 
\end{align}
where $a_{k\ \tiny\yng(1)}^{\mathfrak{so}(4)}(i)$ are some non-negative integers. 
For example, the four-, six- and eight-point functions are given by
\begin{align}
\langle W_{\tiny \yng(1)} W_{\tiny \yng(1)} W_{\tiny \yng(1)} W_{\tiny \yng(1)} \rangle^{SO(4)}_{\textrm{$\frac12$BPS}}(\mathfrak{q})
&=\frac{4+12\mathfrak{q}^2+13\mathfrak{q}^4+6\mathfrak{q}^6+\mathfrak{q}^8}{(1-\mathfrak{q}^4)^2}, \\
\langle W_{\tiny \yng(1)} W_{\tiny \yng(1)} W_{\tiny \yng(1)}W_{\tiny \yng(1)} W_{\tiny \yng(1)} W_{\tiny \yng(1)} \rangle^{SO(4)}_{\textrm{$\frac12$BPS}}(\mathfrak{q})
&=\frac{25+90\mathfrak{q}^2+131\mathfrak{q}^4+100\mathfrak{q}^6+43\mathfrak{q}^8+10\mathfrak{q}^{10}+\mathfrak{q}^{12}}
{(1-\mathfrak{q}^4)^2}, \\
\langle W_{\tiny \yng(1)} W_{\tiny \yng(1)}W_{\tiny \yng(1)} W_{\tiny \yng(1)} W_{\tiny \yng(1)}W_{\tiny \yng(1)} W_{\tiny \yng(1)} W_{\tiny \yng(1)} \rangle^{SO(4)}_{\textrm{$\frac12$BPS}}(\mathfrak{q})
&=\frac{1}{(1-\mathfrak{q}^4)^2}
(
196+784\mathfrak{q}^2+1344\mathfrak{q}^4+1316\mathfrak{q}^6
\nonumber\\
&+820\mathfrak{q}^8+336\mathfrak{q}^{10}+89\mathfrak{q}^{12}+14\mathfrak{q}^{14}+\mathfrak{q}^{16}
). 
\end{align}
The sequence of the first coefficients $a_{k\ \tiny\yng(1)}^{\mathfrak{so}(4)}(0)$ is given by a square of the $k$-th Catalan number
\begin{align}
a_{k\ \tiny\yng(1)}^{\mathfrak{so}(4)}(0)&=C_k^2. 
\end{align}

Next consider the $SO(4)^{-}$ gauge theory, the disconnected component for $SO(4)$ gauge theory. 
In this case, the matrix integral of the two-point function of the Wilson lines in the fundamental representation takes the form 
\begin{align}
\label{so4-_wfund}
&
\langle W_{\tiny \yng(1)} W_{\tiny \yng(1)}\rangle^{SO(4)^{-}}(t;q)
=\frac12 \frac{(q)_{\infty}^2(-q;q)_{\infty}^2}{(q^{\frac12}t^{\pm2};q)_{\infty}(-q^{\frac12}t^{\pm2};q)_{\infty}}
\oint \frac{ds}{2\pi is}
\nonumber\\
&\times 
\frac{(s^{\pm};q)_{\infty}(-s;q)_{\infty}(qs^{\pm};q)_{\infty}(-qs^{\pm};q)_{\infty}}
{(q^{\frac12}t^2 s^{\pm};q)_{\infty}(-q^{\frac12}t^2 s^{\pm};q)_{\infty}
(q^{\frac12}t^{-2} s^{\pm};q)_{\infty}(-q^{\frac12}t^{-2} s^{\pm};q)_{\infty}}
(s+s^{-1})^2. 
\end{align}
We find that it agrees with the two-point function of the S-dual 't Hooft lines with $B=(1)$ for $SO(4)^{-}$ of the form
\begin{align}
\label{so4-_t1}
&\langle T_{(1)} T_{(1)}\rangle^{SO(4)^-}(t;q)
=\frac{(q)_{\infty}^2(-q;q)_{\infty}^2}{(q^{\frac12}t^{\pm2};q)_{\infty}(-q^{\frac12}t^{\pm2};q)_{\infty}}
\oint \frac{ds}{2\pi is}
\nonumber\\
&\times 
\frac{(q^{\frac12}s^{\pm};q)_{\infty}(-q^{\frac12}s;q)_{\infty}(q^{\frac32}s^{\pm};q)_{\infty}(-q^{\frac32}s^{\pm};q)_{\infty}}
{(qt^2 s^{\pm};q)_{\infty}(-qt^2 s^{\pm};q)_{\infty}
(qt^{-2} s^{\pm};q)_{\infty}(-qt^{-2} s^{\pm};q)_{\infty}}. 
\end{align}
The exact form of the two-point functions (\ref{so4-_wfund}) and (\ref{so4-_t1}) can be obtained from the two-point function 
of the fundamental Wilson lines in $SU(2)$ gauge theory
\begin{align}
\label{so4mW1_T1}
\langle W_{\tiny \yng(1)} W_{\tiny \yng(1)}\rangle^{SO(4)^{-}}(t;q)
&=\langle T_{(1)} T_{(1)}\rangle^{SO(4)^-}(t;q)
\nonumber\\
&=\langle W_{\tiny \yng(1)} W_{\tiny \yng(1)}\rangle^{SU(2)}(t^2;q^2). 
\end{align}

The half-BPS limits of the two-point functions of the fundamental Wilson lines is 
\begin{align}
\langle W_{\tiny \yng(1)} W_{\tiny \yng(1)}\rangle^{SO(4)^{-}}_{\textrm{$\frac12$BPS}}(\mathfrak{q})
&=\langle T_{(1)} T_{(1)}\rangle^{SO(4)^-}_{\textrm{$\frac12$BPS}}(\mathfrak{q})
\nonumber\\
&=\frac{1}{1-\mathfrak{q}^4}. 
\end{align}

Furthermore, there exist non-trivial even-point correlation functions of the fundamental Wilson lines. 
In the half-BPS limit they are given by
\begin{align}
\langle \underbrace{W_{\tiny \yng(1)} \cdots W_{\tiny \yng(1)}}_{2k} \rangle^{SO(4)^{-}}_{\textrm{$\frac12$BPS}}(\mathfrak{q})
&=\frac{\sum_{i=0}^{k} a_{k\ \tiny \yng(1)}^{\mathfrak{so}(4)^-}(i) \mathfrak{q}^{4i}}{1-\mathfrak{q}^8}
\end{align}
where 
\begin{align}
a_{k\ \tiny \yng(1)}^{\mathfrak{so}(4)^-}(i)
&=(2i+1)\frac{(2k)!}{(k-i)!(k+i+1)!}, 
\end{align}
which agrees with (\ref{so4_sp_coeff}). 

Also, the correlators in gauge theories of disconnected gauge group $O(4)$ can be obtained by gauging the $\mathbb{Z}_2$ automorphism group
\begin{align}
\langle W_{\lambda_1} \cdots W_{\lambda_k}\rangle^{O(4)^{+}}(t;q)
&=\frac12 \left[
\langle W_{\lambda_1} \cdots W_{\lambda_k}\rangle^{SO(4)}(t;q)
+ \langle W_{\lambda_1} \cdots W_{\lambda_k}\rangle^{SO(4)^{-}}(t;q)
\right]. 
\end{align}
Similarly, we can obtain the 't Hooft line correlators by gauging the $\mathbb{Z}_2$ symmetry. 
It then follows from the equalities (\ref{so4W1_T1}) and (\ref{so4mW1_T1}) that 
\begin{align}
\langle W_{\tiny \yng(1)} W_{\tiny \yng(1)}\rangle^{O(4)^{+}}(t;q)
&=\langle T_{(1)} T_{(1)}\rangle^{O(4)^+}(t;q), 
\end{align}
which supports the S-duality between the fundamental Wilson line and the basic 't Hooft line in $O(4)$ gauge theory. 

The exact closed-form expressions for the half-BPS limits of the correlators can be found. 
They can be written as
\begin{align}
\langle \underbrace{W_{\tiny \yng(1)} \cdots W_{\tiny \yng(1)}}_{2k} \rangle^{O(4)^{+}}_{\textrm{$\frac12$BPS}}(\mathfrak{q})
&=\frac{\sum_{i=0}^{2k+1} a_{k\ \tiny \yng(1)}^{\mathfrak{o}(4)}(i) \mathfrak{q}^{2i}}{(1-\mathfrak{q}^4)(1-\mathfrak{q}^8)}
\end{align}
where $a_{k\ \tiny \yng(1)}^{\mathfrak{o}(4)}(i)$ are some non-negative integers. 
For example, we get
\begin{align}
\langle W_{\tiny \yng(1)} W_{\tiny \yng(1)}\rangle^{O(4)^{+}}_{\textrm{$\frac12$BPS}}(\mathfrak{q})
&=\frac{1+\mathfrak{q}^2+\mathfrak{q}^4+\mathfrak{q}^6}{(1-\mathfrak{q}^4)(1-\mathfrak{q}^8)}
\nonumber\\
&=\frac{1}{(1-\mathfrak{q}^2)(1-\mathfrak{q}^4)}, \\
\langle W_{\tiny \yng(1)} W_{\tiny \yng(1)}W_{\tiny \yng(1)} W_{\tiny \yng(1)}\rangle^{O(4)^{+}}_{\textrm{$\frac12$BPS}}(\mathfrak{q})
&=\frac{3+6\mathfrak{q}^2+9\mathfrak{q}^4+9\mathfrak{q}^6+6\mathfrak{q}^8+3\mathfrak{q}^{10}}
{(1-\mathfrak{q}^4)(1-\mathfrak{q}^8)}, \\
\langle W_{\tiny \yng(1)} W_{\tiny \yng(1)}W_{\tiny \yng(1)} W_{\tiny \yng(1)}W_{\tiny \yng(1)} W_{\tiny \yng(1)}\rangle^{O(4)^{+}}_{\textrm{$\frac12$BPS}}(\mathfrak{q})
&=\frac{1}{(1-\mathfrak{q}^4)(1-\mathfrak{q}^8)}(
15+45\mathfrak{q}^2+80\mathfrak{q}^4+95\mathfrak{q}^6+85\mathfrak{q}^8
\nonumber\\
&+55\mathfrak{q}^{10}+20\mathfrak{q}^{12}+5\mathfrak{q}^{14}
).
\end{align}
We observe that the coefficients $a_{k\ \tiny \yng(1)}^{\mathfrak{o}(4)}(0)$ are given by
\begin{align}
\label{o4_fund_coeff0}
a_{k\ \tiny \yng(1)}^{\mathfrak{o}(4)}(0)&=\frac12 (C_k^2+C_k), 
\end{align}
where $C_k$ is the $k$-th Catalan number. 
The sequence of the coefficients (\ref{o4_fund_coeff0}) counts 
the multiplicity of the trivial representation in the $2k$-th tensor power of the fundamental representation of $O(4)$.

\subsubsection{Adjoint Wilson line}
Next consider the Wilson lines in the higher rank representations. 
In this case odd-point correlation functions are also non-trivial. 
We use the following notation of the highest weight:
\begin{align}
{\scriptsize \yng(1,1)}=(1,1),\qquad \overline{{\scriptsize \yng(1,1)}}=(1,-1). 
\end{align}
It follows that 
\begin{align}
\label{so4wasymk_1pt_fact}
\langle \underbrace{W_{\tiny \yng(1,1)} \cdots W_{\tiny \yng(1,1)}}_{k} \rangle^{SO(4)}(t;q)
&=\langle \underbrace{W_{\overline{{\tiny \yng(1,1)}}} \cdots W_{\overline{{\tiny \yng(1,1)}}}}_{k} \rangle^{SO(4)}(t;q)\notag \\
&=\mathcal{I}^{SU(2)}(t;q) \langle \underbrace{W_{\tiny \yng(2)} \cdots W_{\tiny \yng(2)}}_{k} \rangle^{SU(2)}(t;q). 
\end{align}
This implies that 
the normalized correlation function of the $SO(4)$ adjoint Wilson lines is equivalent to that of the $SU(2)$ adjoint Wilson lines 
or equivalently to that of the $SO(3)$ fundamental Wilson lines. 

In the half-BPS limit, they can be written as
\begin{align}
\langle \underbrace{W_{\tiny \yng(1,1)} \cdots W_{\tiny \yng(1,1)}}_{k} \rangle^{SO(4)}_{\textrm{$\frac12$BPS}}(\mathfrak{q})
&=\frac{\sum_{i=0}^{k} a_{k\ \tiny \yng(1,1)}^{\mathfrak{so}(4)}(i)\mathfrak{q}^{2i} }{(1-\mathfrak{q}^4)^2}, 
\end{align}
with $a_{k\ \tiny \yng(1,1)}^{\mathfrak{so}(4)}(i)$ being non-negative integers. 
For example, we obtain 
\begin{align}
\langle W_{\tiny \yng(1,1)} \rangle^{SO(4)}_{\textrm{$\frac12$BPS}}(\mathfrak{q})
&=\frac{\mathfrak{q}^2}{(1-\mathfrak{q}^4)^2}, \\
\langle W_{\tiny \yng(1,1)} W_{\tiny \yng(1,1)}\rangle^{SO(4)}_{\textrm{$\frac12$BPS}}(\mathfrak{q})
&=\frac{1+\mathfrak{q}^2+\mathfrak{q}^4}{(1-\mathfrak{q}^4)^2}, \\
\langle W_{\tiny \yng(1,1)}W_{\tiny \yng(1,1)} W_{\tiny \yng(1,1)}\rangle^{SO(4)}_{\textrm{$\frac12$BPS}}(\mathfrak{q})
&=\frac{1+3\mathfrak{q}^2+2\mathfrak{q}^4+\mathfrak{q}^6}{(1-\mathfrak{q}^4)^2}, \\
\langle W_{\tiny \yng(1,1)}W_{\tiny \yng(1,1)}W_{\tiny \yng(1,1)} W_{\tiny \yng(1,1)}\rangle^{SO(4)}_{\textrm{$\frac12$BPS}}(\mathfrak{q})
&=\frac{3+6\mathfrak{q}^2+6\mathfrak{q}^4+3\mathfrak{q}^6+\mathfrak{q}^8}{(1-\mathfrak{q}^4)^2}, \\
\langle W_{\tiny \yng(1,1)}W_{\tiny \yng(1,1)}W_{\tiny \yng(1,1)}W_{\tiny \yng(1,1)} W_{\tiny \yng(1,1)}\rangle^{SO(4)}_{\textrm{$\frac12$BPS}}(\mathfrak{q})
&=\frac{6+15\mathfrak{q}^2+15\mathfrak{q}^4+10\mathfrak{q}^6+4\mathfrak{q}^8+\mathfrak{q}^{10}}{(1-\mathfrak{q}^4)^2}, \\
\langle W_{\tiny \yng(1,1)}W_{\tiny \yng(1,1)}W_{\tiny \yng(1,1)}W_{\tiny \yng(1,1)}W_{\tiny \yng(1,1)} W_{\tiny \yng(1,1)}\rangle^{SO(4)}_{\textrm{$\frac12$BPS}}(\mathfrak{q})
&=\frac{15+36\mathfrak{q}^2+40\mathfrak{q}^4+29\mathfrak{q}^6+15\mathfrak{q}^8+5\mathfrak{q}^{10}+\mathfrak{q}^{12}}{(1-\mathfrak{q}^4)^2}. 
\end{align}
As expected, the sequence of the coefficients $a_{k\ \tiny \yng(1,1)}^{\mathfrak{so}(4)}(i)$ precisely coincides with $a_{k\ \tiny \yng(1)}^{\mathfrak{so}(3)}(i)$ 
listed in (\ref{so3fund_coeff_table}).  
We also observe the factorization property
\begin{align}
\langle (W_{\tiny \yng(1,1)})^k (W_{\overline{{\tiny \yng(1,1)}}})^m \rangle^{SO(4)}_{\textrm{$\frac12$BPS}}(\mathfrak{q})
&=(1-\mathfrak{q}^4)^2
\langle (W_{\tiny \yng(1,1)})^k  \rangle^{SO(4)}_{\textrm{$\frac12$BPS}}(\mathfrak{q})
\langle  (W_{\overline{{\tiny \yng(1,1)}}})^m \rangle^{SO(4)}_{\textrm{$\frac12$BPS}}(\mathfrak{q}).
\end{align}

\subsubsection{Symmetric Wilson line}
For the Wilson line in the rank-$l$ symmetric representation, 
the correlation function is factorized as
\begin{align}
\label{so4wsymk_2pt_fact}
\langle \underbrace{W_{(l)} \cdots W_{(l)}}_{k} \rangle^{SO(4)}(t;q)
&=\langle \underbrace{W_{(l)} \cdots W_{(l)}}_{k} \rangle^{SU(2)}(t;q)^2, 
\end{align}
where $\underbrace{\langle W_{(l)} \cdots W_{(l)}}_{k}\rangle^{SU(2)}(t;q)$ is the $k$-point function of the rank-$l$ symmetric Wilson line in $SU(2)$ gauge theory. 
When both $l$ and $k$ are odd, the correlator (\ref{so4wsymk_2pt_fact}) vanishes. 

For rank-$2$ symmetric Wilson lines, the correlation functions are identified with the squares of the correlation functions of the $SO(3)$ fundamental Wilson lines. 
In the half-BPS limit, they take the form
\begin{align}
\langle \underbrace{W_{\tiny \yng(2)} \cdots W_{\tiny \yng(2)}}_{k} \rangle^{SO(4)}_{\textrm{$\frac12$BPS}}(\mathfrak{q})
&=\frac{\sum_{i=0}^{2k} a_{k\ \tiny\yng(2)}^{\mathfrak{so}(4)^-}(i) \mathfrak{q}^{2i}}{(1-\mathfrak{q}^4)^2}, 
\end{align}
with non-negative integers $a_{k\ \tiny\yng(2)}^{\mathfrak{so}(4)^-}(i)$. 
For example, 
\begin{align}
\langle W_{\tiny \yng(2)} \rangle^{SO(4)}_{\textrm{$\frac12$BPS}}(\mathfrak{q})
&=\frac{\mathfrak{q}^4}{(1-\mathfrak{q}^4)^2}, \\
\langle W_{\tiny \yng(2)} W_{\tiny \yng(2)} \rangle^{SO(4)}_{\textrm{$\frac12$BPS}}(\mathfrak{q})
&=\frac{(1+\mathfrak{q}^2+\mathfrak{q}^4)^2}{(1-\mathfrak{q}^4)^2}, \\
\langle W_{\tiny \yng(2)} W_{\tiny \yng(2)} W_{\tiny \yng(2)} \rangle^{SO(4)}_{\textrm{$\frac12$BPS}}(\mathfrak{q})
&=\frac{(1+3\mathfrak{q}^2+2\mathfrak{q}^4+\mathfrak{q}^6)^2}{(1-\mathfrak{q}^4)^2}. 
\end{align}
The first two coefficients are given by
\begin{align}
a_{k\ \tiny\yng(2)}^{\mathfrak{so}(4)^-}(0)&=R_{k}^2, \\
a_{k\ \tiny\yng(2)}^{\mathfrak{so}(4)^-}(1)&=2 R_{k}R_{k+1}, 
\end{align}
where $R_k$ is the $k$-th Riordan number. Since there is an exact relation $\chi_{\tiny\yng(2)}^{\mathfrak{so}(4)}=\chi_{\tiny\yng(1,1)}^{\mathfrak{so}(4)}\chi_{\overline{\tiny\yng(1,1)}}^{\mathfrak{so}(4)}$, we have
\begin{align}
\langle (W_{\tiny\yng(2)})^k \rangle^{SO(4)}
&=\langle (W_{\tiny\yng(1,1)})^k (W_{\overline{\tiny\yng(1,1)}})^k \rangle^{SO(4)}.
\end{align}

In the half-BPS limits they can be written as
\begin{align}
\label{halfBPS_so4_wsym2k}
\langle W_{(2k)} \rangle^{SO(4)}_{\textrm{$\frac12$BPS}}(\mathfrak{q})
&=\frac{\mathfrak{q}^{4k}}{(1-\mathfrak{q}^4)^2}. 
\end{align}

The half-BPS limit of the two-point function is given by
\begin{align}
\langle W_{(k)} W_{(k)}\rangle^{SO(4)}_{\textrm{$\frac12$BPS}}(\mathfrak{q})
&=\frac{(1-\mathfrak{q}^{2k+2})^2}{(1-\mathfrak{q}^2)^2(1-\mathfrak{q}^4)^2}. 
\end{align}
In the large representation limit, we obtain 
\begin{align}
\langle W_{(\infty)} W_{(\infty)}\rangle^{SO(4)}_{\textrm{$\frac12$BPS}}(\mathfrak{q})
&=\frac{1}{(1-\mathfrak{q}^2)^2(1-\mathfrak{q}^4)^2}. 
\end{align}

Let us consider the disconnected component for $SO(4)$ gauge theory. 
Again the correlators of the rank-$l$ symmetric Wilson lines are related to those for $SU(2)$ gauge theory
\begin{align}
\langle \underbrace{W_{(l)}\cdots W_{(l)}}_{k} \rangle^{SO(4)^{-}}(t;q)
&=\langle \underbrace{W_{(l)}\cdots W_{(l)}}_{k} \rangle^{SU(2)}(t^2;q^2). 
\end{align}
Hence the exact form can be obtained from those for the $SU(2)$ correlators by rescaling the fugacities. 
For the Wilson lines in the even rank symmetric representation, there exist non-vanishing one-point functions. 
For example, in the half-BPS limit, the one- and two-point functions are evaluated as
\begin{align}
\langle W_{(2l)} \rangle^{SO(4)^{-}}_{\textrm{$\frac12$BPS}}(\mathfrak{q})
&=\frac{\mathfrak{q}^{4l}}{1-\mathfrak{q}^8}, \\
\langle W_{(l)} W_{(l)}\rangle^{SO(4)^{-}}_{\textrm{$\frac12$BPS}}(\mathfrak{q})
&=\frac{1-\mathfrak{q}^{4l+4}}{(1-\mathfrak{q}^4)(1-\mathfrak{q}^8)}. 
\end{align}

Furthermore, gauging the $\mathbb{Z}_2$ automorphism group in the $SO(4)$ gauge theory, one finds the correlators in $O(4)$ gauge theory. 
For example, the one- and two-point function in the half-BPS limits are 
\begin{align}
\langle W_{(2l)}\rangle^{O(4)^{+}}_{\textrm{$\frac12$BPS}}(\mathfrak{q})
&=\frac{\mathfrak{q}^{4l}}{(1-\mathfrak{q}^4)(1-\mathfrak{q}^8)}, \\
\langle W_{(l)} W_{(l)}\rangle^{O(4)^{+}}_{\textrm{$\frac12$BPS}}(\mathfrak{q})
&=\frac{1-\mathfrak{q}^2+\mathfrak{q}^4-\mathfrak{q}^{2l+2}-\mathfrak{q}^{2l+6}+\mathfrak{q}^{4l+6}}
{(1-\mathfrak{q}^2)^2(1-\mathfrak{q}^4)(1-\mathfrak{q}^8)}. 
\end{align}
In the large representation limit, while the one-point function vanishes, 
the two-point function become 
\begin{align}
\langle W_{(\infty)} W_{(\infty)}\rangle^{O(4)^{+}}_{\textrm{$\frac12$BPS}}(\mathfrak{q})
&=\frac{1-\mathfrak{q}^2+\mathfrak{q}^4}
{(1-\mathfrak{q}^2)^2(1-\mathfrak{q}^4)(1-\mathfrak{q}^8)}. 
\end{align}

\subsubsection{Rectangular Wilson line}
We can also evaluate the correlators of the Wilson lines associated with the rectangular Young diagram from the $SU(2)$ correlators. 
We have
\begin{align}
\langle \underbrace{W_{(l,l)}\cdots W_{(l,l)}}_{k} \rangle^{SO(4)}(t;q)
&=\langle \underbrace{W_{(l,-l)}\cdots W_{(l,-l)}}_{k} \rangle^{SO(4)}(t;q)\notag \\
&=\mathcal{I}^{SU(2)}(t;q) \langle \underbrace{W_{(2l)}\cdots W_{(2l)}}_{k} \rangle^{SU(2)}(t;q). 
\end{align}
For example, in the half-BPS limit, we find
\begin{align}
\langle W_{(l,l)} \rangle^{SO(4)}_{\textrm{$\frac12$BPS}}(\mathfrak{q})
&=\langle W_{(l,-l)} \rangle^{SO(4)}_{\textrm{$\frac12$BPS}}(\mathfrak{q})
=\frac{\mathfrak{q}^{2l}}{(1-\mathfrak{q}^4)^2}, \\
\langle W_{(l, l)} W_{(l, l)}\rangle^{SO(4)}_{\textrm{$\frac12$BPS}}(\mathfrak{q})
&=\langle W_{(l, -l)} W_{(l, -l)}\rangle^{SO(4)}_{\textrm{$\frac12$BPS}}(\mathfrak{q})
=\frac{1-\mathfrak{q}^{4l+2}}{(1-\mathfrak{q}^2)(1-\mathfrak{q}^{4})^2}. 
\end{align}
On the other hand, we find $\chi_{(l,l)}^{\mathfrak{so}(4)}\chi_{(l,-l)}^{\mathfrak{so}(4)}=\chi_{(2l)}^{\mathfrak{so}(4)}$. Therefore we have
\begin{align}
\langle W_{(l,l)}^k W_{(l,-l)}^k \rangle^{SO(4)}
&=\langle W_{(2l)}^k \rangle^{SO(4)},
\end{align}
where $W_{(l,l)}^k$ stands for $k$ insertions of the Wilson line $W_{(l,l)}$.

\subsection{$\mathfrak{so}(6)$}
The closed-form expression for the $SO(6)$ index can be obtained by means of the formula in \cite{Hatsuda:2022xdv} as it is equal to the $SU(4)$ index. 
We have
\begin{align}
&\mathcal{I}^{SO(6)}(t;q)=\mathcal{I}^{SU(4)}(t;q)
\nonumber\\
&=-\frac{(q^{\frac12}t^{\pm2};q)_{\infty}}{(q;q)_{\infty}^2}
\sum_{\begin{smallmatrix}
p_1, p_2, p_3, p_4\in \mathbb{Z}\\
p_1<p_2<p_3<p_4\\
\end{smallmatrix}}
\frac{(q^{\frac12}t^{-2})^{p_1+p_2+p_3+p_4-8}}
{(1-q^{p_1-1}t^4) (1-q^{p_2-1}t^4)(1-q^{p_3-1}t^4) (1-q^{p_4-1}t^4)}. 
\end{align}

\subsubsection{Spinor Wilson line}
The $Spin(6)$ gauge theory has the Wilson line in the (anti)chiral spinor representation. 
In this case, the two spinor representations are complex conjugates. 
The non-trivial two-point function is obtained for such a pair of conjugate spinor Wilson lines. 
It can be computed from the matrix integral 
\begin{align}
\label{so6_wsp}
&
\langle W_{\textrm{sp}} W_{\overline{\textrm{sp}}}\rangle^{Spin(6)}(t;q)
=\frac{1}{24}
\frac{(q)_{\infty}^{6}}{(q^{\frac12}t^{\pm2};q)_{\infty}^{3}}
\oint \prod_{i=1}^{3}\frac{ds_i}{2\pi is_i}
\nonumber\\
&\times 
\prod_{i\neq j}
\frac{(s_i^{\pm}s_j^{\mp};q)_{\infty} (s_i^{\pm}s_j^{\pm};q)_{\infty}(qs_i^{\pm}s_j^{\mp};q)_{\infty}(qs_i^{\pm}s_j^{\pm};q)_{\infty}}
{(q^{\frac12}t^2 s_i^{\pm}s_j^{\mp};q)_{\infty} (q^{\frac12}t^2s_i^{\pm}s_j^{\pm};q)_{\infty}
(q^{\frac12}t^{-2}s_i^{\pm}s_j^{\mp};q)_{\infty}(q^{\frac12}t^{-2}s_i^{\pm}s_j^{\pm};q)_{\infty}}
\nonumber\\
&\times (1+s_1 s_2+s_1 s_3+s_2 s_3)(1+s_1^{-1} s_2^{-1}+s_1^{-1} s_3^{-1}+s_2^{-1} s_3^{-1}). 
\end{align}
The Wilson line in the spinor representation for $Spin(6)$ gauge theory is 
S-dual to the 't Hooft lines with $B=(\frac12,\frac12,\frac12)$ in $(SO(6)/\mathbb{Z}_2)_{0}$ gauge theory. 
The two-point function of the S-dual 't Hooft lines can be evaluated as 
\begin{align}
\label{so6_tsp}
&
\langle T_{(\frac12,\frac12,\frac12)} T_{(\frac12,\frac12,\frac12)}\rangle^{SO(6)/\mathbb{Z}_2}(t;q)
=\frac16 \frac{(q)_{\infty}^6}{(q^{\frac12}t^{\pm2};q)_{\infty}^3}
\oint \prod_{i=1}^3 \frac{ds_i}{2\pi is_i}
\nonumber\\
&\times 
\prod_{i<j}
\frac{(s_i^{\pm}s_j^{\mp};q)_{\infty}(q^{\frac12}s_i^{\pm}s_j^{\pm};q)_{\infty}
(qs_i^{\pm}s_j^{\mp};q)_{\infty}(q^{\frac32}s_i^{\pm}s_j^{\pm};q)_{\infty}}
{(q^{\frac12}t^2s_i^{\pm}s_j^{\mp};q)_{\infty}(qt^2s_i^{\pm}s_j^{\pm};q)_{\infty}
(q^{\frac12}t^{-2}s_i^{\pm}s_j^{\mp};q)_{\infty}(qt^{-2}s_i^{\pm}s_j^{\pm};q)_{\infty}}. 
\end{align}
The expressions (\ref{so6_wsp}) and (\ref{so6_tsp}) agree with each other. 

According to the isomorphism $Spin(6)$ $\cong$ $SU(4)$, 
the exact closed-form expressions can be addressed from the results for $SU(4)$ gauge theory. 
One finds that the two-point functions (\ref{so6_wsp}) and (\ref{so6_tsp}) are equal to the two-point function of the (anti)fundamental Wilson lines for $SU(4)$ SYM theory 
\begin{align}
\langle W_{\textrm{sp}} W_{\overline{\textrm{sp}}}\rangle^{Spin(6)}(t;q)
&=\langle T_{(\frac12,\frac12,\frac12)} T_{(\frac12,\frac12,\frac12)}\rangle^{SO(6)/\mathbb{Z}_2}(t;q)
\nonumber\\
&=\langle W_{\tiny \yng(1)} W_{\overline{\tiny \yng(1)}}\rangle^{SU(4)}(t;q). 
\end{align}
It then follows from the exact results \cite{Hatsuda:2023iwi} for the $SU(4)$ SYM theory 
that the half-BPS limits of the two-point functions (\ref{so6_wsp}) and (\ref{so6_tsp}) are given by
\begin{align}
\langle W_{\textrm{sp}} W_{\overline{\textrm{sp}}}\rangle^{Spin(6)}_{\textrm{$\frac12$BPS}}(\mathfrak{q})
&=\langle T_{(\frac12,\frac12,\frac12)} T_{(\frac12,\frac12,\frac12)}\rangle^{SO(6)/\mathbb{Z}_2}_{\textrm{$\frac12$BPS}}(\mathfrak{q})
\nonumber\\
&=\frac{1}{(1-\mathfrak{q}^2)(1-\mathfrak{q}^{4}) (1-\mathfrak{q}^6)}. 
\end{align}

There exist non-trivial multi-point correlation functions of pairs of the spinor Wilson line and its conjugate. 
They can be also computed from the correlators of the $SU(4)$ (anti)fundamental Wilson lines. 
We have 
\begin{align}
\underbrace{\langle W_{\textrm{sp}} W_{\overline{\textrm{sp}}} \cdots W_{\textrm{sp}} W_{\overline{\textrm{sp}}} }_{2k} \rangle^{Spin(6)}(t;q)
&=\langle \underbrace{W_{\tiny \yng(1)} W_{\overline{\tiny \yng(1)}}\cdots W_{\tiny \yng(1)} W_{\overline{\tiny \yng(1)}}}_{2k} \rangle^{SU(4)}(t;q). 
\end{align}
In the half-BPS limit the correlators take the form
\begin{align}
\langle \underbrace{W_{\textrm{sp}} W_{\overline{\textrm{sp}}} \cdots W_{\textrm{sp}} W_{\overline{\textrm{sp}}}}_{2k} \rangle^{Spin(6)}_{\textrm{$\frac12$BPS}}(\mathfrak{q})
&=\frac{\sum_{i=0}^{3k} a_{k\ \textrm{sp}}^{\mathfrak{so}(6)}(i) \mathfrak{q}^{2i}}{(1-\mathfrak{q}^4)(1-\mathfrak{q}^6)(1-\mathfrak{q}^8)}, 
\end{align}
with $a_{k\ \textrm{sp}}^{\mathfrak{so}(6)}(i)$ being non-negative integers. 
For example, one finds
\begin{align}
\langle W_{\textrm{sp}} W_{\overline{\textrm{sp}}} W_{\textrm{sp}} W_{\overline{\textrm{sp}}} \rangle^{Spin(6)}_{\textrm{$\frac12$BPS}}(\mathfrak{q})
&=\frac{2+4\mathfrak{q}^2+6\mathfrak{q}^4+7\mathfrak{q}^6+5\mathfrak{q}^8+3\mathfrak{q}^{10}+\mathfrak{q}^{12}}
{(1-\mathfrak{q}^4)(1-\mathfrak{q}^{6}) (1-\mathfrak{q}^8)}, \\
\langle W_{\textrm{sp}} W_{\overline{\textrm{sp}}} W_{\textrm{sp}} W_{\overline{\textrm{sp}}} W_{\textrm{sp}} W_{\overline{\textrm{sp}}} \rangle^{Spin(6)}_{\textrm{$\frac12$BPS}}(\mathfrak{q})
&=\frac{1}{(1-\mathfrak{q}^4)(1-\mathfrak{q}^{6}) (1-\mathfrak{q}^8)}
(6+18\mathfrak{q}^2+35\mathfrak{q}^4+50\mathfrak{q}^6
\nonumber\\
&+53\mathfrak{q}^8+45\mathfrak{q}^{10}
+29\mathfrak{q}^{12}+14\mathfrak{q}^{14}+5\mathfrak{q}^{16}+\mathfrak{q}^{18}
). 
\end{align}
The sequence of the first coefficients $a_{k\ \textrm{sp}}^{\mathfrak{so}(6)}(0)$ agrees with 
the multiplicity of the trivial representation in the $k$-th tensor power of a product of the spinor representation and its conjugate of $Spin(6)$, 
or equivalently that in the $k$-th tensor power of a product of the fundamental and antifundamental representation of $SU(4)$ \cite{MR1235279}. 
It is found from the generating function
\begin{align}
\det \left(
\begin{matrix}
I_{0}(2x)&I_{1}(2x)&I_{2}(2x)&I_{3}(2x)\\
I_{1}(2x)&I_{0}(2x)&I_{1}(2x)&I_{2}(2x)\\
I_{2}(2x)&I_{1}(2x)&I_{0}(2x)&I_{1}(2x)\\
I_{3}(2x)&I_{2}(2x)&I_{1}(2x)&I_{0}(2x)\\
\end{matrix}
\right)&=\sum_{k=0}^{\infty}\frac{a_{k\ \textrm{sp}}^{\mathfrak{so}(6)}(0)}{(k!)^2}x^{2k}, 
\end{align}
where $I_k(2x)$ is the modified Bessel function (\ref{modBessel1st}) of the first kind of order $k$.

\subsubsection{Fundamental Wilson line}
The Wilson line transforming in the fundamental representation is allowed in $SO(6)$ gauge theories. 
While the one-point function of the fundamental Wilson line vanishes, 
we get the non-trivial two-point function of the fundamental Wilson lines for $SO(6)$ SYM theory. 
It is evaluated from 
\begin{align}
\label{so6_wfund}
&
\langle W_{\tiny \yng(1)} W_{\tiny \yng(1)}\rangle^{SO(6)}(t;q)
=\frac{1}{24}
\frac{(q)_{\infty}^{6}}{(q^{\frac12}t^{\pm2};q)_{\infty}^{3}}
\oint \prod_{i=1}^{3}\frac{ds_i}{2\pi is_i}
\nonumber\\
&\times 
\prod_{i\neq j}
\frac{(s_i^{\pm}s_j^{\mp};q)_{\infty} (s_i^{\pm}s_j^{\pm};q)_{\infty}(qs_i^{\pm}s_j^{\mp};q)_{\infty}(qs_i^{\pm}s_j^{\pm};q)_{\infty}}
{(q^{\frac12}t^2 s_i^{\pm}s_j^{\mp};q)_{\infty} (q^{\frac12}t^2s_i^{\pm}s_j^{\pm};q)_{\infty}
(q^{\frac12}t^{-2}s_i^{\pm}s_j^{\mp};q)_{\infty}(q^{\frac12}t^{-2}s_i^{\pm}s_j^{\pm};q)_{\infty}}
\nonumber\\
&\times (s_1+s_2+s_3+s_1^{-1}+s_2^{-1}+s_3^{-1})^2. 
\end{align}
The fundamental Wilson line is S-dual to the 't Hooft lines of $B=(1,0,0)$. 
The two-point function of the dual 't Hooft lines can  be evaluated as
\begin{align}
\label{so6_t1}
&
\langle T_{(1,0,0)} T_{(1,0,0)}\rangle^{SO(6)}(t;q)
=\frac{1}{4}\frac{(q)_{\infty}^6}{(q^{\frac12}t^{\pm2};q)_{\infty}^3}
\oint \prod_{i=1}^3 \frac{ds_i}{2\pi is_i}
\nonumber\\
&\times 
\prod_{i<j}
\frac{
(q^{\frac12\delta_{i+j,1}}s_i^{\pm}s_j^{\mp};q)_{\infty}
(q^{\frac12\delta_{i+j,1}}s_i^{\pm}s_j^{\pm};q)_{\infty}
}
{
(q^{\frac12(1+\delta_{i+j,1})}t^2s_i^{\pm}s_j^{\mp};q)_{\infty}
(q^{\frac12(1+\delta_{i+j,1})}t^2s_i^{\pm}s_j^{\pm};q)_{\infty}
}
\nonumber\\
&\times 
\frac{
(q^{1+\frac12\delta_{i+j,1}}s_i^{\pm}s_j^{\mp};q)_{\infty}
(q^{1+\frac12\delta_{i+j,1}}s_i^{\pm}s_j^{\pm};q)_{\infty}
}
{
(q^{\frac12(1+\delta_{i+j,1})}t^{-2}s_i^{\pm}s_j^{\mp};q)_{\infty}
(q^{\frac12(1+\delta_{i+j,1})}t^{-2}s_i^{\pm}s_j^{\pm};q)_{\infty}
}. 
\end{align}
The expressions (\ref{so6_wfund}) and (\ref{so6_t1}) coincide with one another. 
The closed-form expressions for the two-point functions 
(\ref{so6_wfund}) and (\ref{so6_t1}) can be obtained from the exact results for $SU(4)$ gauge theory. 
We find that 
\begin{align}
\langle W_{\tiny \yng(1)} W_{\tiny \yng(1)}\rangle^{SO(6)}(t;q)
&=\langle T_{(1,0,0)} T_{(1,0,0)}\rangle^{SO(6)}(t;q)
\nonumber\\
&=\langle W_{\tiny \yng(1,1)} W_{\tiny \yng(1,1)}\rangle^{SU(4)}(t;q). 
\end{align}
In the half-BPS limit, the two-point functions become
\begin{align}
\langle W_{\tiny \yng(1)} W_{\tiny \yng(1)}\rangle^{SO(6)}_{\textrm{$\frac12$BPS}}(\mathfrak{q})
&=\langle T_{(1,0,0)} T_{(1,0,0)}\rangle^{SO(6)}_{\textrm{$\frac12$BPS}}(\mathfrak{q})
\nonumber\\
&=\frac{1+\mathfrak{q}^2+2\mathfrak{q}^4+\mathfrak{q}^6+\mathfrak{q}^8}
{(1-\mathfrak{q}^4)(1-\mathfrak{q}^6)(1-\mathfrak{q}^8)}
\nonumber\\
&=\frac{1}{(1-\mathfrak{q}^2)(1-\mathfrak{q}^4)^2}. 
\end{align}

More generally, the even-point correlation functions of the fundamental Wilson lines can be computed from the those of the $SU(4)$ antisymmetric Wilson lines
\begin{align}
\langle \underbrace{W_{\tiny \yng(1)} \cdots W_{\tiny \yng(1)}}_{2k} \rangle^{SO(6)}(t;q)
&=\langle \underbrace{W_{\tiny \yng(1,1)} \cdots W_{\tiny \yng(1,1)}}_{2k}\rangle^{SU(4)}(t;q). 
\end{align}
In the half-BPS limit they can be written as
\begin{align}
\langle \underbrace{W_{\tiny \yng(1)}\cdots W_{\tiny \yng(1)}}_{2k} \rangle^{SO(6)}_{\textrm{$\frac12$BPS}}(\mathfrak{q})
&=\frac{\sum_{i=0}^{4k} a_{k\ \tiny \yng(1)}^{\mathfrak{so}(6)}(i) \mathfrak{q}^{2i}}
{(1-\mathfrak{q}^4)(1-\mathfrak{q}^6)(1-\mathfrak{q}^8)}, 
\end{align}
with $a_{k\ \tiny \yng(1)}^{\mathfrak{so}(6)}(i)$ being positive definite integers. 
For example, the four- and six-point functions are given by
\begin{align}
\langle W_{\tiny \yng(1)} W_{\tiny \yng(1)}W_{\tiny \yng(1)} W_{\tiny \yng(1)} \rangle^{SO(6)}_{\textrm{$\frac12$BPS}}(\mathfrak{q})
&=\frac{1}{(1-\mathfrak{q}^4)(1-\mathfrak{q}^6)(1-\mathfrak{q}^8)}
(3+7\mathfrak{q}^2+15\mathfrak{q}^{4}+18\mathfrak{q}^6+20\mathfrak{q}^8
\nonumber\\
&+14\mathfrak{q}^{10}+9\mathfrak{q}^{12}+3\mathfrak{q}^{14}+\mathfrak{q}^{16}), \\
\langle W_{\tiny \yng(1)} W_{\tiny \yng(1)}W_{\tiny \yng(1)} W_{\tiny \yng(1)}W_{\tiny \yng(1)} W_{\tiny \yng(1)} \rangle^{SO(6)}_{\textrm{$\frac12$BPS}}(\mathfrak{q})
&=\frac{1}{(1-\mathfrak{q}^4)(1-\mathfrak{q}^6)(1-\mathfrak{q}^8)}
(16+60\mathfrak{q}^2+149\mathfrak{q}^{4}+249\mathfrak{q}^6
\nonumber\\
&+334\mathfrak{q}^{8}+347\mathfrak{q}^{10}+301\mathfrak{q}^{12}+206\mathfrak{q}^{14}+119\mathfrak{q}^{16}
\nonumber\\
&+53\mathfrak{q}^{18}+20\mathfrak{q}^{20}+5\mathfrak{q}^{22}+\mathfrak{q}^{24}). 
\end{align}
We see that the sequence of the first coefficients $a_{k\ \tiny \yng(1)}^{\mathfrak{so}(6)}(0)$ is 
the multiplicity of the trivial representation in the $2k$-th power of the fundamental representation of $SO(6)$ 
or equivalently that of the rank-$2$ antisymmetric representation of $SU(4)$. 

For $SO(6)^{-}$ gauge theory, the disconnected part of $SO(6)$ gauge theory, 
the two-point function of the fundamental Wilson lines can be evaluated as
\begin{align}
\label{so6-_wfund}
&
\langle W_{\tiny \yng(1)} W_{\tiny \yng(1)}\rangle^{SO(6)^{-}}(t;q)
=\frac{1}{8}
\frac{(q)_{\infty}^{4}(-q;q)_{\infty}^{2}}
{(q^{\frac12}t^{\pm2};q)_{\infty}^{2}(-q^{\frac12}t^{\pm2};q)_{\infty}}
\nonumber\\
&\times 
\oint \prod_{i=1}^{2}\frac{ds_i}{2\pi is_i}
\frac{(s_i^{\pm};q)_{\infty}(-s_i^{\pm};q)_{\infty}(qs_i^{\pm};q)_{\infty}(-qs_i^{\pm};q)_{\infty}}
{(q^{\frac12}t^2s_i^{\pm};q)_{\infty}(-q^{\frac12}t^2s_i^{\pm};q)_{\infty}
(q^{\frac12}t^{-2}s_i^{\pm};q)_{\infty}(-q^{\frac12}t^{-2}s_i^{\pm};q)_{\infty}}
\nonumber\\
&\times 
\frac{(s_1^{\pm}s_2^{\mp};q)_{\infty}(s_1^{\pm}s_2^{\pm};q)_{\infty}
(qs_1^{\pm}s_2^{\mp};q)_{\infty}(qs_1^{\pm}s_2^{\pm};q)_{\infty}}
{(q^{\frac12}t^2s_1^{\pm}s_2^{\mp};q)_{\infty}(q^{\frac12}t^2s_1^{\pm}s_2^{\pm};q)_{\infty}
(q^{\frac12}t^{-2}s_1^{\pm}s_2^{\mp};q)_{\infty}(q^{\frac12}t^{-2}s_1^{\pm}s_2^{\pm};q)_{\infty}
}
\nonumber\\
&\times (s_1+s_2+s_1^{-1}+s_2^{-1})^2. 
\end{align}
We find that the two-point function (\ref{so6-_wfund}) agrees with
\begin{align}
\label{so6-_t1}
&
\langle T_{(1,0)} T_{(1,0)}\rangle^{SO(6)^{-}}(t;q)
=\frac12 \frac{(q)_{\infty}^4 (-q;q)_{\infty}^2}
{(q^{\frac12}t^{\pm2};q)_{\infty}^2 (-q^{\frac12}t^{\pm};q)_{\infty}}
\oint \prod_{i=1}^2 \frac{ds_i}{2\pi is_i}
\nonumber\\
&\times 
\frac{
(q^{\frac12}s_1^{\pm};q)_{\infty}(s_2^{\pm};q)_{\infty}
(-q^{\frac12}s_1^{\pm};q)_{\infty}(-s_2^{\pm};q)_{\infty}
}
{
(qt^{2}s_1^{\pm};q)_{\infty}(q^{\frac12}t^2s_2^{\pm};q)_{\infty}
(-qt^2s_1^{\pm};q)_{\infty}(-q^{\frac12}t^2s_2^{\pm};q)_{\infty}
}
\nonumber\\
&\times 
\frac{
(q^{\frac32}s_1^{\pm};q)_{\infty}(qs_2^{\pm};q)_{\infty}
(-q^{\frac32}s_1^{\pm};q)_{\infty}(-qs_2^{\pm};q)_{\infty}
}{
(qt^{-2}s_1^{\pm};q)_{\infty}(q^{\frac12}t^{-2}s_2^{\pm};q)_{\infty}
(-qt^{-2}s_1^{\pm};q)_{\infty}(-q^{\frac12}t^{-2}s_2^{\pm};q)_{\infty}
}
\nonumber\\
&\times 
\frac{
(q^{\frac12}s_1^{\pm}s_2^{\mp};q)_{\infty}(q^{\frac12}s_1^{\pm}s_2^{\pm};q)_{\infty}
(q^{\frac32}s_1^{\pm}s_2^{\mp};q)_{\infty}(q^{\frac32}s_1^{\pm}s_2^{\pm};q)_{\infty}
}
{
(qt^2s_1^{\pm}s_2^{\mp};q)_{\infty}(qt^2s_1^{\pm}s_2^{\pm};q)_{\infty}
(qt^{-2}s_1^{\pm}s_2^{\mp};q)_{\infty}(qt^{-2}s_1^{\pm}s_2^{\pm};q)_{\infty}
}, 
\end{align}
which is the two-point function of the S-dual 't Hooft lines of $B=(1,0)$ for $SO(6)^-$ gauge theory. 
In the half-BPS limit we obtain the closed-form expression 
\begin{align}
\langle W_{\tiny \yng(1)} W_{\tiny \yng(1)}\rangle^{SO(6)^{-}}_{\textrm{$\frac12$BPS}}(\mathfrak{q})
&=\langle T_{(1,0)} T_{(1,0)}\rangle^{SO(6)^{-}}_{\textrm{$\frac12$BPS}}(\mathfrak{q})
\nonumber\\
&=\frac{1+\mathfrak{q}^2+\mathfrak{q}^6+\mathfrak{q}^8}
{(1+\mathfrak{q}^6)(1-\mathfrak{q}^4)(1-\mathfrak{q}^8)}
\nonumber\\
&=\frac{1}{(1-\mathfrak{q}^2)(1-\mathfrak{q}^8)}. 
\end{align}

There exist even-point correlation functions of the fundamental Wilson lines. 
In the half-BPS limit, they can be expressed as
\begin{align}
\langle \underbrace{W_{\tiny \yng(1)} \cdots W_{\tiny \yng(1)}}_{2k} \rangle^{SO(6)^{-}}_{\textrm{$\frac12$BPS}}(\mathfrak{q})
&=\frac{\sum_{i=0}^{4k} a_{k\ \tiny\yng(1)}^{\mathfrak{so}(6)^-}(i) \mathfrak{q}^{2i}}
{(1+\mathfrak{q}^6)(1-\mathfrak{q}^4)(1-\mathfrak{q}^8)}, 
\end{align}
where $a_{k\ \tiny\yng(1)}^{\mathfrak{so}(6)^-}(i)$ are some positive definite integers. 
For example, we find
\begin{align}
\langle W_{\tiny \yng(1)} W_{\tiny \yng(1)} W_{\tiny \yng(1)} W_{\tiny \yng(1)} \rangle^{SO(6)^{-}}_{\textrm{$\frac12$BPS}}(\mathfrak{q})
&=\frac{3+5\mathfrak{q}^2+3\mathfrak{q}^4+6\mathfrak{q}^6+8\mathfrak{q}^{8}
+4\mathfrak{q}^{10}+3\mathfrak{q}^{12}+3\mathfrak{q}^{14}+\mathfrak{q}^{16}}
{(1+\mathfrak{q}^6)(1-\mathfrak{q}^4)(1-\mathfrak{q}^8)}, \\
\langle W_{\tiny \yng(1)} W_{\tiny \yng(1)}W_{\tiny \yng(1)} W_{\tiny \yng(1)} W_{\tiny \yng(1)} W_{\tiny \yng(1)} \rangle^{SO(6)^{-}}_{\textrm{$\frac12$BPS}}(\mathfrak{q})
&=\frac{1}{(1+\mathfrak{q}^6)(1-\mathfrak{q}^4)(1-\mathfrak{q}^8)}
(14+30\mathfrak{q}^2+31\mathfrak{q}^4+49\mathfrak{q}^6
\nonumber\\
&+66\mathfrak{q}^8+55\mathfrak{q}^{10}+49\mathfrak{q}^{12}+46\mathfrak{q}^{14}+29\mathfrak{q}^{16}+15\mathfrak{q}^{18}
\nonumber\\
&+10\mathfrak{q}^{20}+5\mathfrak{q}^{22}+\mathfrak{q}^{24}). 
\end{align}
The first coefficients are given by
\begin{align}
a_{k\ \tiny\yng(1)}^{\mathfrak{so}(6)^-}(0)&=C_{k}C_{k+2}-C_{k+1}^2, 
\end{align}
which coincides with the coefficient (\ref{spin5_spmult1}) of the normalized correlators in the $Spin(5)$ spinor Wilson lines. 

The correlators for the $O(6)$ gauge theory can be computed by gauging the $\mathbb{Z}_2$ symmetry of the $SO(6)$ gauge theory
\begin{align}
\langle W_{\lambda_1}\cdots W_{\lambda_k}\rangle^{O(6)^{+}}(t;q)
&=\frac12
\left[
\langle W_{\lambda_1}\cdots W_{\lambda_k}\rangle^{SO(6)^{+}}(t;q)
+
\langle W_{\lambda_1}\cdots W_{\lambda_k}\rangle^{SO(6)^{-}}(t;q)
\right]. 
\end{align}
For example, the half-BPS limits of the two- and four-point functions are given by 
\begin{align}
\langle W_{\tiny \yng(1)} W_{\tiny \yng(1)} \rangle^{O(6)^{+}}_{\textrm{$\frac12$BPS}}(\mathfrak{q})
&=\frac{1+\mathfrak{q}^2+\mathfrak{q}^4+\mathfrak{q}^6+\mathfrak{q}^8+\mathfrak{q}^{10}}{(1-\mathfrak{q}^4)(1-\mathfrak{q}^8)(1-\mathfrak{q}^{12})}
\nonumber\\
&=\frac{1}{(1-\mathfrak{q}^2)(1-\mathfrak{q}^4)(1-\mathfrak{q}^8)}, \\
\langle W_{\tiny \yng(1)} W_{\tiny \yng(1)} W_{\tiny \yng(1)} W_{\tiny \yng(1)} \rangle^{O(6)^{+}}_{\textrm{$\frac12$BPS}}(\mathfrak{q})
&=\frac{1}{(1-\mathfrak{q}^2)(1-\mathfrak{q}^4)(1-\mathfrak{q}^8)}
(3+6\mathfrak{q}^2+9\mathfrak{q}^4+12\mathfrak{q}^6
\nonumber\\
&+15\mathfrak{q}^8+15\mathfrak{q}^{10}+12\mathfrak{q}^{12}+9\mathfrak{q}^{14}+6\mathfrak{q}^{16}+3\mathfrak{q}^{18}
). 
\end{align}

\subsubsection{Antisymmetric Wilson line}
Let us turn to the correlation functions of Wilson lines transforming in the higher rank representations for $SO(6)$ gauge theory. 
For the antisymmetric representations, we have
\begin{align}
\langle \underbrace{W_{\tiny \yng(1,1)} \cdots W_{\tiny \yng(1,1)}}_{k}\rangle^{SO(6)}(t;q)
&=\langle \underbrace{W_{\tiny \yng(2,1,1)} W_{\tiny \yng(2,1,1)}}_{k} \rangle^{SU(4)}(t;q), \\
\langle \underbrace{W_{\tiny \yng(1,1,1)} \cdots W_{\overline{\tiny \yng(1,1,1)}}}_{2k} \rangle^{SO(6)}(t;q)
&=\langle \underbrace{W_{\tiny \yng(2)} \cdots W_{\tiny \yng(2)}}_{2k}\rangle^{SU(4)}(t;q). 
\end{align}
Note that the odd-point correlation function vanishes for the rank-$3$ antisymmetric Wilson lines. 
For example, in the half-BPS limit, we obtain
\begin{align}
\langle W_{\tiny \yng(1,1)}\rangle^{SO(6)}_{\textrm{$\frac12$BPS}}(\mathfrak{q})
&=\frac{\mathfrak{q}^2}{(1-\mathfrak{q}^2)(1-\mathfrak{q}^4)(1-\mathfrak{q}^8)}, \\
\langle W_{\tiny \yng(1,1)} W_{\tiny \yng(1,1)}\rangle^{SO(6)}_{\textrm{$\frac12$BPS}}(\mathfrak{q})
&=\frac{1}{(1-\mathfrak{q}^4)(1-\mathfrak{q}^8)(1-\mathfrak{q}^{12})}
\nonumber\\
&\times 
(1+2\mathfrak{q}^2+4\mathfrak{q}^4+6\mathfrak{q}^6
+7\mathfrak{q}^8+7\mathfrak{q}^{10}
\nonumber\\
&+6\mathfrak{q}^{12}+5\mathfrak{q}^{14}+3\mathfrak{q}^{16}+\mathfrak{q}^{18}
), \\
\langle W_{\tiny \yng(1,1,1)} W_{\tiny \yng(1,1,1)}\rangle^{SO(6)}_{\textrm{$\frac12$BPS}}(\mathfrak{q})
&=\frac{1}{(1-\mathfrak{q}^4)(1-\mathfrak{q}^8)(1-\mathfrak{q}^{12})}
\nonumber\\
&\times 
(1+\mathfrak{q}^2+2\mathfrak{q}^4+3\mathfrak{q}^6+3\mathfrak{q}^8
+3\mathfrak{q}^{10}
\nonumber\\
&+3\mathfrak{q}^{12}+2\mathfrak{q}^{14}+\mathfrak{q}^{16}+\mathfrak{q}^{18}). 
\end{align}

For the disconnected component of $SO(6)$ gauge theory, we obtain 
\begin{align}
\langle W_{\tiny \yng(1,1)} \rangle^{SO(6)^{-}}_{\textrm{$\frac12$BPS}}(\mathfrak{q})
&=\frac{\mathfrak{q}^2}{(1+\mathfrak{q}^2)(1-\mathfrak{q}^4)(1-\mathfrak{q}^8)}, \\
\langle W_{\tiny \yng(1,1)}W_{\tiny \yng(1,1)} \rangle^{SO(6)^{-}}_{\textrm{$\frac12$BPS}}(\mathfrak{q})
&=\frac{1+\mathfrak{q}^2+2\mathfrak{q}^4+\mathfrak{q}^8}{(1+\mathfrak{q}^2)(1-\mathfrak{q}^4)(1-\mathfrak{q}^8)}. 
\end{align}

Furthermore, the correlators of $O(6)$ gauge theory can be computed by gauging the $\mathbb{Z}_2$ symmetry. 
For example, we get
\begin{align}
\langle W_{\tiny \yng(1,1)} \rangle^{O(6)^{+}}_{\textrm{$\frac12$BPS}}(\mathfrak{q})
&=\frac{\mathfrak{q}^2}{(1-\mathfrak{q}^4)^2(1-\mathfrak{q}^8)}, \\
\langle W_{\tiny \yng(1,1)} W_{\tiny \yng(1,1)} \rangle^{O(6)^{+}}_{\textrm{$\frac12$BPS}}(\mathfrak{q})
&=\frac{1+\mathfrak{q}^2+2\mathfrak{q}^4+\mathfrak{q}^6+2\mathfrak{q}^8}
{(1-\mathfrak{q}^4)^2(1-\mathfrak{q}^8)}. 
\end{align}

\subsubsection{Symmetric Wilson line}
For the symmetric representations, one finds that 
\begin{align}
\langle \underbrace{W_{(l)} \cdots W_{(l)}}_{2k}\rangle^{SO(6)}(t;q)
&=\langle \underbrace{W_{(l^2)} \cdots W_{\overline{(l^2)}}}_{k}\rangle^{SU(4)}(t;q). 
\end{align}
We can also compute the exact forms for the half-BPS limits of the two-point functions. 
For example, we get
\begin{align}
\langle W_{\tiny \yng(2)}\rangle^{SO(6)}_{\textrm{$\frac12$BPS}}(\mathfrak{q})
&=\frac{\mathfrak{q}^4+\mathfrak{q}^8}
{(1+\mathfrak{q}^4)(1-\mathfrak{q}^6)(1-\mathfrak{q}^8)}, \\
\langle W_{\tiny \yng(2)} W_{\tiny \yng(2)}\rangle^{SO(6)}_{\textrm{$\frac12$BPS}}(\mathfrak{q})
&=\frac{1
}{(1-\mathfrak{q}^4)(1-\mathfrak{q}^8)(1-\mathfrak{q}^{12})}
\nonumber\\
&\times 
(1+\mathfrak{q}^2+3\mathfrak{q}^4+4\mathfrak{q}^6+6\mathfrak{q}^8+6\mathfrak{q}^{10}+7\mathfrak{q}^{12}
\nonumber\\
&+6\mathfrak{q}^{14}+4\mathfrak{q}^{16}+4\mathfrak{q}^{18}+\mathfrak{q}^{20}+\mathfrak{q}^{22}
). 
\end{align}

Next consider the Wilson lines in the symmetric representations. 
For even rank, the one-point function is also non-trivial. 
For example, we find the half-BPS limits of the correlators of the rank-$2$ Wilson lines with the forms 
\begin{align}
\langle W_{\tiny \yng(2)} \rangle^{SO(6)^{-}}_{\textrm{$\frac12$BPS}}(\mathfrak{q})
&=\frac{\mathfrak{q}^4+\mathfrak{q}^8}{(1+\mathfrak{q}^6)(1-\mathfrak{q}^4)(1-\mathfrak{q}^8)}, \\
\langle W_{\tiny \yng(2)}W_{\tiny \yng(2)} \rangle^{SO(6)^{-}}_{\textrm{$\frac12$BPS}}(\mathfrak{q})
&=\frac{1+2\mathfrak{q}^8-2\mathfrak{q}^{10}+\mathfrak{q}^{12}-\mathfrak{q}^{14}-\mathfrak{q}^{18}}
{(1+\mathfrak{q}^6)(1-\mathfrak{q}^2)(1-\mathfrak{q}^4)(1-\mathfrak{q}^8)}. 
\end{align}

For example, for the Wilson lines in the representation of rank-$1$ and $2$, 
the exact forms of the half-BPS limits of the one- and two-point functions are given by
\begin{align}
\langle W_{\tiny \yng(2)} \rangle^{O(6)^{+}}_{\textrm{$\frac12$BPS}}(\mathfrak{q})
&=\frac{\mathfrak{q}^4+\mathfrak{q}^8}{(1+\mathfrak{q}^6)(1-\mathfrak{q}^4)(1-\mathfrak{q}^8)}, \\
\langle W_{\tiny \yng(2)}W_{\tiny \yng(2)} \rangle^{O(6)^{+}}_{\textrm{$\frac12$BPS}}(\mathfrak{q})
&=\frac{1+\mathfrak{q}^2+2\mathfrak{q}^4+2\mathfrak{q}^6+4\mathfrak{q}^8+3\mathfrak{q}^{10}
+4\mathfrak{q}^{12}+2\mathfrak{q}^{14}+2\mathfrak{q}^{16}+\mathfrak{q}^{18}
}{(1-\mathfrak{q}^4)(1-\mathfrak{q}^8)(1-\mathfrak{q}^{12})}. 
\end{align}

\subsection{$\mathfrak{so}(2N)$}
Let us discuss the line defect correlators of line defects in gauge theories with $\mathfrak{g}$ $=$ $\mathfrak{so}(2N)$ of general rank. 

\subsubsection{Spinor Wilson line}
The Wilson lines transforming as the (anti)chiral spinor representation exist in the $Spin(2N)$ gauge theories. 
When $N$ is even, the chiral spinor representation and the antichiral spinor representation are not complex conjugate, while when $N$ is odd, they are. 
There is no non-trivial one-point function of the spinor Wilson line. 

For even $N$, the two-point function of Wilson lines in the chiral spinor representation for $Spin(2N)$ gauge theory reads 
\begin{align}
\label{so2N_wsp1}
&
\langle W_{\textrm{sp}} W_{\textrm{sp}}\rangle^{Spin(2N)}(t;q)
=\frac{1}{2^{N-1}N!}
\frac{(q)_{\infty}^{2N}}{(q^{\frac12}t^{\pm2};q)_{\infty}^{N}}
\oint \prod_{i=1}^{N}\frac{ds_i}{2\pi is_i}
\nonumber\\
&\times 
\prod_{i\neq j}
\frac{(s_i^{\pm}s_j^{\mp};q)_{\infty} (s_i^{\pm}s_j^{\pm};q)_{\infty}(qs_i^{\pm}s_j^{\mp};q)_{\infty}(qs_i^{\pm}s_j^{\pm};q)_{\infty}}
{(q^{\frac12}t^2 s_i^{\pm}s_j^{\mp};q)_{\infty} (q^{\frac12}t^2s_i^{\pm}s_j^{\pm};q)_{\infty}
(q^{\frac12}t^{-2}s_i^{\pm}s_j^{\mp};q)_{\infty}(q^{\frac12}t^{-2}s_i^{\pm}s_j^{\pm};q)_{\infty}}
\nonumber\\
&\times 
\frac14 \left[
\prod_{i=1}^N (s_i^{\frac12}+s_i^{-\frac12})+
\prod_{i=1}^N (s_i^{\frac12}-s_i^{-\frac12})
\right]^2. 
\end{align}
Similarly, we have the two-point function of the Wilson lines in the antichiral spinor representation, which is equivalent to (\ref{so2N_wsp1}). 
However, the two-point function of the chiral spinor Wilson line and antichiral one vanishes. 

For $Spin(2N)$ gauge theory with $N$ odd, we have the two-point function of the Wilson line in the chiral spinor representation and the Wilson line in the antichiral representation with the form 
\begin{align}
\label{so2N_wsp2}
&
\langle W_{\textrm{sp}} W_{\overline{\textrm{sp}}}\rangle^{Spin(2N)}(t;q)
=\frac{1}{2^{N-1}N!}
\frac{(q)_{\infty}^{2N}}{(q^{\frac12}t^{\pm2};q)_{\infty}^{N}}
\oint \prod_{i=1}^{N}\frac{ds_i}{2\pi is_i}
\nonumber\\
&\times 
\prod_{i\neq j}
\frac{(s_i^{\pm}s_j^{\mp};q)_{\infty} (s_i^{\pm}s_j^{\pm};q)_{\infty}(qs_i^{\pm}s_j^{\mp};q)_{\infty}(qs_i^{\pm}s_j^{\pm};q)_{\infty}}
{(q^{\frac12}t^2 s_i^{\pm}s_j^{\mp};q)_{\infty} (q^{\frac12}t^2s_i^{\pm}s_j^{\pm};q)_{\infty}
(q^{\frac12}t^{-2}s_i^{\pm}s_j^{\mp};q)_{\infty}(q^{\frac12}t^{-2}s_i^{\pm}s_j^{\pm};q)_{\infty}}
\nonumber\\
&\times 
\frac14 \left[
\prod_{i=1}^N (s_i^{\frac12}+s_i^{-\frac12})+
\prod_{i=1}^N (s_i^{\frac12}-s_i^{-\frac12})
\right]
\left[
\prod_{i=1}^N (s_i^{\frac12}+s_i^{-\frac12})-
\prod_{i=1}^N (s_i^{\frac12}-s_i^{-\frac12})
\right]. 
\end{align}

According to S-duality, the two-point function (\ref{so2N_wsp1}) or (\ref{so2N_wsp2}) agrees with 
the two-point function of the dual 't Hooft lines with $B=(\frac12^{N})$ for $SO(2N)/\mathbb{Z}_2$ SYM theory of the form
\begin{align}
\label{so2N_tsp}
&
\langle T_{(\frac12^{N})} T_{(\frac12^{N})}\rangle^{SO(2N)/\mathbb{Z}_2}(t;q)
=\frac{1}{N!}\frac{(q)_{\infty}^{2N}}{(q^{\frac12}t^{\pm2};q)_{\infty}^N}
\oint \prod_{i=1}^N 
\frac{ds_i}{2\pi is_i}
\nonumber\\
&\times 
\prod_{i<j}
\frac{(s_i^{\pm}s_j^{\mp};q)_{\infty}(q^{\frac12}s_i^{\pm}s_j^{\pm};q)_{\infty}
(qs_i^{\pm}s_j^{\mp};q)_{\infty}(q^{\frac32}s_i^{\pm}s_j^{\pm};q)_{\infty}
}
{(q^{\frac12}t^2s_i^{\pm}s_j^{\mp};q)_{\infty}(qt^2s_i^{\pm}s_j^{\pm};q)_{\infty}
(q^{\frac12}t^{-2}s_i^{\pm}s_j^{\mp};q)_{\infty}(qt^{-2}s_i^{\pm}s_j^{\pm};q)_{\infty}}. 
\end{align}

The exact form of the half-BPS limit of the two-point functions can be obtained by observing that 
the two-point function (\ref{so2N_tsp}) reduces to the half-BPS index of $U(N)$ gauge theory. 
We get
\begin{align}
\langle W_{\textrm{sp}} W_{\textrm{sp}}\rangle^{Spin(2N)}_{\textrm{$\frac12$BPS}}(\mathfrak{q})
&=\langle T_{(\frac12^{N})} T_{(\frac12^{N})}\rangle^{SO(2N)/\mathbb{Z}_2}_{\textrm{$\frac12$BPS}}(\mathfrak{q})
\nonumber\\
&=\prod_{n=1}^{N}\frac{1}{1-\mathfrak{q}^{2n}}
\end{align}
for even $N$ and 
\begin{align}
\langle W_{\textrm{sp}} W_{\overline{\textrm{sp}}}\rangle^{Spin(2N)}_{\textrm{$\frac12$BPS}}(\mathfrak{q})
&=\langle T_{(\frac12^{N})} T_{(\frac12^{N})}\rangle^{SO(2N)/\mathbb{Z}_2}_{\textrm{$\frac12$BPS}}(\mathfrak{q})
\nonumber\\
&=\prod_{n=1}^{N}\frac{1}{1-\mathfrak{q}^{2n}}
\end{align}
for odd $N$. 
As we have the half-BPS index of $Spin(2N)$ gauge theory of the form 
\begin{align}
\mathcal{I}^{Spin(2N)}_{\textrm{$\frac12$BPS}}(\mathfrak{q})
&=\frac{1}{1-\mathfrak{q}^{2N}}\prod_{n=1}^{N-1}\frac{1}{1-\mathfrak{q}^{4n}}, 
\end{align}
the normalized two-point function is given by
\begin{align}
\label{nhalfBPS_so2Nsp}
\langle \mathcal{W}_{\textrm{sp}} \mathcal{W}_{\textrm{sp}}\rangle^{Spin(2N)}_{\textrm{$\frac12$BPS}}(\mathfrak{q})
&=\langle \mathcal{W}_{\textrm{sp}} \mathcal{W}_{\overline{\textrm{sp}}}\rangle^{Spin(2N)}_{\textrm{$\frac12$BPS}}(\mathfrak{q})
\nonumber\\
&=\prod_{n=1}^{N-1}(1+\mathfrak{q}^{2n}). 
\end{align}

More generally, there exist even-point correlation functions of the spinor Wilson lines. 
In the half-BPS limit they take the form
\begin{align}
&\langle \underbrace{W_{\textrm{sp}} \cdots W_{\textrm{sp}}}_{2k} \rangle^{Spin(2N=4n)}_{\textrm{$\frac12$BPS}}(\mathfrak{q})
=\langle \underbrace{W_{\textrm{sp}} \cdots W_{\overline{\textrm{sp}}}}_{2k} \rangle^{Spin(2N=4n+2)}_{\textrm{$\frac12$BPS}}(\mathfrak{q})
\nonumber\\
&=\frac{\sum_{i=0}^{\frac{N(N+1)k}{2}} a_{k\ \mathfrak{sp}}^{\mathfrak{so}(2N)}(i) \mathfrak{q}^{2i} }{1-\mathfrak{q}^{2N}\prod_{n=1}^{N-1} 1-\mathfrak{q}^{4n}}, 
\end{align}
where $a_{k\ \mathfrak{sp}}^{\mathfrak{so}(2N)}(i)$ are some non-negative integer coefficients. 
The sequence of the first coefficients $a_{k\ \mathfrak{sp}}^{\mathfrak{so}(2N)}(0)$ agrees with 
the multiplicity of the trivial representation in the $2k$-th tensor power of the spinor representation of $Spin(2N)$. 

\subsubsection{Fundamental Wilson line}
For $SO(2N)$ gauge theories based on the gauge algebra $\mathfrak{so}(2N)$, 
we have also the Wilson line operator transforming in the fundamental representation 
as a basic electric line operator with minimal charge. 

The two-point function of the fundamental Wilson lines for $SO(2N)$ SYM theory can be computed from the matrix integral
\begin{align}
\label{so2N_wfund}
&
\langle W_{\tiny \yng(1)} W_{\tiny \yng(1)}\rangle^{SO(2N)}(t;q)
=\frac{1}{2^{N-1}N!}
\frac{(q)_{\infty}^{2N}}{(q^{\frac12}t^{\pm2};q)_{\infty}^{N}}
\oint \prod_{i=1}^{N}\frac{ds_i}{2\pi is_i}
\nonumber\\
&\times 
\prod_{i\neq j}
\frac{(s_i^{\pm}s_j^{\mp};q)_{\infty} (s_i^{\pm}s_j^{\pm};q)_{\infty}(qs_i^{\pm}s_j^{\mp};q)_{\infty}(qs_i^{\pm}s_j^{\pm};q)_{\infty}}
{(q^{\frac12}t^2 s_i^{\pm}s_j^{\mp};q)_{\infty} (q^{\frac12}t^2s_i^{\pm}s_j^{\pm};q)_{\infty}
(q^{\frac12}t^{-2}s_i^{\pm}s_j^{\mp};q)_{\infty}(q^{\frac12}t^{-2}s_i^{\pm}s_j^{\pm};q)_{\infty}}
\left[\sum_{i=1}^N (s_i+s_i^{-1})\right]^2. 
\end{align}
The two-point function of the S-dual 't Hooft lines of $B=(1,0^{N-1})$ for $SO(2N)$ SYM theory takes the form 
\begin{align}
\label{so2N_t1}
&
\langle T_{(1,0^{N-1})} T_{(1,0^{N-1})}\rangle^{SO(2N)}(t;q)
=\frac{1}{2^{N-2}(N-1)!}
\frac{(q)_{\infty}^{2N}}{(q^{\frac12}t^{\pm2};q)_{\infty}^N}
\nonumber\\
&\times 
\oint \prod_{i=1}^N \frac{ds_i}{2\pi is_i}
\prod_{i<j}
\frac{
(q^{\frac12\delta_{i+j,1}}s_i^{\pm}s_j^{\mp};q)_{\infty}
(q^{\frac12\delta_{i+j,1}}s_i^{\pm}s_j^{\pm};q)_{\infty}
}
{
(q^{\frac12(1+\delta_{i+j,1})}t^2s_i^{\pm}s_j^{\mp};q)_{\infty}
(q^{\frac12(1+\delta_{i+j,1})}t^2s_i^{\pm}s_j^{\pm};q)_{\infty}
}
\nonumber\\
&\times 
\frac{
(q^{1+\frac12\delta_{i+j,1}}s_i^{\pm}s_j^{\mp};q)_{\infty}
(q^{1+\frac12\delta_{i+j,1}}s_i^{\pm}s_j^{\pm};q)_{\infty}
}
{
(q^{\frac12(1+\delta_{i+j,1})}t^{-2}s_i^{\pm}s_j^{\mp};q)_{\infty}
(q^{\frac12(1+\delta_{i+j,1})}t^{-2}s_i^{\pm}s_j^{\pm};q)_{\infty}
}. 
\end{align}
As a consequence of S-duality, 
the two-point function (\ref{so2N_wfund}) is conjectured to be equivalent to the two-point function (\ref{so2N_t1}).  

The half-BPS limit of the two-point function of the fundamental Wilson lines is given by 
\begin{align}
\label{halfBPS_so2N_wfund}
\langle W_{\tiny \yng(1)} W_{\tiny \yng(1)}\rangle^{SO(2N)}_{\textrm{$\frac12$BPS}}(\mathfrak{q})
&=\frac{1+\mathfrak{q}^{2}+\cdots \mathfrak{q}^{2N-4}+2\mathfrak{q}^{2N-2}+\mathfrak{q}^{2N}+\cdots \mathfrak{q}^{4N-4}}
{(1-\mathfrak{q}^{2N})\prod_{n=1}^{N-1}(1-\mathfrak{q}^{4n})}
\nonumber\\
&=\frac{1}{(1-\mathfrak{q}^2)(1-\mathfrak{q}^{2(N-1)}) \prod_{n=1}^{N-2}(1-\mathfrak{q}^{4n})}. 
\end{align}

There exist even-point correlation functions of the fundamental Wilson lines. 
They can be written as
\begin{align}
\langle \underbrace{W_{\tiny \yng(1)} \cdots W_{\tiny \yng(1)}}_{2k} \rangle^{SO(2N)}_{\textrm{$\frac12$BPS}}(\mathfrak{q})
&=\frac{\sum_{i=0}^{(2N-2)k} a_{k\ \tiny\yng(1)}^{\mathfrak{so}(2N)}(i) \mathfrak{q}^{2i}}
{(1-\mathfrak{q}^{2N})\prod_{n=1}^{N-1}(1-\mathfrak{q}^{4n})}. 
\end{align}

For the disconnected part $SO(2N)^{-}$, 
the two-point function of the Wilson line in the fundamental representation is given by
\begin{align}
\label{so2N-_wfund}
&
\langle W_{\tiny \yng(1)} W_{\tiny \yng(1)}\rangle^{SO(2N)^{-}}(t;q)
=\frac{1}{2^{N-1}(N-1)!}
\frac{(q)_{\infty}^{2N-2}(-q;q)_{\infty}^{2}}
{(q^{\frac12}t^{\pm2};q)_{\infty}^{N-1}(-q^{\frac12}t^{\pm2};q)_{\infty}}
\nonumber\\
&\times 
\oint \prod_{i=1}^{N-1}\frac{ds_i}{2\pi is_i}
\frac{(s_i^{\pm};q)_{\infty}(-s_i^{\pm};q)_{\infty}(qs_i^{\pm};q)_{\infty}(-qs_i^{\pm};q)_{\infty}}
{(q^{\frac12}t^2s_i^{\pm};q)_{\infty}(-q^{\frac12}t^2s_i^{\pm};q)_{\infty}
(q^{\frac12}t^{-2}s_i^{\pm};q)_{\infty}(-q^{\frac12}t^{-2}s_i^{\pm};q)_{\infty}}
\nonumber\\
&\times 
\prod_{i<j}
\frac{(s_i^{\pm}s_j^{\mp};q)_{\infty}(s_i^{\pm}s_j^{\pm};q)_{\infty}
(qs_i^{\pm}s_j^{\mp};q)_{\infty}(qs_i^{\pm}s_j^{\pm};q)_{\infty}}
{(q^{\frac12}t^2s_i^{\pm}s_j^{\mp};q)_{\infty}(q^{\frac12}t^2s_i^{\pm}s_j^{\pm};q)_{\infty}
(q^{\frac12}t^{-2}s_i^{\pm}s_j^{\mp};q)_{\infty}(q^{\frac12}t^{-2}s_i^{\pm}s_j^{\pm};q)_{\infty}
}
\left[
\sum_{i=1}^{N-1}(s_i+s_i^{-1})
\right]^2. 
\end{align}
We conjecture that the two-point function (\ref{so2N-_wfund}) is equal to 
the two-point function of the S-dual 't Hooft lines of $B=(1,0^{N-2})$ with the form 
\begin{align}
\label{so2N-_t1}
&
\langle T_{(1,0^{N-2})} T_{(1,0^{N-2})}\rangle^{SO(2N)^{-}}(t;q)
\nonumber\\
&=\frac{1}{2^{N-2}(N-2)!}
\frac{(q)_{\infty}^{2N-2}(-q;q)_{\infty}^{2}}
{(q^{\frac12}t^{\pm2};q)_{\infty}^{N-1}(-q^{\frac12}t^{\pm2};q)_{\infty}}
\oint \prod_{i=1}^{N-1}\frac{ds_i}{2\pi is_i}
\nonumber\\
&\times 
\frac{
(q^{\frac12\delta_{i,1}}s_i^{\pm};q)_{\infty}
(-q^{\frac12\delta_{i,1}}s_i^{\pm};q)_{\infty}
(q^{1+\frac12\delta_{i,1}}s_i^{\pm};q)_{\infty}
(-q^{1+\frac12\delta_{i,1}}s_i^{\pm};q)_{\infty}
}
{
(q^{\frac12(1+\delta_{i,1})}t^2s_i^{\pm};q)_{\infty}
(-q^{\frac12(1+\delta_{i,1})}t^2s_i^{\pm};q)_{\infty}
(q^{\frac12(1+\delta_{i,1})}t^{-2}s_i^{\pm};q)_{\infty}
(-q^{\frac12(1+\delta_{i,1})}t^{-2}s_i^{\pm};q)_{\infty}
}
\nonumber\\
&\times 
\prod_{i<j}
\frac{
(q^{\frac12\delta_{i+j,1}}s_{i}^{\pm}s_{j}^{\mp};q)_{\infty}
(q^{\frac12\delta_{i+j,1}}s_{i}^{\pm}s_{j}^{\pm};q)_{\infty}
}
{
(q^{\frac12(1+\delta_{i+j,1})}t^2s_{i}^{\pm}s_{j}^{\mp};q)_{\infty}
(q^{\frac12(1+\delta_{i+j,1})}t^2s_{i}^{\pm}s_{j}^{\pm};q)_{\infty}
}
\nonumber\\
&\times 
\frac{
(q^{1+\frac12\delta_{i+j,1}}s_{i}^{\pm}s_{j}^{\mp};q)_{\infty}
(q^{1+\frac12\delta_{i+j,1}}s_{i}^{\pm}s_{j}^{\pm};q)_{\infty}
}
{
(q^{\frac12(1+\delta_{i+j,1})}t^{-2}s_{i}^{\pm}s_{j}^{\mp};q)_{\infty}
(q^{\frac12(1+\delta_{i+j,1})}t^{-2}s_{i}^{\pm}s_{j}^{\pm};q)_{\infty}
}. 
\end{align}

In the half-BPS limit the two-point functions (\ref{so2N-_wfund}) and (\ref{so2N-_t1}) become 
\begin{align}
\label{halfBPS_so2N-_wfund}
\langle W_{\tiny \yng(1)} W_{\tiny \yng(1)}\rangle^{SO(2N)^{-}}_{\textrm{$\frac12$BPS}}(\mathfrak{q})
&=\langle T_{(1,0^{N-2})} T_{(1,0^{N-2})}\rangle^{SO(2N)^{-}}_{\textrm{$\frac12$BPS}}(\mathfrak{q})
\nonumber\\
&=\frac{1+\mathfrak{q}^2+\cdots+\mathfrak{q}^{2N-4}}
{\prod_{n=1}^{N-1}(1-\mathfrak{q}^{4n})}
\nonumber\\
&=\frac{1-\mathfrak{q}^{2(N-1)}}{1-\mathfrak{q}^{2}}
\prod_{n=1}^{N-1}\frac{1}{1-\mathfrak{q}^{4n}}. 
\end{align}

It follows from (\ref{halfBPS_so2N_wfund}) and (\ref{halfBPS_so2N-_wfund}) that 
the half-BPS limits of the two-point functions of the fundamental Wilson lines in $O(2N)$ gauge theories are given by 
\begin{align}
\langle W_{\tiny \yng(1)} W_{\tiny \yng(1)}\rangle^{O(2N)^{+}}_{\textrm{$\frac12$BPS}}(\mathfrak{q})
&=\frac{1+\mathfrak{q}^2+\cdots +\mathfrak{q}^{4N-2}}
{\prod_{n=1}^{N}(1-\mathfrak{q}^{4n})}
\nonumber\\
&=\frac{1}{1-\mathfrak{q}^2}\prod_{n=1}^{N-1}\frac{1}{1-\mathfrak{q}^{4n}}. 
\end{align}
As one has the half-BPS indices for $\mathcal{N}=4$ $O(2N)$ gauge theories of the forms 
\begin{align}
\mathcal{I}^{O(2N)^+}_{\textrm{$\frac12$BPS}}(\mathfrak{q})&=
\prod_{n=1}^{N}\frac{1}{1-\mathfrak{q}^{4n}},
\end{align}
the normalized two-point functions can be written as
\begin{align}
\label{nhalfBPS_o2N+_wfund}
\langle \mathcal{W}_{\tiny \yng(1)} \mathcal{W}_{\tiny \yng(1)}\rangle^{O(2N)^{+}}_{\textrm{$\frac12$BPS}}(\mathfrak{q})
&=\frac{1-\mathfrak{q}^{4N}}{1-\mathfrak{q}^2}
\end{align}
The expression (\ref{nhalfBPS_o2N+_wfund}) agrees with 
the normalized two-point functions (\ref{nhalfBPS_so2N+1_wfund}) and (\ref{nhalfBPS_sp2N_wfund}) of the fundamental Wilson lines for $SO(2N+1)$ and $USp(2N)$ gauge theories. 

\section{Large $N$ limits}
\label{sec_largeN}
The KK spectrum on $AdS_{5}\times \mathbb{RP}^{5}$ can be obtained 
by expanding the linearized ten-dimensional fields \cite{Gunaydin:1984fk} in harmonics on the $\mathbb{RP}^{5}$, 
the smooth $\mathbb{Z}_{2}$ quotient of a five-sphere, and keeping the corresponding modes that remain invariant under the smooth quotient. 
The harmonics are hence organized according to representations of the corresponding isometry group after the quotient. 
The corresponding modes in this spectrum are characterized by KK levels, 
which correspond to oscillator numbers \cite{Gunaydin:1984qu,Gunaydin:1984fk}. 
According to the AdS/CFT correspondence, 
the KK spectrum can be obtained from the large $N$ limit of the supersymmetric index of $\mathcal{N}=4$ SYM theory with orthogonal and symplectic gauge groups. 
It takes the form \cite{Hatsuda:2024lcc}
\begin{align}
&\mathcal{I}^{SO(2\infty+1)}(t;q)=\mathcal{I}^{USp(2\infty)}(t;q)=\mathcal{I}^{SO(2\infty)}(t;q)=\mathcal{I}^{O(2\infty)^{+}}(t;q)
\nonumber\\
&=\prod_{n,m,l=0}^{\infty}
\frac{(1-q^{n+m+l+\frac32}t^{-4m+4l\pm2})^2}
{(1-q^{n+m+l+1}t^{-4m+4l\pm4})(1-q^{n+m++1}t^{-4m+4l})(1-q^{n+m+l+3}t^{-4m+4l})}. 
\end{align}
Taking the plethystic logarithm \cite{MR1601666}, 
we obtain the single gravity index as a generating function for the KK spectrum 
\begin{align}
i^{AdS_5\times\mathbb{RP}^5}(t;q)&=
\frac{q^{\frac12}(t^2+t^{-2})-q-q^2}{(1-qt^4)(1-qt^{-4})}
-\frac{q^{\frac12}(t^2+t^{-2})}{(1+q^{\frac12}t^2)(1+q^{\frac12}t^{-2})(1-q)}. 
\end{align}

Here we examine the large $N$ limits of the line defect correlation functions in $\mathcal{N}=4$ SYM theory with orthogonal and symplectic gauge groups. 
They are expected to capture the spectra of the quantum fluctuation modes of various holographically dual geometries with $AdS_2$ factor. 
The spectrum of the quantum excitations around the gravity dual geometry $X$ is encoded by the single gravity index defined by
\begin{align}
\label{gravity_ind}
i^{X}(t;q)&:=\Tr (-1)^F q^{\frac{h+j}{2}} t^{2(q_2-q_3)}, 
\end{align}
where the trace is taken over the Hilbert space of the BPS states with the condition $h=j+q_2+q_3$. 
Here $F$, $h$, $j$ and $q_{\alpha}$, $\alpha=1,2,3$ stand for the Fermion number operator, the scaling dimension, the $SO(3)$ spin and the $SO(6)$ Cartan generators respectively. 
By computing the large $N$ limits of the (connected) two-point functions of the line defects, 
we address the gravity index for the spectra of the quantum fluctuations on the holographically dual geometries. 
We also note that the half-BPS limit of the line defect correlator corresponds to the gravity index 
for the spectrum of the BPS states which saturate the BPS bound $h=q_2$ and $j=-q_3$. 
Thus we have 
\begin{align}
\label{gravity_hind}
i^{X}_{\textrm{$\frac12$BPS}}(\mathfrak{q})&:=\Tr (-1)^F \mathfrak{q}^{2(q_2-q_3)}.  
\end{align}
This encodes the fluctuation modes carrying the non-trivial quantum numbers associated with the isometry of the compact direction inside the $\mathbb{RP}^5$. 

\subsection{Fundamental Wilson line}
The Wilson line in the fundamental representation in $\mathcal{N}=4$ SYM theory is holographically dual 
to the fundamental string wrapping $AdS_2$ $\subset$ $AdS_5$. 
The fundamental string wrapping the $AdS_2$ in the global $AdS_5$ encounters the boundary of the $AdS_5$ at two points. 
Hence it should correspond to the superconformal line that map to pairs of half-lines localized at the north pole and the south pole of $S^3$. 
Accordingly, the fluctuation modes of the gravity dual configuration is expected to be calculated from the normalized two-point functions of the fundamental Wilson lines. 

The half-BPS limits of the normalized two-point functions of the fundamental Wilson lines in $SO(2N+1)$, $USp(2N)$, $SO(2N)$ and $O(2N)^+$ gauge theories are given by 
(\ref{nhalfBPS_so2N+1_wfund}), (\ref{nhalfBPS_sp2N_wfund}), (\ref{halfBPS_so2N_wfund}) and (\ref{nhalfBPS_o2N+_wfund}) respectively. 
In the large $N$ limit, they become equal to 
\footnote{Note that the unconnected part of the two-point function for $SO(2N+1)$ gauge theory is equal to the connected part in the large $N$ limit since the one-point function vanishes. }
\begin{align}
&\langle \mathcal{W}_{\tiny \yng(1)} \mathcal{W}_{\tiny \yng(1)}\rangle^{SO(2\infty+1)}_{\textrm{$\frac12$BPS},c}(\mathfrak{q})
=\langle \mathcal{W}_{\tiny \yng(1)} \mathcal{W}_{\tiny \yng(1)}\rangle^{USp(2\infty)}_{\textrm{$\frac12$BPS}}(\mathfrak{q})
\nonumber\\
&=\langle \mathcal{W}_{\tiny \yng(1)} \mathcal{W}_{\tiny \yng(1)}\rangle^{SO(2\infty)}_{\textrm{$\frac12$BPS},c}(\mathfrak{q})
=\langle \mathcal{W}_{\tiny \yng(1)} \mathcal{W}_{\tiny \yng(1)}\rangle^{O(2\infty)^{+}}_{\textrm{$\frac12$BPS}}(\mathfrak{q})
=\frac{1}{1-\mathfrak{q}^2}. 
\end{align}

Furthermore, we find that the large $N$ normalized two-point functions of the fundamental Wilson lines in these theories are given by
\begin{align}
\label{large_wfund}
&\langle \mathcal{W}_{\tiny \yng(1)} \mathcal{W}_{\tiny \yng(1)}\rangle^{SO(2\infty+1)}(t;q)
=\langle \mathcal{W}_{\tiny \yng(1)} \mathcal{W}_{\tiny \yng(1)}\rangle^{USp(2\infty)}(t;q)
\nonumber\\
&=\langle \mathcal{W}_{\tiny \yng(1)} \mathcal{W}_{\tiny \yng(1)}\rangle^{SO(2\infty)}(t;q)
=\langle \mathcal{W}_{\tiny \yng(1)} \mathcal{W}_{\tiny \yng(1)}\rangle^{O(2\infty)^{+}}(t;q)
=\frac{1-q}{(1-q^{\frac12}t^2) (1-q^{\frac12}t^{-2})}. 
\end{align}
This exactly agrees with the large $N$ normalized two-point functions of the fundamental Wilson lines in $U(N)$ gauge theory \cite{Gang:2012yr,Hatsuda:2023iwi}. 
Taking the plethystic logarithm of the expression (\ref{large_wfund}), we get 
\begin{align}
i^{\textrm{string}}(t;q)&=-q+q^{\frac12}t^2+q^{\frac12}t^{-2}. 
\end{align}
This is identified with the single particle gravity index describing the spectrum of the quantum fluctuations of the
gravity dual configuration where the fundamental string is wrapped on the $AdS_2$. 

\subsection{Adjoint Wilson line}
In the large $N$ limit, the expansion coefficients of the normalized one- and two-point functions of the Wilson lines transforming in the adjoint representation are stabilized 
for $SO(2N+1)$, $USp(2N)$ and $O(2N)$ gauge theories. 

In the half-BPS limits, the one-point functions become equal to 
\begin{align}
&\langle \mathcal{W}_{\tiny \yng(1,1)} \rangle^{SO(2\infty+1)}_{\textrm{$\frac12$BPS}}(\mathfrak{q})
=\langle \mathcal{W}_{\tiny \yng(2)} \rangle^{USp(2\infty)}_{\textrm{$\frac12$BPS}}(\mathfrak{q})
=\langle \mathcal{W}_{\tiny \yng(1,1)} \rangle^{O(2\infty)^+}_{\textrm{$\frac12$BPS}}(\mathfrak{q})
\nonumber\\
&=\frac{\mathfrak{q}^2}{(1-\mathfrak{q}^4)}
\nonumber\\
&=\mathfrak{q}^2+\mathfrak{q}^6+\mathfrak{q}^{10}+\mathfrak{q}^{14}+\mathfrak{q}^{18}+\cdots. 
\end{align}
Also the half-BPS limits of the two-point functions also become equal in the large $N$ limit. 
They are given by
\begin{align}
&\langle \mathcal{W}_{\tiny \yng(1,1)} \mathcal{W}_{\tiny \yng(1,1)}\rangle^{SO(2\infty+1)}_{\textrm{$\frac12$BPS}}(\mathfrak{q})
=\langle \mathcal{W}_{\tiny \yng(2)} \mathcal{W}_{\tiny \yng(2)}\rangle^{USp(2\infty)}_{\textrm{$\frac12$BPS}}(\mathfrak{q})
=\langle \mathcal{W}_{\tiny \yng(1,1)} \mathcal{W}_{\tiny \yng(1,1)}\rangle^{O(2\infty)^+}_{\textrm{$\frac12$BPS}}(\mathfrak{q})
\nonumber\\
&=\frac{1+\mathfrak{q}^2+\mathfrak{q}^4}{(1-\mathfrak{q}^4)^2}
\nonumber\\
&=1+\mathfrak{q}^2+3\mathfrak{q}^4+2\mathfrak{q}^6+5\mathfrak{q}^8+3\mathfrak{q}^{10}+7\mathfrak{q}^{12}+4\mathfrak{q}^{14}+9\mathfrak{q}^{16}+\cdots. 
\end{align}
Accordingly, the large $N$ connected two-point functions in the half-BPS limit can be written as
\begin{align}
\label{halfBPS_large_wadj}
&\langle \mathcal{W}_{\tiny \yng(1,1)} \mathcal{W}_{\tiny \yng(1,1)}\rangle^{SO(2\infty+1)}_{\textrm{$\frac12$BPS},c}(\mathfrak{q})
=\langle \mathcal{W}_{\tiny \yng(2)} \mathcal{W}_{\tiny \yng(2)}\rangle^{USp(2\infty)}_{\textrm{$\frac12$BPS},c}(\mathfrak{q})
=\langle \mathcal{W}_{\tiny \yng(1,1)} \mathcal{W}_{\tiny \yng(1,1)}\rangle^{O(2\infty)^+}_{\textrm{$\frac12$BPS},c}(\mathfrak{q})
\nonumber\\
&=\frac{1}{(1-\mathfrak{q}^2)(1-\mathfrak{q}^4)}
\nonumber\\
&=1+\mathfrak{q}^2+2\mathfrak{q}^4+2\mathfrak{q}^6+3\mathfrak{q}^8+3\mathfrak{q}^{10}+4\mathfrak{q}^{12}+4\mathfrak{q}^{14}+5\mathfrak{q}^{16}+5\mathfrak{q}^{18}+\cdots, 
\end{align}
where we have defined
\begin{align}
\langle \mathcal{W}_{\lambda} \mathcal{W}_{\lambda}\rangle^{G}_{\textrm{$\frac12$BPS},c}(\mathfrak{q})
&:=\langle \mathcal{W}_{\lambda} \mathcal{W}_{\lambda}\rangle^{G}_{\textrm{$\frac12$BPS}}(\mathfrak{q})
-\langle  \mathcal{W}_{\lambda}\rangle^{G}_{\textrm{$\frac12$BPS}}(\mathfrak{q})^2. 
\end{align}

More generally, the large $N$ Schur line defect correlators of the adjoint Wilson lines can be evaluated. 
We obtain
\begin{align}
&\langle \mathcal{W}_{\tiny \yng(1,1)} \rangle^{SO(2\infty+1)}(t;q)
=\langle \mathcal{W}_{\tiny \yng(2)}\rangle^{USp(2\infty)}(t;q)
=\langle \mathcal{W}_{\tiny \yng(1,1)} \rangle^{O(2\infty)^+}(t;q)
\nonumber\\
&=\frac{q^{\frac12}(t^2+t^{-2})-q-q^2}{(1-qt^4)(1-qt^{^4})}
\end{align}
and
\begin{align}
&\langle \mathcal{W}_{\tiny \yng(1,1)} \mathcal{W}_{\tiny \yng(1,1)} \rangle^{SO(2\infty+1)}(t;q)
=\langle \mathcal{W}_{\tiny \yng(2)} \mathcal{W}_{\tiny \yng(2)} \rangle^{USp(2\infty)}(t;q)
=\langle \mathcal{W}_{\tiny \yng(1,1)}\mathcal{W}_{\tiny \yng(1,1)} \rangle^{O(2\infty)^+}(t;q)
\nonumber\\
&=\frac{1}{(1-qt^4)(1-qt^{-4})}
\Bigl(1+(t^2+t^{-2})q^{\frac12}+(3+t^4+t^{-4})q-3(t^2+t^{-2})q^{\frac32}
\nonumber\\
&-(t^2+t^{-2})q^2-3(t^2+t^{-2})q^{\frac52}+(3+t^4+t^{-4})q^3+(t^2+t^{-2})q^{\frac72}+q^4
\Bigr). 
\end{align}
Hence the large $N$ connected two-point functions of the adjoint Wilson lines are given by
\begin{align}
\label{largec_wadj}
&\langle \mathcal{W}_{\tiny \yng(1,1)} \mathcal{W}_{\tiny \yng(1,1)} \rangle^{SO(2\infty+1)}_{c}(t;q)
=\langle \mathcal{W}_{\tiny \yng(2)} \mathcal{W}_{\tiny \yng(2)} \rangle^{USp(2\infty)}_{c}(t;q)
=\langle \mathcal{W}_{\tiny \yng(1,1)}\mathcal{W}_{\tiny \yng(1,1)} \rangle^{O(2\infty)^+}_{c}(t;q)
\nonumber\\
&=\frac{(1-q)(1+q-q^{\frac32}(t^2+t^{-2}))}{(1-q^{\frac12}t^2)(1-q^{\frac12}t^{-2})(1-qt^4)(1-qt^{-4})}. 
\end{align}

\subsection{Spinor Wilson line}
The Wilson line in the spinor representation for $Spin(2N+1)$ or $Spin(2N)$ gauge theory is holographically dual to a fat string, 
that is a D5-brane wrapping $AdS_2$ $\times$ $\mathbb{RP}^4$ 
where $AdS_2$ is subspace of $AdS_5$ and $\mathbb{RP}^4$ is subspace of $\subset$ $\mathbb{RP}^5$ \cite{Witten:1998xy} 
(see \cite{Giombi:2020kvo} for more recent discussion). 

In the large $N$ limit, the normalized two-point functions (\ref{nhalfBPS_so2N+1sp}) and (\ref{nhalfBPS_so2Nsp}) 
of the Wilson lines transforming in the spinor representation become 
\begin{align}
\label{halfBPS_large_wsp}
&
\langle \mathcal{W}_{\textrm{sp}} \mathcal{W}_{\textrm{sp}}\rangle^{Spin(2\infty+1)}_{\textrm{$\frac12$BPS}}(\mathfrak{q})
=\langle \mathcal{W}_{\textrm{sp}} \mathcal{W}_{\textrm{sp}}\rangle^{Spin(4\infty)}_{\textrm{$\frac12$BPS}}(\mathfrak{q})
=\langle \mathcal{W}_{\textrm{sp}} \mathcal{W}_{\overline{\textrm{sp}}}\rangle^{Spin(4\infty+2)}_{\textrm{$\frac12$BPS}}(\mathfrak{q})
\nonumber\\
&=\prod_{n=1}^{\infty}\frac{1}{1-\mathfrak{q}^{4n-2}}. 
\end{align}
When we expand the normalized two-point function (\ref{halfBPS_large_wsp}) as
\begin{align}
\langle \mathcal{W}_{\textrm{sp}} \mathcal{W}_{\textrm{sp}}\rangle^{Spin(\infty)}_{\textrm{$\frac12$BPS}}(\mathfrak{q})
&=\sum_{n\ge0} d^{(H)}_{\{ \textrm{sp},\textrm{sp} \}}(n) \mathfrak{q}^{2n}, 
\end{align}
the number $d^{(H)}_{\{ \textrm{sp},\textrm{sp} \}}(n)$ of states is identified with the number of partitions of $n$ into distinct parts 
or equivalently the number of partitions of $n$ into odd parts. 
The half-BPS limit of the single particle gravity index is obtained by taking the plethystic logarithm of the expression (\ref{halfBPS_large_wsp})
\begin{align}
\label{fatstring_hind}
i^{\textrm{fat string}}_{\textrm{$\frac12$BPS}}(\mathfrak{q})
&=\frac{\mathfrak{q}^2}{1-\mathfrak{q}^4}
=\mathfrak{q}^2+\mathfrak{q}^6+\mathfrak{q}^{10}+\cdots. 
\end{align}

Here we provide an alternative way to obtain the above result by analyzing the fluctuations of a D5-brane wrapped on $AdS_2$ $\times$ $\mathbb{RP}^4$, 
which is similar to the methods in \cite{Faraggi:2011bb,Faraggi:2011ge}.
In the convention of Einstein frame, the bosonic part of the Euclidean action of D5-brane takes the form
\begin{align}
S_{\text{D5}}=T_{5}\int d^{6}\sigma \sqrt{\det \left( g+2\pi \alpha ^{\prime
}F\right) }-iT_{5}\int 2\pi \alpha ^{\prime }F\wedge C_{(4)},  \label{eqn 82}
\end{align}
where $\sigma ^{a}$, $a=1,\ldots ,6$ are world-volume coordinates, 
$g_{ab}$ is the induced metric on the brane, and $F_{ab}$ is the field strength on the world-volume gauge field. 
In the case where the D5-brane wrapped on $AdS_2\times \mathbb{RP}^4$ with vanishing NS discrete torsion, 
the world-volume electric flux is turned off, i.e. $F=0$. 
Hence the action for the D5-brane wrapped on $AdS_2$ $\times$ $\mathbb{RP}^4$ is given by
\begin{align}
S_{\text{D5 $AdS_2\times \mathbb{RP}^4$}}=T_{5}\int d^{6}\sigma \sqrt{\det g}=T_{5}\mathrm{vol}(AdS_{2})%
\mathrm{vol}(\mathbb{RP}^{4})  \label{eqn 128}
\end{align}
where $\mathrm{vol}(\mathbb{RP}^{4})=\frac{4}{3}\pi ^{2}$. The brane is on
the $\mathbb{RP}^{4}$ located at the equator of $\mathbb{RP}^{5}$, similar
to a maximal giant located at equator. \footnote{See \cite{Giombi:2020kvo} for the related discussion. }
This equator is denoted as $\theta _{0}=\frac{\pi }{2}$, 
where we denote the metric of $\mathbb{RP}^{5}$ by $ds^{2}(\mathbb{RP}^{5})=d^{2}\theta +\sin ^{2}\theta \ ds^{2}(\mathbb{RP}^{4})$.

The above D5 action, expanded around the $AdS_{2}\times\mathbb{RP}^{4}$ background configuration, 
contains the quadratic action of the BPS fluctuation modes. 
In the half-BPS limit, the fluctuation modes which can contribute to the gravity index (\ref{gravity_hind}) 
should carry the quantum numbers associated to the isometry of the four-dimensional space. 
Let us denote the part of metric of $AdS_{2}$ and of $\mathbb{RP}^{4}$ by
\begin{align}
ds_{AdS_{2}}^{2}=\frac{1}{r^{2}}(-dt^{2}+dr^{2}),\;\;ds_{4}^{2}=g_{ij}d%
\sigma ^{i}d\sigma ^{j},\qquad i,j=1,2,3,4
\end{align}
in the unit $L=1$ of unit AdS radius. 
The $ds_{4}^{2}$ denotes the line element of $\mathbb{RP}^{4}$. 
We use $\mu ,\nu =0,1$ to denote the 2d
coordinates along $AdS_{2}$.

Expanding the action (\ref{eqn 82}) around the background configuration
associated to (\ref{eqn 128}), the quadratic action for this set of
fluctuation fields is
\begin{equation}
S=T_{5}\int d^{6}\sigma \sqrt{g^{(4)}}\frac{1}{2}\frac{1}{r^{2}}%
[r^{2}(\partial _{t}\phi )^{2}-r^{2}(\partial _{r}\phi )^{2}+(\nabla
_{i}\phi \nabla ^{i}\phi -4\phi ^{2})].  \label{eqn 116}
\end{equation}%
Here $g^{(4)}$ is the determinant of 4d $g_{ij}$ on the internal four
dimensional space. 

We have the spherical harmonics from the covering space and make a $\mathbb{Z%
}_{2}$ projection to the quotient space. Denoting the spherical harmonics $%
Y^{i}$ in the fundamental representation of $SO(5)$, we have the spherical
harmonics by symmetric traceless products $Y^{w}(\Theta ):=Y^{i_{1}\ldots
i_{w}}(\Theta )=Y^{(i_{1}}\ldots Y^{i_{w})}$, where $\Theta $ denotes the
internal four directions and $\nabla ^{2}Y^{i}=-4Y^{i}$. The above
fluctuation fields can be expanded,
\begin{equation}
\phi (t,r,\Theta )=\sum_{w}\phi _{w}(t,r)Y^{w}(\Theta ).  \label{eqn 117}
\end{equation}%
Then the equations of motion of (\ref{eqn 116}) reduce to decoupled 2d
Laplace equations for eigenmodes. We see that the eigenmodes from (\ref{eqn
117}), that we are interested in, have the quadratic action of $\phi _{w}$
reduced from (\ref{eqn 116}).

By compactifying the action (\ref{eqn 116}) on the compact four-dimensional space, 
we get the 2d quadratic action 
\begin{align}
\label{2d_fatstring}
S=T_{5}\sum_{w}\frac{2}{3}\pi ^{2}\int d^{2}\sigma \,\frac{1}{r^{2}}\left(
r^{2}(\partial _{t}\phi _{w})^{2}-r^{2}(\partial _{r}\phi
_{w})^{2}-w(w+1)\phi _{w}^{2}\right) .
\end{align}%
Accordingly, these BPS bosonic fluctuation modes $\phi _{w}$, satisfies the Klein-Gordon equations on $AdS_{2}$. 
\footnote{We note that our $\phi_w$ are identified with $\eta_l$ in \cite{Faraggi:2011ge} with $w=l$}

The scaling dimensions obey the relations
\begin{align}
h=\frac{1}{2}+\sqrt{\frac{1}{4}+m^{2}},  \label{eqn 112}
\end{align}%
corresponding to a set of massive fields on $AdS_{2}$. 
From the quadratic action (\ref{2d_fatstring}) we see that the fluctuation modes have masses $m^{2}=w(w+1)$. 
So they have the scaling dimension $h=w+1$. 
However, by the $\mathbb{Z}_{2}$ projection, we should keep only the modes with $w=2n$, where $n=0,1,2,\cdots$. 
Hence $w$ is an even integer so that we have $h=1,3,5,\cdots$. 
For these fluctuation modes we have $j=0$. 
The $h=1$ corresponds to the lowest order term $\mathfrak{q}^{2}$ in (\ref{fatstring_hind}). 
The subsequent modes $h=3,5,\cdots$ correspond to the terms $\mathfrak{q}^{6}$, $\mathfrak{q}^{10}$, $\cdots$. 
These are the BPS states that saturate the BPS bound condition, discussed near (\ref{gravity_hind}). 

Furthermore, we find that the large $N$ normalized Schur line defect two-point function of the spinor Wilson lines is given by
\begin{align}
\label{large_wsp}
&
\langle \mathcal{W}_{\textrm{sp}} \mathcal{W}_{\textrm{sp}}\rangle^{Spin(2\infty+1)}(t;q)
=\langle \mathcal{W}_{\textrm{sp}} \mathcal{W}_{\textrm{sp}}\rangle^{Spin(4\infty)}(t;q)
=\langle \mathcal{W}_{\textrm{sp}} \mathcal{W}_{\overline{\textrm{sp}}}\rangle^{Spin(4\infty+2)}(t;q)
\nonumber\\
&=\prod_{n=0}^{\infty}\prod_{m=0}^{\infty}
\frac{(1-q^{1+n+m}t^{4n-4m}) (1-q^{2+n+m}t^{4n-4m})}
{(1-q^{\frac12+n+m}t^{2+4n-4m}) (1-q^{\frac12+n+m}t^{-2+4n+4m})}. 
\end{align}
In the unflavored limit it becomes
\begin{align}
&
\langle \mathcal{W}_{\textrm{sp}} \mathcal{W}_{\textrm{sp}}\rangle^{Spin(2\infty+1)}(q)
=\langle \mathcal{W}_{\textrm{sp}} \mathcal{W}_{\textrm{sp}}\rangle^{Spin(4\infty)}(q)
=\langle \mathcal{W}_{\textrm{sp}} \mathcal{W}_{\overline{\textrm{sp}}}\rangle^{Spin(4\infty+2)}(q)
\nonumber\\
&=\prod_{n=1}^{\infty}
\frac{(1-q^{n})^{2n-1}}
{(1-q^{n-\frac12})^{2n}}
\nonumber\\
&=1+2q^{1/2}+2q^2+6q^{3/2}+7q^2+10q^{5/2}+21q^3+22q^{7/2}+\cdots. 
\end{align}

From the plethystic logarithm of the expression (\ref{large_wsp}) we get
\begin{align}
\label{fatstring_index}
i^{\textrm{fat string}}(t;q)
&=\frac{q^{\frac12}(t^2+t^{-2})-q-q^2}{(1-qt^4)(1-qt^{-4})}. 
\end{align}
We conclude that the expression (\ref{fatstring_index}) should be the single particle gravity index 
for the spectrum of the fluctuation modes of the gravity dual configuration 
in which the D5-brane is wrapped on the $AdS_2\times \mathbb{RP}^4$ 
or equivalently the fat string is wrapped on the $AdS_2$. 

\subsection{Other Wilson lines}
We find that the normalized one- and two-point functions of the $USp(2N)$ rank-$2$ antisymmetric Wilson lines, 
those of the $SO(2N+1)$ rank-$2$ symmetric Wilson lines and those of the $O(2N)$ rank-$2$ symmetric Wilson lines coincide in the large $N$ limit. 

In the half-BPS limit we find 
\begin{align}
&\langle \mathcal{W}_{\tiny \yng(2)} \rangle^{SO(2\infty+1)}_{\textrm{$\frac12$BPS}}(\mathfrak{q})
=\langle \mathcal{W}_{\tiny \yng(1,1)} \rangle^{USp(2\infty)}_{\textrm{$\frac12$BPS}}(\mathfrak{q})
=\langle \mathcal{W}_{\tiny \yng(2)} \rangle^{O(2\infty)^+}_{\textrm{$\frac12$BPS}}(\mathfrak{q})
\nonumber\\
&=\frac{\mathfrak{q}^4}{(1-\mathfrak{q}^4)}
\end{align}
and 
\begin{align}
&\langle \mathcal{W}_{\tiny \yng(2)} \mathcal{W}_{\tiny \yng(2)} \rangle^{SO(2\infty+1)}_{\textrm{$\frac12$BPS}}(\mathfrak{q})
=\langle \mathcal{W}_{\tiny \yng(1,1)}\mathcal{W}_{\tiny \yng(1,1)} \rangle^{USp(2\infty)}_{\textrm{$\frac12$BPS}}(\mathfrak{q})
=\langle \mathcal{W}_{\tiny \yng(2)}  \mathcal{W}_{\tiny \yng(2)} \rangle^{O(2\infty)^+}_{\textrm{$\frac12$BPS}}(\mathfrak{q})
\nonumber\\
&=\frac{1+\mathfrak{q}^2+\mathfrak{q}^8}{(1-\mathfrak{q}^4)^2}. 
\end{align}
Thus the large $N$ connected two-point functions in the half-BPS limit are given by
\begin{align}
&\langle \mathcal{W}_{\tiny \yng(2)} \mathcal{W}_{\tiny \yng(2)} \rangle^{SO(2\infty+1)}_{\textrm{$\frac12$BPS},c}(\mathfrak{q})
=\langle \mathcal{W}_{\tiny \yng(1,1)}\mathcal{W}_{\tiny \yng(1,1)} \rangle^{USp(2\infty)}_{\textrm{$\frac12$BPS},c}(\mathfrak{q})
=\langle \mathcal{W}_{\tiny \yng(2)}  \mathcal{W}_{\tiny \yng(2)} \rangle^{O(2\infty)^+}_{\textrm{$\frac12$BPS},c}(\mathfrak{q})
\nonumber\\
&=\frac{1}{(1-\mathfrak{q}^2)(1-\mathfrak{q}^4)}, 
\end{align}
which agrees with the expression (\ref{halfBPS_large_wadj}) for the large $N$ connected two-point functions of the adjoint Wilson lines in the half-BPS limit. 

For the large $N$ Schur line defect two-point functions, we get 
\begin{align}
&\langle \mathcal{W}_{\tiny \yng(2)} \rangle^{SO(2\infty+1)}(t;q)
=\langle \mathcal{W}_{\tiny \yng(1,1)} \rangle^{USp(2\infty)}(t;q)
=\langle \mathcal{W}_{\tiny \yng(2)} \rangle^{O(2\infty)^+}(t;q)
\nonumber\\
&=\frac{q(1+t^4+t^{-4})-q^{\frac32}(t^2+t^{-2})-q^2}{(1-qt^4)(1-qt^{-4})}
\end{align}
and 
\begin{align}
&\langle \mathcal{W}_{\tiny \yng(2)} \mathcal{W}_{\tiny \yng(2)} \rangle^{SO(2\infty+1)}(t;q)
=\langle \mathcal{W}_{\tiny \yng(1,1)} \mathcal{W}_{\tiny \yng(1,1)} \rangle^{USp(2\infty)}(t;q)
=\langle \mathcal{W}_{\tiny \yng(2)} \mathcal{W}_{\tiny \yng(2)} \rangle^{O(2\infty)^+}(t;q)
\nonumber\\
&=\frac{1}{(1-qt^4)(1-qt^{-4})}
\Bigl(
1+(t^2+t^{-2})q^{\frac12}+q-(t^2+t^{-2})q^{\frac32}+(t^8+t^4+t^{-4}+t^{-8})q^2
\nonumber\\
&-(2t^6+5t^2+5t^{-2}+2t^{-6})q^{\frac52}
+q^3+3(t^2+t^{-2})q^{\frac72}+q^4
\Bigr). 
\end{align}
Then we find the large $N$ normalized connected two-point functions of the form
\begin{align}
&\langle \mathcal{W}_{\tiny \yng(2)} \mathcal{W}_{\tiny \yng(2)} \rangle^{SO(2\infty+1)}_{c}(t;q)
=\langle \mathcal{W}_{\tiny \yng(1,1)} \mathcal{W}_{\tiny \yng(1,1)} \rangle^{USp(2\infty)}_{c}(t;q)
=\langle \mathcal{W}_{\tiny \yng(2)} \mathcal{W}_{\tiny \yng(2)} \rangle^{O(2\infty)^+}_{c}(t;q)
\nonumber\\
&=\frac{(1-q)(1+q-q^{\frac32}(t^2+t^{-2}))}{(1-q^{\frac12}t^2)(1-q^{\frac12}t^{-2})(1-qt^4)(1-qt^{-4})}. 
\end{align}
Again this agrees with the large $N$ normalized connected two-point functions (\ref{largec_wadj}) of the adjoint Wilson lines.

\subsection*{Acknowledgements}
We thank Jin Chen, Diego Correa and Masatoshi Noumi for useful discussion. 
The work of Y.H. was supported in part by JSPS KAKENHI Grant Nos. 22K03641 and 22K03594. 
The work of H.L. was supported in part by National Key R\&D Program of China grant 2020YFA0713000, 
by Overseas high-level talents program, by Fundamental Research Funds for the Central Universities of China, and by Grant No. 3207012204. 
The work of T.O. was supported by the Startup Funding no.\ 4007012317 of the Southeast University. 

\appendix

\section{Character formulae}\label{app:character}
In this appendix, we summarize a way to express the characters for $\mathfrak{usp}$ and $\mathfrak{so}$ in terms of the power sum function.
Let $h_k(x_1,\dots, x_n)$ and $e_k(x_1,\dots, x_n)$ be completely and elementary symmetric polynomials with $n$ variables, respectively.
The characters for $\mathfrak{usp}(2N)$ and $\mathfrak{so}(2N+1)$ are given by determinant formulae \cite{MR1153249}:
\begin{align}
\chi_{\lambda}^{\mathfrak{usp}(2N)}&=\det(E_{\lambda_i'-i+j}-E_{\lambda_i'-i-j})_{1\leq i, j \leq l(\lambda')}\\
\chi_{\lambda}^{\mathfrak{so}(2N+1)}&=\det(\overline{H}_{\lambda_i-i+j}-\overline{H}_{\lambda_i-i-j})_{1\leq i, j \leq l(\lambda)}
\end{align}
where $\lambda'$ is the transpose partition of $\lambda$, and
\begin{align}
E_k&=e_k(s_1,\dots, s_N, s_1^{-1}, \dots, s_N^{-1})\\
\overline{H}_k&=h_k(s_1,\dots, s_N, s_1^{-1}, \dots, s_N^{-1},1)
\end{align}
The power sum is also written as
\begin{align}
P_k&=p_k(s_1,\dots, s_N, s_1^{-1},\dots, s_N^{-1}) \\
\overline{P}_k&=p_k(s_1,\dots, s_N, s_1^{-1},\dots, s_N^{-1},1)
\end{align}
where
\begin{align}
p_k(x_1,\dots, x_n)=\sum_{i=1}^n x_i^k
\end{align}
Newton's identity relates $e_k$ and $h_k$ to $p_k$. Combining these results, we get the character in terms of the power sum.
Here we show them,
\begin{align}
\chi_{\tiny \yng(1)}^{\mathfrak{usp}(2N)}&=P_1,& 
\chi_{\tiny \yng(1)}^{\mathfrak{so}(2N+1)}&=\overline{P}_1,\\
\chi_{\tiny \yng(2)}^{\mathfrak{usp}(2N)}&=\frac{P_2}{2}+\frac{P_1^2}{2},& 
\chi_{\tiny \yng(2)}^{\mathfrak{so}(2N+1)}&=\frac{\overline{P}_2}{2}+\frac{\overline{P}_1^2}{2}-1, \\
\chi_{\tiny \yng(1,1)}^{\mathfrak{usp}(2N)}&=-\frac{P_2}{2}+\frac{P_1^2}{2}-1,&
\chi_{\tiny \yng(1,1)}^{\mathfrak{so}(2N+1)}&=-\frac{\overline{P}_2}{2}+\frac{\overline{P}_1^2}{2},\\
\chi_{\tiny \yng(3)}^{\mathfrak{usp}(2N)}&=\frac{P_3}{3}+\frac{P_2P_1}{2}+\frac{P_1^3}{6},&
\chi_{\tiny \yng(3)}^{\mathfrak{so}(2N+1)}&=\frac{\overline{P}_3}{3}+\frac{\overline{P}_2\overline{P}_1}{2}+\frac{\overline{P}_1^3}{6}-\overline{P}_1, \\
\chi_{\tiny \yng(2,1)}^{\mathfrak{usp}(2N)}&=-\frac{P_3}{3}+\frac{P_1^3}{3}-P_1, &
\chi_{\tiny \yng(2,1)}^{\mathfrak{so}(2N+1)}&=-\frac{\overline{P}_3}{3}+\frac{\overline{P}_1^3}{3}-\overline{P}_1,\\
\chi_{\tiny \yng(1,1,1)}^{\mathfrak{usp}(2N)}&=\frac{P_3}{3}-\frac{P_2P_1}{2}+\frac{P_1^3}{6}-P_1,&
\chi_{\tiny \yng(1,1,1)}^{\mathfrak{so}(2N+1)}&=\frac{\overline{P}_3}{3}-\frac{\overline{P}_2\overline{P}_1}{2}+\frac{\overline{P}_1^3}{6}.
\end{align}
For $\mathfrak{so}(2N)$, the character $\chi_\lambda^{\mathfrak{so}(2N)}$ with $l(\lambda) <N$ can be expressed by $P_k$ in the same way.

\bibliographystyle{utphys}
\bibliography{ref}

\providecommand{\href}[2]{#2}\begingroup\raggedright\begin{thebibliography}{10}

\bibitem{Goddard:1976qe}
P.~Goddard, J.~Nuyts, and D.~I. Olive, ``{Gauge Theories and Magnetic
  Charge},''
\href{http://dx.doi.org/10.1016/0550-3213(77)90221-8}{{\em Nucl.Phys.}
  {\bfseries B125} (1977) 1}.

\bibitem{Kapustin:2005py}
A.~Kapustin, ``{Wilson-'t Hooft operators in four-dimensional gauge theories
  and S-duality},'' \href{http://dx.doi.org/10.1103/PhysRevD.74.025005}{{\em
  Phys. Rev. D} {\bfseries 74} (2006) 025005},
  \href{http://arxiv.org/abs/hep-th/0501015}{{\ttfamily arXiv:hep-th/0501015}}.

\bibitem{Aharony:2013hda}
O.~Aharony, N.~Seiberg, and Y.~Tachikawa, ``{Reading between the lines of
  four-dimensional gauge theories},''
  \href{http://dx.doi.org/10.1007/JHEP08(2013)115}{{\em JHEP} {\bfseries 08}
  (2013) 115}, \href{http://arxiv.org/abs/1305.0318}{{\ttfamily arXiv:1305.0318
  [hep-th]}}.

\bibitem{Maldacena:1998im}
J.~M. Maldacena, ``{Wilson loops in large N field theories},''
  \href{http://dx.doi.org/10.1103/PhysRevLett.80.4859}{{\em Phys. Rev. Lett.}
  {\bfseries 80} (1998) 4859--4862},
  \href{http://arxiv.org/abs/hep-th/9803002}{{\ttfamily arXiv:hep-th/9803002}}.

\bibitem{Rey:1998ik}
S.-J. Rey and J.-T. Yee, ``{Macroscopic strings as heavy quarks in large N
  gauge theory and anti-de Sitter supergravity},''
  \href{http://dx.doi.org/10.1007/s100520100799}{{\em Eur. Phys. J. C}
  {\bfseries 22} (2001) 379--394},
  \href{http://arxiv.org/abs/hep-th/9803001}{{\ttfamily arXiv:hep-th/9803001}}.

\bibitem{Drukker:2005kx}
N.~Drukker and B.~Fiol, ``{All-genus calculation of Wilson loops using
  D-branes},'' \href{http://dx.doi.org/10.1088/1126-6708/2005/02/010}{{\em
  JHEP} {\bfseries 02} (2005) 010},
  \href{http://arxiv.org/abs/hep-th/0501109}{{\ttfamily arXiv:hep-th/0501109}}.

\bibitem{Gomis:2006sb}
J.~Gomis and F.~Passerini, ``{Holographic Wilson Loops},''
  \href{http://dx.doi.org/10.1088/1126-6708/2006/08/074}{{\em JHEP} {\bfseries
  08} (2006) 074}, \href{http://arxiv.org/abs/hep-th/0604007}{{\ttfamily
  arXiv:hep-th/0604007}}.

\bibitem{Gomis:2006im}
J.~Gomis and F.~Passerini, ``{Wilson Loops as D3-Branes},''
  \href{http://dx.doi.org/10.1088/1126-6708/2007/01/097}{{\em JHEP} {\bfseries
  01} (2007) 097}, \href{http://arxiv.org/abs/hep-th/0612022}{{\ttfamily
  arXiv:hep-th/0612022}}.

\bibitem{Rodriguez-Gomez:2006fmx}
D.~Rodriguez-Gomez, ``{Computing Wilson lines with dielectric branes},''
  \href{http://dx.doi.org/10.1016/j.nuclphysb.2006.06.037}{{\em Nucl. Phys. B}
  {\bfseries 752} (2006) 316--326},
  \href{http://arxiv.org/abs/hep-th/0604031}{{\ttfamily arXiv:hep-th/0604031}}.

\bibitem{Yamaguchi:2007ps}
S.~Yamaguchi, ``{Semi-classical open string corrections and symmetric Wilson
  loops},'' \href{http://dx.doi.org/10.1088/1126-6708/2007/06/073}{{\em JHEP}
  {\bfseries 06} (2007) 073},
  \href{http://arxiv.org/abs/hep-th/0701052}{{\ttfamily arXiv:hep-th/0701052}}.

\bibitem{Yamaguchi:2006tq}
S.~Yamaguchi, ``{Wilson loops of anti-symmetric representation and
  D5-branes},'' \href{http://dx.doi.org/10.1088/1126-6708/2006/05/037}{{\em
  JHEP} {\bfseries 05} (2006) 037},
  \href{http://arxiv.org/abs/hep-th/0603208}{{\ttfamily arXiv:hep-th/0603208}}.

\bibitem{Hartnoll:2006hr}
S.~A. Hartnoll and S.~P. Kumar, ``{Multiply wound Polyakov loops at strong
  coupling},'' \href{http://dx.doi.org/10.1103/PhysRevD.74.026001}{{\em Phys.
  Rev. D} {\bfseries 74} (2006) 026001},
  \href{http://arxiv.org/abs/hep-th/0603190}{{\ttfamily arXiv:hep-th/0603190}}.

\bibitem{Witten:1998xy}
E.~Witten, ``{Baryons and branes in anti-de Sitter space},''
  \href{http://dx.doi.org/10.1088/1126-6708/1998/07/006}{{\em JHEP} {\bfseries
  07} (1998) 006}, \href{http://arxiv.org/abs/hep-th/9805112}{{\ttfamily
  arXiv:hep-th/9805112}}.

\bibitem{Fiol:2014fla}
B.~Fiol, B.~Garolera, and G.~Torrents, ``{Exact probes of orientifolds},''
  \href{http://dx.doi.org/10.1007/JHEP09(2014)169}{{\em JHEP} {\bfseries 09}
  (2014) 169}, \href{http://arxiv.org/abs/1406.5129}{{\ttfamily arXiv:1406.5129
  [hep-th]}}.

\bibitem{Giombi:2020kvo}
S.~Giombi and B.~Offertaler, ``{Wilson loops in $ \mathcal{N} $ = 4 SO(N) SYM
  and D-branes in AdS$_{5}$ \texttimes{}
  \ensuremath{\mathbb{R}}\ensuremath{\mathbb{P}}$^{5}$},''
  \href{http://dx.doi.org/10.1007/JHEP10(2021)016}{{\em JHEP} {\bfseries 10}
  (2021) 016}, \href{http://arxiv.org/abs/2006.10852}{{\ttfamily
  arXiv:2006.10852 [hep-th]}}.

\bibitem{Okuyama:2022lbn}
K.~Okuyama, ``{\textquoteright{}t Hooft expansion of SO(N) and Sp(N) $
  \mathcal{N} $ = 4 SYM revisited},''
  \href{http://dx.doi.org/10.1007/JHEP09(2022)064}{{\em JHEP} {\bfseries 09}
  (2022) 064}, \href{http://arxiv.org/abs/2207.09191}{{\ttfamily
  arXiv:2207.09191 [hep-th]}}.

\bibitem{Zhang:2023yus}
H.-Z. Zhang, W.-Z. Feng, and J.-B. Wu, ``{Holographic operator product
  expansion of loop operators in the super Yang-Mills theory*},''
  \href{http://dx.doi.org/10.1088/1674-1137/acd364}{{\em Chin. Phys. C}
  {\bfseries 47} no.~8, (2023) 083101},
  \href{http://arxiv.org/abs/2303.13225}{{\ttfamily arXiv:2303.13225
  [hep-th]}}.

\bibitem{Bergman:2022otk}
O.~Bergman and S.~Hirano, ``{The holography of duality in $ \mathcal{N} $ = 4
  Super-Yang-Mills theory},''
  \href{http://dx.doi.org/10.1007/JHEP11(2022)069}{{\em JHEP} {\bfseries 11}
  (2022) 069}, \href{http://arxiv.org/abs/2208.09396}{{\ttfamily
  arXiv:2208.09396 [hep-th]}}.

\bibitem{Dimofte:2011py}
T.~Dimofte, D.~Gaiotto, and S.~Gukov, ``{3-Manifolds and 3d Indices},''
  \href{http://dx.doi.org/10.4310/ATMP.2013.v17.n5.a3}{{\em Adv. Theor. Math.
  Phys.} {\bfseries 17} no.~5, (2013) 975--1076},
\href{http://arxiv.org/abs/1112.5179}{{\ttfamily arXiv:1112.5179 [hep-th]}}.

\bibitem{Gang:2012yr}
D.~Gang, E.~Koh, and K.~Lee, ``{Line Operator Index on $S^{1}\times S^{3}$},''
  \href{http://dx.doi.org/10.1007/JHEP05(2012)007}{{\em JHEP} {\bfseries 05}
  (2012) 007},
\href{http://arxiv.org/abs/1201.5539}{{\ttfamily arXiv:1201.5539 [hep-th]}}.

\bibitem{Drukker:2015spa}
N.~Drukker, ``{The $\mathcal{N}=4 $ Schur index with Polyakov loops},''
  \href{http://dx.doi.org/10.1007/JHEP12(2015)012}{{\em JHEP} {\bfseries 12}
  (2015) 012}, \href{http://arxiv.org/abs/1510.02480}{{\ttfamily
  arXiv:1510.02480 [hep-th]}}.

\bibitem{Cordova:2016uwk}
C.~Cordova, D.~Gaiotto, and S.-H. Shao, ``{Infrared Computations of Defect
  Schur Indices},'' \href{http://dx.doi.org/10.1007/JHEP11(2016)106}{{\em JHEP}
  {\bfseries 11} (2016) 106}, \href{http://arxiv.org/abs/1606.08429}{{\ttfamily
  arXiv:1606.08429 [hep-th]}}.

\bibitem{Hatsuda:2023iwi}
Y.~Hatsuda and T.~Okazaki, ``{Exact $ \mathcal{N} $ = 2$^{*}$ Schur line defect
  correlators},'' \href{http://dx.doi.org/10.1007/JHEP06(2023)169}{{\em JHEP}
  {\bfseries 06} (2023) 169}, \href{http://arxiv.org/abs/2303.14887}{{\ttfamily
  arXiv:2303.14887 [hep-th]}}.

\bibitem{Hatsuda:2023imp}
Y.~Hatsuda and T.~Okazaki, ``{Large N and large representations of Schur line
  defect correlators},'' \href{http://dx.doi.org/10.1007/JHEP01(2024)096}{{\em
  JHEP} {\bfseries 01} (2024) 096},
  \href{http://arxiv.org/abs/2309.11712}{{\ttfamily arXiv:2309.11712
  [hep-th]}}.

\bibitem{Hatsuda:2023iof}
Y.~Hatsuda and T.~Okazaki, ``{Excitations of bubbling geometries for line
  defects},'' \href{http://dx.doi.org/10.1103/PhysRevD.109.066013}{{\em Phys.
  Rev. D} {\bfseries 109} no.~6, (2024) 066013},
  \href{http://arxiv.org/abs/2311.13740}{{\ttfamily arXiv:2311.13740
  [hep-th]}}.

\bibitem{Garcia-Etxebarria:2015wns}
I.~n. Garcia-Etxebarria and D.~Regalado, ``{$ \mathcal{N}=3 $ four dimensional
  field theories},'' \href{http://dx.doi.org/10.1007/JHEP03(2016)083}{{\em
  JHEP} {\bfseries 03} (2016) 083},
  \href{http://arxiv.org/abs/1512.06434}{{\ttfamily arXiv:1512.06434
  [hep-th]}}.

\bibitem{Aharony:2016kai}
O.~Aharony, Y.~Tachikawa, and K.~Gomi, ``{S-folds and 4d N=3 superconformal
  field theories},'' \href{http://dx.doi.org/10.1007/JHEP06(2016)044}{{\em
  JHEP} {\bfseries 06} (2016) 044},
  \href{http://arxiv.org/abs/1602.08638}{{\ttfamily arXiv:1602.08638
  [hep-th]}}.

\bibitem{Drukker:2000ep}
N.~Drukker, D.~J. Gross, and A.~A. Tseytlin, ``{Green-Schwarz string in $AdS_5
  \times S^5$: Semiclassical partition function},''
  \href{http://dx.doi.org/10.1088/1126-6708/2000/04/021}{{\em JHEP} {\bfseries
  04} (2000) 021}, \href{http://arxiv.org/abs/hep-th/0001204}{{\ttfamily
  arXiv:hep-th/0001204}}.

\bibitem{Faraggi:2011bb}
A.~Faraggi and L.~A. Pando~Zayas, ``{The Spectrum of Excitations of Holographic
  Wilson Loops},'' \href{http://dx.doi.org/10.1007/JHEP05(2011)018}{{\em JHEP}
  {\bfseries 05} (2011) 018}, \href{http://arxiv.org/abs/1101.5145}{{\ttfamily
  arXiv:1101.5145 [hep-th]}}.

\bibitem{Faraggi:2011ge}
A.~Faraggi, W.~Mueck, and L.~A. Pando~Zayas, ``{One-loop Effective Action of
  the Holographic Antisymmetric Wilson Loop},''
  \href{http://dx.doi.org/10.1103/PhysRevD.85.106015}{{\em Phys. Rev. D}
  {\bfseries 85} (2012) 106015},
  \href{http://arxiv.org/abs/1112.5028}{{\ttfamily arXiv:1112.5028 [hep-th]}}.

\bibitem{Frobenius:1882uber}
G.~Frobenius, ``{\"{U}ber die elliptischen Funktionen zweiter},'' {\em Art, J.
  Reine Angew. Math} {\bfseries 93} (1882) 53--68.

\bibitem{MR0335789}
J.~D. Fay, {\em Theta functions on {R}iemann surfaces}.
\newblock Lecture Notes in Mathematics, Vol. 352. Springer-Verlag, Berlin-New
  York, 1973.

\bibitem{Du:2023kfu}
B.-n. Du, M.-x. Huang, and X.~Wang, ``{Schur indices for $ \mathcal{N} $ = 4
  super-Yang-Mills with more general gauge groups},''
  \href{http://dx.doi.org/10.1007/JHEP03(2024)009}{{\em JHEP} {\bfseries 03}
  (2024) 009}, \href{http://arxiv.org/abs/2311.08714}{{\ttfamily
  arXiv:2311.08714 [hep-th]}}.

\bibitem{Hayashi:2020ofu}
H.~Hayashi, T.~Okuda, and Y.~Yoshida, ``{ABCD of 't Hooft operators},''
  \href{http://dx.doi.org/10.1007/JHEP04(2021)241}{{\em JHEP} {\bfseries 04}
  (2021) 241}, \href{http://arxiv.org/abs/2012.12275}{{\ttfamily
  arXiv:2012.12275 [hep-th]}}.

\bibitem{Gaiotto:2008ak}
D.~Gaiotto and E.~Witten, ``{S-Duality of Boundary Conditions In N=4 Super
  Yang-Mills Theory},''
  \href{http://dx.doi.org/10.4310/ATMP.2009.v13.n3.a5}{{\em Adv. Theor. Math.
  Phys.} {\bfseries 13} no.~3, (2009) 721--896},
\href{http://arxiv.org/abs/0807.3720}{{\ttfamily arXiv:0807.3720 [hep-th]}}.

\bibitem{Hatsuda:2024lcc}
Y.~Hatsuda, H.~Lin, and T.~Okazaki, ``{Orbifold ETW brane and half-indices},''
  \href{http://dx.doi.org/10.1007/JHEP12(2024)227}{{\em JHEP} {\bfseries 12}
  (2024) 227}, \href{http://arxiv.org/abs/2409.16841}{{\ttfamily
  arXiv:2409.16841 [hep-th]}}.

\bibitem{Imamura:2024lkw}
Y.~Imamura, ``{Giant Graviton Expansions for the Line Operator Index},''
  \href{http://dx.doi.org/10.1093/ptep/ptae084}{{\em PTEP} {\bfseries 2024}
  no.~6, (2024) 063B03}, \href{http://arxiv.org/abs/2403.11543}{{\ttfamily
  arXiv:2403.11543 [hep-th]}}.

\bibitem{Imamura:2024pgp}
Y.~Imamura and M.~Inoue, ``{Brane expansions for anti-symmetric line operator
  index},'' \href{http://dx.doi.org/10.1007/JHEP08(2024)020}{{\em JHEP}
  {\bfseries 08} (2024) 020}, \href{http://arxiv.org/abs/2404.08302}{{\ttfamily
  arXiv:2404.08302 [hep-th]}}.

\bibitem{Beccaria:2024oif}
M.~Beccaria, ``{Schur line defect correlators and giant graviton expansion},''
  \href{http://dx.doi.org/10.1007/JHEP06(2024)088}{{\em JHEP} {\bfseries 06}
  (2024) 088}, \href{http://arxiv.org/abs/2403.14553}{{\ttfamily
  arXiv:2403.14553 [hep-th]}}.

\bibitem{Hatsuda:2024uwt}
Y.~Hatsuda, H.~Lin, and T.~Okazaki, ``{Giant graviton expansions and ETW
  brane},'' \href{http://dx.doi.org/10.1007/JHEP09(2024)181}{{\em JHEP}
  {\bfseries 09} (2024) 181}, \href{http://arxiv.org/abs/2405.14564}{{\ttfamily
  arXiv:2405.14564 [hep-th]}}.

\bibitem{Imamura:2024zvw}
Y.~Imamura, A.~Sei, and D.~Yokoyama, ``{Giant graviton expansion for general
  Wilson line operator indices},''
  \href{http://dx.doi.org/10.1007/JHEP09(2024)202}{{\em JHEP} {\bfseries 09}
  (2024) 202}, \href{http://arxiv.org/abs/2406.19777}{{\ttfamily
  arXiv:2406.19777 [hep-th]}}.

\bibitem{Lewis-Brown:2018dje}
C.~Lewis-Brown and S.~Ramgoolam, ``{BPS operators in $\mathcal{N}=4$ $SO(N)$
  super Yang-Mills theory: plethysms, dominoes and words},''
  \href{http://dx.doi.org/10.1007/JHEP11(2018)035}{{\em JHEP} {\bfseries 11}
  (2018) 035}, \href{http://arxiv.org/abs/1804.11090}{{\ttfamily
  arXiv:1804.11090 [hep-th]}}.

\bibitem{deMelloKoch:2024sdf}
R.~de~Mello~Koch, M.~Kim, and A.~L. Mahu, ``{A pedagogical introduction to
  restricted Schur polynomials with applications to heavy operators},''
  \href{http://dx.doi.org/10.1142/S0217751X24300035}{{\em Int. J. Mod. Phys. A}
  {\bfseries 39} no.~31, (2024) 2430003},
  \href{http://arxiv.org/abs/2409.15751}{{\ttfamily arXiv:2409.15751
  [hep-th]}}.

\bibitem{Berenstein:2022srd}
D.~Berenstein and S.~Wang, ``{BPS coherent states and localization},''
  \href{http://dx.doi.org/10.1007/JHEP08(2022)164}{{\em JHEP} {\bfseries 08}
  (2022) 164}, \href{http://arxiv.org/abs/2203.15820}{{\ttfamily
  arXiv:2203.15820 [hep-th]}}.

\bibitem{Yamaguchi:2006te}
S.~Yamaguchi, ``{Bubbling geometries for half BPS Wilson lines},''
  \href{http://dx.doi.org/10.1142/S0217751X07035070}{{\em Int. J. Mod. Phys. A}
  {\bfseries 22} (2007) 1353--1374},
  \href{http://arxiv.org/abs/hep-th/0601089}{{\ttfamily arXiv:hep-th/0601089}}.

\bibitem{Lunin:2006xr}
O.~Lunin, ``{On gravitational description of Wilson lines},''
  \href{http://dx.doi.org/10.1088/1126-6708/2006/06/026}{{\em JHEP} {\bfseries
  06} (2006) 026}, \href{http://arxiv.org/abs/hep-th/0604133}{{\ttfamily
  arXiv:hep-th/0604133}}.

\bibitem{DHoker:2007mci}
E.~D'Hoker, J.~Estes, and M.~Gutperle, ``{Gravity duals of half-BPS Wilson
  loops},'' \href{http://dx.doi.org/10.1088/1126-6708/2007/06/063}{{\em JHEP}
  {\bfseries 06} (2007) 063}, \href{http://arxiv.org/abs/0705.1004}{{\ttfamily
  arXiv:0705.1004 [hep-th]}}.

\bibitem{Hatsuda:2022xdv}
Y.~Hatsuda and T.~Okazaki, ``{$\mathcal{N} $ = 2$^{*}$ Schur indices},''
  \href{http://dx.doi.org/10.1007/JHEP01(2023)029}{{\em JHEP} {\bfseries 01}
  (2023) 029}, \href{http://arxiv.org/abs/2208.01426}{{\ttfamily
  arXiv:2208.01426 [hep-th]}}.

\bibitem{MR1659828}
N.~M. Katz and P.~Sarnak, \href{http://dx.doi.org/10.1090/coll/045}{{\em Random
  matrices, {F}robenius eigenvalues, and monodromy}}, vol.~45 of {\em American
  Mathematical Society Colloquium Publications}.
\newblock American Mathematical Society, Providence, RI, 1999.
\newblock \url{https://doi.org/10.1090/coll/045}.

\bibitem{MR1794267}
J.~P. Keating and N.~C. Snaith, ``Random matrix theory and {$L$}-functions at
  {$s=1/2$},'' \href{http://dx.doi.org/10.1007/s002200000262}{{\em Comm. Math.
  Phys.} {\bfseries 214} no.~1, (2000) 91--110}.
  \url{https://doi.org/10.1007/s002200000262}.

\bibitem{Ito:2011ea}
Y.~Ito, T.~Okuda, and M.~Taki, ``{Line operators on $S^1 \times \mathbb{R}^3$
  and quantization of the Hitchin moduli space},''
  \href{http://dx.doi.org/10.1007/JHEP03(2016)085}{{\em JHEP} {\bfseries 04}
  (2012) 010}, \href{http://arxiv.org/abs/1111.4221}{{\ttfamily arXiv:1111.4221
  [hep-th]}}. [Erratum: JHEP 03, 085 (2016)].

\bibitem{Brennan:2018yuj}
T.~D. Brennan, A.~Dey, and G.~W. Moore, ``{On \textquoteright{}t Hooft defects,
  monopole bubbling and supersymmetric quantum mechanics},''
  \href{http://dx.doi.org/10.1007/JHEP09(2018)014}{{\em JHEP} {\bfseries 09}
  (2018) 014}, \href{http://arxiv.org/abs/1801.01986}{{\ttfamily
  arXiv:1801.01986 [hep-th]}}.

\bibitem{Brennan:2018rcn}
T.~D. Brennan, A.~Dey, and G.~W. Moore, ``{\textquoteright{}t Hooft defects and
  wall crossing in SQM},''
  \href{http://dx.doi.org/10.1007/JHEP10(2019)173}{{\em JHEP} {\bfseries 10}
  (2019) 173}, \href{http://arxiv.org/abs/1810.07191}{{\ttfamily
  arXiv:1810.07191 [hep-th]}}.

\bibitem{Hayashi:2019rpw}
H.~Hayashi, T.~Okuda, and Y.~Yoshida, ``{Wall-crossing and operator ordering
  for 't Hooft operators in $\mathcal{N} $ = 2 gauge theories},''
  \href{http://dx.doi.org/10.1007/JHEP11(2019)116}{{\em JHEP} {\bfseries 11}
  (2019) 116}, \href{http://arxiv.org/abs/1905.11305}{{\ttfamily
  arXiv:1905.11305 [hep-th]}}.

\bibitem{MR1153249}
W.~Fulton and J.~Harris,
  \href{http://dx.doi.org/10.1007/978-1-4612-0979-9}{{\em Representation
  theory}}, vol.~129 of {\em Graduate Texts in Mathematics}.
\newblock Springer-Verlag, New York, 1991.
\newblock \url{https://doi.org/10.1007/978-1-4612-0979-9}.
\newblock A first course, Readings in Mathematics.

\bibitem{Dolan:2007rq}
F.~A. Dolan, ``{Counting BPS operators in N=4 SYM},''
  \href{http://dx.doi.org/10.1016/j.nuclphysb.2007.07.026}{{\em Nucl. Phys. B}
  {\bfseries 790} (2008) 432--464},
  \href{http://arxiv.org/abs/0704.1038}{{\ttfamily arXiv:0704.1038 [hep-th]}}.

\bibitem{Sei:2023fjk}
A.~Sei, ``{Character Expansion Methods for $\mathrm{USp}(2N)$,
  $\mathrm{SO}(n)$, and $\mathrm{O}(n)$ using the Characters of the Symmetric
  Group},'' \href{http://arxiv.org/abs/2303.03674}{{\ttfamily arXiv:2303.03674
  [hep-th]}}.

\bibitem{MR1507257}
F.~D. Murnaghan, ``On the {R}epresentations of the {S}ymmetric {G}roup,''
  \href{http://dx.doi.org/10.2307/2371574}{{\em Amer. J. Math.} {\bfseries 59}
  no.~3, (1937) 437--488}. \url{https://doi.org/10.2307/2371574}.

\bibitem{MR5729}
T.~Nakayama, ``On some modular properties of irreducible representations of a
  symmetric group. {I},'' {\em Jpn. J. Math.} {\bfseries 17} (1941) 165--184.

\bibitem{MR106711}
E.~P. Wigner, {\em Group theory and its application to the quantum mechanics of
  atomic spectra}.
\newblock Pure and Applied Physics, Vol. 5. Academic Press, New York-London,
  1959.

\bibitem{MR2761939}
G.~I. Lehrer and R.~B. Zhang,
  \href{http://dx.doi.org/10.1007/978-0-8176-4697-4￥_7}{``A {T}emperley-{L}ieb
  analogue for the {BMW} algebra,''} in {\em Representation theory of algebraic
  groups and quantum groups}, vol.~284 of {\em Progr. Math.}, pp.~155--190.
\newblock Birkh\"{a}user/Springer, New York, 2010.
\newblock \url{https://doi.org/10.1007/978-0-8176-4697-4_7}.

\bibitem{euler1767observationes}
L.~Euler, ``Observationes analyticae,'' {\em Novi Comm. Acad. Sci.
  Petropolitanae} no.~11, (1767) 124--143.

\bibitem{MR4023517}
A.~M. Perelomov, ``Euler's triangle and the decomposition of tensor powers of
  the adjoint {$\mathfrak{sl}(2)$}-module,''
  \href{http://dx.doi.org/10.1080/14029251.2020.1684001}{{\em J. Nonlinear
  Math. Phys.} {\bfseries 27} no.~1, (2020) 1--6}.
  \url{https://doi.org/10.1080/14029251.2020.1684001}.

\bibitem{MR1691863}
F.~R. Bernhart, ``Catalan, {M}otzkin, and {R}iordan numbers,''
  \href{http://dx.doi.org/10.1016/S0012-365X(99)00054-0}{{\em Discrete Math.}
  {\bfseries 204} no.~1-3, (1999) 73--112}.
  \url{https://doi.org/10.1016/S0012-365X(99)00054-0}.

\bibitem{MR2697358}
A.~Mihailovs, {\em A combinatorial approach to representations of {L}ie groups
  and algebras}.
\newblock ProQuest LLC, Ann Arbor, MI, 1998.
\newblock Thesis (Ph.D.)--University of Pennsylvania.

\bibitem{MR927758}
M.~de~Sainte-Catherine and G.~Viennot,
  \href{http://dx.doi.org/10.1007/BFb0072509}{``Enumeration of certain {Y}oung
  tableaux with bounded height,''} in {\em Combinatoire \'{e}num\'{e}rative
  ({M}ontreal, {Q}ue., 1985/{Q}uebec, {Q}ue., 1985)}, vol.~1234 of {\em Lecture
  Notes in Math.}, pp.~58--67.
\newblock Springer, Berlin, 1986.
\newblock \url{https://doi.org/10.1007/BFb0072509}.

\bibitem{MR2388243}
B.~W. Westbury, ``Invariant tensors for the spin representation of
  {$\mathfrak{so}(7)$},''
  \href{http://dx.doi.org/10.1017/S0305004107000722}{{\em Math. Proc. Cambridge
  Philos. Soc.} {\bfseries 144} no.~1, (2008) 217--240}.
  \url{https://doi.org/10.1017/S0305004107000722}.

\bibitem{MR3959676}
S.-j. Oh and T.~Scrimshaw, ``Identities from representation theory,''
  \href{http://dx.doi.org/10.1016/j.disc.2019.05.020}{{\em Discrete Math.}
  {\bfseries 342} no.~9, (2019) 2493--2541}.
  \url{https://doi.org/10.1016/j.disc.2019.05.020}.

\bibitem{MR1235279}
D.~J. Grabiner and P.~Magyar, ``Random walks in {W}eyl chambers and the
  decomposition of tensor powers,''
  \href{http://dx.doi.org/10.1023/A:1022499531492}{{\em J. Algebraic Combin.}
  {\bfseries 2} no.~3, (1993) 239--260}.
  \url{https://doi.org/10.1023/A:1022499531492}.

\bibitem{MR2555991}
K.~S. Kedlaya and A.~V. Sutherland,
  \href{http://dx.doi.org/10.1090/conm/487/09529}{``Hyperelliptic curves,
  {$L$}-polynomials, and random matrices,''} in {\em Arithmetic, geometry,
  cryptography and coding theory}, vol.~487 of {\em Contemp. Math.},
  pp.~119--162.
\newblock Amer. Math. Soc., Providence, RI, 2009.
\newblock \url{https://doi.org/10.1090/conm/487/09529}.

\bibitem{MR3502944}
G.~Lachaud, \href{http://dx.doi.org/10.1090/conm/663/13355}{``Frobenius
  distributions: {L}ang-{T}rotter and {S}ato-{T}ate conjectures,''} vol.~663 of
  {\em Contemp. Math.}, pp.~185--221.
\newblock Amer. Math. Soc., Providence, RI, 2016.
\newblock \url{https://doi.org/10.1090/conm/663/13355}.

\bibitem{Gunaydin:1984fk}
M.~Gunaydin and N.~Marcus, ``{The Spectrum of the $S^5$ Compactification of the
  Chiral N=2, D=10 Supergravity and the Unitary Supermultiplets of U(2,
  2/4)},'' \href{http://dx.doi.org/10.1088/0264-9381/2/2/001}{{\em Class.
  Quant. Grav.} {\bfseries 2} (1985) L11}.

\bibitem{Gunaydin:1984qu}
M.~Gunaydin, L.~J. Romans, and N.~P. Warner, ``{Gauged N=8 Supergravity in
  Five-Dimensions},''
  \href{http://dx.doi.org/10.1016/0370-2693(85)90361-2}{{\em Phys. Lett. B}
  {\bfseries 154} (1985) 268--274}.

\bibitem{MR1601666}
E.~Getzler and M.~M. Kapranov, ``Modular operads,''
  \href{http://dx.doi.org/10.1023/A:1000245600345}{{\em Compositio Math.}
  {\bfseries 110} no.~1, (1998) 65--126}.
  \url{https://doi.org/10.1023/A:1000245600345}.

\end{thebibliography}\endgroup

\end{document}